%% file: main.tex
\documentclass[12pt,authoryear]{elsarticle}

\usepackage{amssymb}
\usepackage{amsthm}
\usepackage{newtxmath}
\usepackage{subfiles}

\usepackage{mathtools}

\usepackage{enumerate}
\usepackage[shortlabels]{enumitem}

\usepackage{graphicx}
\usepackage{caption}
\usepackage{subcaption}
\usepackage{float}

\usepackage{tikz}
\usetikzlibrary{positioning}

\usepackage{algorithm}
\usepackage{algpseudocode}

\usepackage{hyperref}
\hypersetup{
    colorlinks=true,  % false: boxed links; true: colored links
    linkcolor=blue,    % color of internal links 
    citecolor=blue,   % color of links to bibliography
    filecolor=blue,   % color of file links
    urlcolor=blue,   % color of external links
    }
\usepackage{multibib}
\newcites{supp}{Reference}

\journal{Computational Statistics and Data Analysis}

\DeclareMathOperator{\Var}{Var}
\DeclareMathOperator{\Cov}{Cov}

\DeclareMathOperator{\tr}{tr}

\DeclareMathOperator{\Ga}{Ga}

\DeclareMathOperator{\etr}{etr}

\DeclareMathOperator{\Exp}{Exp}

\begin{document}

% Some self-defined abbreviations
\newcommand{\vzero}{\underline{0}}
\newcommand{\mzero}{\mathbf{0}}
\newcommand{\mZ}{\underline{Z}}
\newcommand{\mz}{\underline{z}}
\newcommand{\mtZ}{\tilde{\underline{Z}}}
\newcommand{\mtz}{\tilde{\underline{z}}}
\newcommand{\mbfZ}{\underline{\mathbf{Z}}}
\newcommand{\mbfz}{\underline{\mathbf{z}}}
\newcommand{\mbftZ}{\tilde{\underline{\mathbf{Z}}}}
\newcommand{\mbftz}{\tilde{\underline{\mathbf{z}}}}
\newcommand{\mX}{\underline{X}}
\newcommand{\mx}{\underline{x}}
\newcommand{\mbfX}{\underline{\mathbf{X}}}
\newcommand{\mbfY}{\underline{\mathbf{Y}}}
\newcommand{\tC}{\tilde{\boldsymbol{C}}}
\newcommand{\tD}{\tilde{\boldsymbol{D}}}
\newcommand{\tF}{\tilde{F}}
\newcommand{\me}{\underline{e}}
\newcommand{\mf}{\boldsymbol{f}}
\newcommand{\mft}{\tilde{\boldsymbol{f}}}
\newcommand{\mfC}{\underline{f}_{C}}
\newcommand{\mc}{\boldsymbol{c}}
\newcommand{\mctrue}{\mc_{0}}
\newcommand{\mct}{\underline{\mc}}
\newcommand{\mcttrue}{\mct_{0}}
\newcommand{\mcttil}{\underline{\tilde{\mc}}}
\newcommand{\mcttiltrue}{\underline{\tilde{\mc}}_{0}}
\newcommand{\mce}{\mc_{\eta}}
\newcommand{\mcetrue}{\mc_{0,\eta}}
\newcommand{\Cnull}{\bsC_{0,nd}}
\newcommand{\Calt}{\bsC_{nd}}
\newcommand{\Cone}{\bsC_{1,nd}}
\newcommand{\bbInd}{\varmathbb{1}}
\newcommand{\bbA}{\varmathbb{A}}
\newcommand{\bbB}{\varmathbb{B}}
\newcommand{\bbC}{\varmathbb{C}}
\newcommand{\bbE}{\varmathbb{E}}
\newcommand{\bbI}{\varmathbb{I}}
\newcommand{\bbL}{\varmathbb{L}}
\newcommand{\bbN}{\varmathbb{N}}
\newcommand{\bbR}{\varmathbb{R}}
\newcommand{\bbS}{\varmathbb{S}}
\newcommand{\bbbS}{\bar{\varmathbb{S}}}
\newcommand{\bbZ}{\varmathbb{Z}}
\newcommand{\bfa}{\mathbf{a}}
\newcommand{\bfZ}{\mathbf{Z}}
\newcommand{\bfz}{\mathbf{z}}
\newcommand{\bfW}{\mathbf{W}}
\newcommand{\bfGa}{\mathbf{\Gamma}}
\newcommand{\bfLa}{\mathbf{\Lambda}}
\newcommand{\bfPhi}{\mathbf{\Phi}}
\newcommand{\bfPsi}{\mathbf{\Psi}}
\newcommand{\bfSig}{\mathbf{\Sigma}}
\newcommand{\bfThe}{\mathbf{\Theta}}
\newcommand{\calB}{\mathcal{B}}
\newcommand{\bmcalB}{\bm{\mathcal{B}}}
\newcommand{\calA}{\mathcal{A}}
\newcommand{\calC}{\mathcal{C}}
\newcommand{\calD}{\mathcal{D}}
\newcommand{\calG}{\mathcal{G}}
\newcommand{\calH}{\mathcal{H}}
\newcommand{\calI}{\mathcal{I}}
\newcommand{\calJ}{\mathcal{J}}
\newcommand{\calL}{\mathcal{L}}
\newcommand{\calM}{\mathcal{M}}
\newcommand{\calS}{\mathcal{S}}
\newcommand{\bcalS}{\bar{\mathcal{S}}}
\newcommand{\calU}{\mathcal{U}}
\newcommand{\calX}{\mathcal{X}}
\newcommand{\calY}{\mathcal{Y}}
\newcommand{\bsa}{\boldsymbol{a}}
\newcommand{\bsta}{\tilde{\boldsymbol{a}}}
\newcommand{\bsA}{\boldsymbol{A}}
\newcommand{\bstA}{\tilde{\boldsymbol{A}}}
\newcommand{\bsB}{\boldsymbol{B}}
\newcommand{\bstB}{\tilde{\boldsymbol{B}}}
\newcommand{\bsC}{\boldsymbol{C}}
\newcommand{\bsD}{\boldsymbol{D}}
\newcommand{\bsE}{\boldsymbol{E}}
\newcommand{\bsF}{\boldsymbol{F}}
\newcommand{\bstF}{\tilde{\boldsymbol{F}}}
\newcommand{\bsH}{\boldsymbol{H}}
\newcommand{\bsI}{\boldsymbol{I}}
\newcommand{\bsJ}{\boldsymbol{J}}
\newcommand{\bsK}{\boldsymbol{K}}
\newcommand{\bsL}{\boldsymbol{L}}
\newcommand{\bsM}{\boldsymbol{M}}
\newcommand{\bsQ}{\boldsymbol{Q}}
\newcommand{\bsS}{\boldsymbol{S}}
\newcommand{\bsT}{\boldsymbol{T}}
\newcommand{\bsU}{\boldsymbol{U}}
\newcommand{\bsV}{\boldsymbol{V}}
\newcommand{\bsW}{\boldsymbol{W}}
\newcommand{\bsx}{\boldsymbol{x}}
\newcommand{\bsX}{\boldsymbol{X}}
\newcommand{\bsY}{\boldsymbol{Y}}
\newcommand{\bsZ}{\boldsymbol{Z}}
\newcommand{\bsThe}{\boldsymbol{\mathit{\Theta}}}

\newcommand{\frakR}{\mathfrak{R}}
\newcommand{\frakI}{\mathfrak{I}}
\newcommand{\frakB}{\mathfrak{B}}
\newcommand{\pa}{\text{pa}}
\newcommand{\Cauchy}{\text{Cauchy}}
\newcommand{\LN}{\text{LN}}
\newcommand{\diag}{\text{diag}}
\newcommand{\Unif}{\text{Unif}}
\newcommand{\Lip}{\text{Lip}}
\newcommand{\ualpha}{\underline{\alpha}}
\newcommand{\ubeta}{\underline{\beta}}
\newcommand{\uvphi}{\underline{\varphi}}
\newcommand{\calF}{\mathcal{F}}
\newcommand{\fp}{f_{\text{param}}}
\newcommand{\tilGap}{\tilde{\Gamma}_{nd,\pa}}
\newcommand{\lambdamax}{\lambda_{\max}}
\newcommand{\lambdamin}{\lambda_{\min}}

\newcommand{\lr}{\left(}
\newcommand{\rr}{\right)}
\newcommand{\ls}{\left[}
\newcommand{\rs}{\right]}
\newcommand{\lb}{\left\lbrace}
\newcommand{\rb}{\right\rbrace}
\newcommand{\la}{\left|}
\newcommand{\ra}{\right|}
\newcommand{\lV}{\left\Vert}
\newcommand{\rV}{\right\Vert}
\newcommand{\lv}{\left\vert}
\newcommand{\rv}{\right\vert}
\newcommand{\lf}{\left\lfloor}
\newcommand{\rf}{\right\rfloor}

%% New theorem, assumption, etc.
\newtheorem{mydef}{Definition}
\newenvironment{mandef}[1]
    {\renewcommand\themydef{#1}\mydef}{\endmydef}
\AfterEndEnvironment{mandef}{\noindent\ignorespaces}

\newtheorem{myassump}{Assumption}
\newenvironment{manassump}[1]
    {\renewcommand\themyassump{#1}\myassump}{\endmyassump}
\AfterEndEnvironment{manassump}{\noindent\ignorespaces}

\newtheorem{myprop}{Proposition}
\newenvironment{manprop}[1]
    {\renewcommand\themyprop{#1}\myprop}{\endmyprop}

\newtheorem{mylemma}{Lemma}
\newenvironment{manlemma}[1]
    {\renewcommand\themylemma{#1}\mylemma}{\endmylemma}
\AfterEndEnvironment{manlemma}{\noindent\ignorespaces}

\newtheorem{mycoro}{Corollary}
\newenvironment{mancoro}[1]
    {\renewcommand\themycoro{#1}\mycoro}{\endmycoro}
    
\newtheorem{myremark}{Remark}
\newenvironment{manremark}[1]
    {\renewcommand\themyremark{#1}\myremark}{\endmyremark}

\newenvironment{mypf}[2]{\proof[\textbf{Proof of {#1} {#2}.}]}{\endproof}
\AfterEndEnvironment{mypf}{\noindent\ignorespaces}

\newtheorem{myth}{Theorem}
\newtheorem{manualtheoreminner}{Theorem}
\newenvironment{manualtheorem}[1]{%
  \renewcommand\themanualtheoreminner{#1}%
  \manualtheoreminner
}{\endmanualtheoreminner}

\newcommand{\supplementarysection}{%
  \setcounter{figure}{0}% Reset figure counter
  \let\oldthefigure\thefigure% Capture figure numbering scheme
  \renewcommand{\thefigure}{S\oldthefigure}% Prefix figure number with S
  \section{Supplementary section}% Set supplementary section
\setcounter{table}{0}% Reset figure counter
  \let\oldthetable\thetable% Capture figure numbering scheme
  \renewcommand{\thetable}{S\oldthetable}% Prefix figure number with S
  }

\begin{frontmatter}



%\title{A nonparametrically corrected likelihood for Bayesian spectral analysis of multivariate time series}
\title{A nonparametrically corrected likelihood for Bayesian spectral analysis of multivariate time series}


\author[UoA]{Yixuan Liu}
\author[UM]{Claudia Kirch}
\author[UoA]{Jeong Eun Lee}
\author[UoA]{Renate Meyer}

\affiliation[UoA]{organization={Department of Statistics, The University of Auckland},%Department and Organization
            addressline={38 Princes Street}, 
            city={Auckland},
            postcode={1010}, 
            %state={a},
            country={New Zealand}}

\affiliation[UM]{organization={Institute of Mathematical Stochastics, Otto-von-Guericke University Magdeburg},%Department and Organization
            addressline={Universitatsplatz 2}, 
            city={Magdeburg},
            postcode={39106}, 
            %state={a},
            country={Germany}}

\begin{abstract}
This paper presents a novel approach to Bayesian nonparametric spectral analysis of stationary multivariate time series. Starting with a parametric vector-autoregressive model, the parametric likelihood is nonparametrically adjusted in the frequency domain to account for potential deviations from parametric assumptions. We show mutual contiguity of the nonparametrically corrected likelihood, the multivariate Whittle likelihood approximation and the exact likelihood for Gaussian time series. 
A multivariate extension of the nonparametric Bernstein-Dirichlet process prior for univariate spectral densities to  the space of Hermitian positive definite spectral density matrices is specified directly on the correction matrices. 
An infinite series representation of this prior is then used to develop a Markov chain Monte Carlo algorithm to sample from the posterior distribution. The code is made publicly available for ease of use and reproducibility. 
With this novel approach we provide a generalization of the multivariate Whittle-likelihood-based method of \cite{meier2020} as well as an extension of the nonparametrically corrected likelihood for univariate stationary time series of \cite{kirch2019} to the multivariate case. 
We demonstrate that the nonparametrically corrected likelihood  combines the efficiencies of a parametric with the robustness of a nonparametric model. Its numerical accuracy is illustrated  in a comprehensive simulation study. We illustrate its practical advantages by a spectral analysis of two environmental time series data sets: a bivariate time series of the Southern Oscillation Index and fish recruitment and  time series of windspeed data at six locations in California.

\end{abstract}




\end{frontmatter}


\section{Introduction}
\label{Se1}
\subfile{section1}

\section{Extension of the Multivariate Whittle Likelihood}
\label{Se2}
\subfile{section2}

\section{Bayesian spectral inference}
\label{Se3}
\subfile{section3}

\section{Simulation study}
\label{Se4}
\subfile{section4}

\section{Case studies}
\label{Se5}
\subfile{section5}

\section{Conclusion}
\label{Se6}
\subfile{section6}

\section*{Acknowledgements}

Y.L., C.K and R.M. acknowledge funding by DFG Grant KI 1443/3-2. R.M. gratefully acknowledges support by the Marsden grant MFP-UOA2131 from New Zealand Government funding, administered by the Royal Society Te Ap\={a}rangi. Y.L., J.L and R.M. thank the Centre for eResearch at the University of Auckland for their technical support.





%\clearpage 

\bibliographystyle{elsarticle-harv} 
\bibliography{reference}



%%\end{thebibliography}

\appendix

\section{More simulation results}\label{Append1}
\subfile{section1_supp}

\section{Proofs}\label{Append2}
\subfile{section2_supp}

\end{document}

%% file: section1.tex
Statistical models can be classified into \textit{parametric} and \textit{nonparametric} models. Parametric techniques assume a pre-specified model with a finite number of parameters from a particular family of distributions for the data. When the specification is correct, those techniques are powerful and  efficient. However, the results of parametric approaches are easily affected by misspecification and their advantages significantly diminished in this case. In contrast, nonparametric techniques do not need to make stringent parametric assumptions about the data generating process. As a result, nonparametric models which can have an infinite number of parameters are better at adapting to varying complexities of the data than parametric approaches. 

Parametric approaches -- in particular  vector autoregressive (VAR) models (Chapter 11 in \cite{brockwell1991}; Part I in \cite{lutkepohl2005}) -- have been broadly applied in both frequentist (\cite{meyer2015}; \cite{krampe2019}) and  Bayesian (\cite{koop2010}; \cite{kastner2020}) inference of multivariate time series. Many nonparametric frequentist approaches have been developed for multivariate time series analysis, especially in the frequency domain (\cite{berkowitz1998}; \cite{jentsch2010}; \cite{fiecas2014};  \cite{jentsch2015}), by extending various univariate bootstrap and resampling approaches. For instance \cite{franke1992}, \cite{kreiss2003}, \cite{kirch2011}  \cite{mcmurry2010} and  \cite{dai2004} generate bootstrap samples with a spectral estimate obtained by smoothing its Cholesky decomposition.

Bayesian nonparametric time series analysis has drawn more attention over the last couple of decades. Many techniques applied in the time domain are based on Hidden Markov models (\cite{fox2011}; \cite{fox2014}; \cite{aicher2019}) and factor models (\cite{carvalho2008}; \cite{rodriguez2011}; \cite{barigozzi2016}; \cite{kalli2018}). In the frequency domain, numerous methods have been developed that  use the Whittle likelihood (\cite{whittle1957})  for univariate stationary Gaussian time series such as \cite{choudhuri2004b}, \cite{rousseau2012}, \cite{cadonna2017}, \cite{kirch2019}, \cite{edwards2019}, \cite{rao2021} and \cite{russel2021}. The Whittle likelihood approximates the Gaussian likelihood by a pseudo-likelihood, making use of the asymptotically independent discrete Fourier coefficients that have variances equal to the spectral density at the corresponding Fourier frequencies.  The merits of using the Whittle likelihood  are its direct dependence on the spectral density and reduced computational costs compared to the true Gaussian likelihood. The Whittle likelihood  has also been extensively applied for the analysis of  multivariate stationary time series by diverse approaches such as \cite{rosen2007}, \cite{krafty2013}, \cite{cadonna2019}, \cite{meier2020} and \cite{hu2023}.

As per the preceding discussion, parametric approaches are more effective than nonparametric ones when the specified model closely matches the true data generating process. However, the performance of parametric methods will significantly degenerate when misspecifications occur. Therefore, we propose a novel method which combines the advantages of both parametric and nonparametric approaches for multivariate spectral density estimation of stationary Gaussian time series using the Bayesian framework. To elaborate, we specify a parametric model for the observations in the time domain (e.g.\ a VAR model) and propose a  nonparametric correction of the parametric likelihood in the frequency domain.
We show mutual contiguity of the multivariate Gaussian, the Whittle and the nonparametrically corrected likelihood.
In particular, when the parametric working model is independent and identical (i.i.d.) Gaussian white noise, the nonparametrically corrected likelihood reduces to the multivariate Whittle likelihood. Thus, with this new corrected likelihood, we effectively propose a novel generalisation of the multivariate Whittle likelihood. This idea has been applied in the univariate case by \cite{kirch2019}. Following the approach by \cite{choudhuri2004b}, they specified a Bernstein-Dirichlet process  prior on the spectral density. However, no direct extension of the Dirichlet process to the multivariate scenario \citep{meier2018} is available. Therefore, \cite{meier2020} proposed a Gamma process prior on the space of Hermitian positive definite (Hpd) matrix-valued functions. This prior has been successfully implemented along with the Whittle likelihood in their work for nonparametric Bayesian spectral inference.
Nonetheless, the case study in \cite{meier2020} showed that spectral density estimates based on the multivariate Whittle likelihood and the Hpd Gamma process are generally too smooth to capture peaks and sharp  features of the spectra. To address this shortcoming,  we adapt the Hpd Gamma process prior and  update it with our proposed generalised multivariate Whittle likelihood. This yields significant improvements. In particular, when the parametric likelihood specified in the time domain is sufficiently close to the true model, our procedure performs almost  as efficiently as the pure parametric procedure. On the other hand, in the case of model misspecification, our approach is almost as robust as the pure nonparametric approach. More importantly, in the partially misspecified case, our approach outperforms both nonparametric and parametric models.% when using a VAR order which is smaller than the order of the best-fitting VAR model.

The remainder of this article is organised as follows. Section \ref{Se2} proposes the nonparametrically corrected likelihood used for Bayesian spectral analysis of stationary multivariate time series and sates its asymptotic properties. Section \ref{Se3} specifies the prior and describes the posterior computation. Section \ref{Se4} provides the simulation result. Section \ref{Se5} demonstrates the application of our approach to two real case studies. Section \ref{Se6} gives the conclusion and an outlook for future work. The supplementary material contains the proofs and additional results of the simulation study. 

%% file: section2.tex
In this section, we will introduce a likelihood for multivariate stationary time series which nonparametrically corrects a parametric likelihood. It can be regarded as an extension of the multivariate Whittle likelihood. To this end, we first revisit the multivariate version of the Whittle likelihood, which was initially proposed by \cite{whittle1957} in the univariate case. 

\subsection{Review of the Whittle likelihood}
Let $\lbrace \mZ_{t}=(Z_{t1},\ldots,Z_{td})^T, t=0, 1,...\rbrace$ be a Gaussian stationary time series in $\bbR^{d}$ with zero mean and autocovariance matrix $\bfGa(h):=\bbE(\mZ_{t+h}\mZ_{t}^{T})=[\gamma_{ij}(h)]^{d}_{i,j=1}\in\bbR^{d\times d}$ for $h\in\bbZ$. Under the assumption that all autocovariance functions are absolutely summable, i.e., $\sum_{h=-\infty}^{\infty}|\gamma_{ij}(h)|<\infty$ for all $i$ and $j$, the spectral density matrix of $\lbrace\mZ_{t}\rbrace$ is given by the Fourier transform of the autocovariance matrix: 
\begin{equation}\label{eq2.1}
    \mf(\omega)=\left( \begin{array}{cccc}
    f_{11}(\omega) & f_{12}(\omega) & \ldots & f_{1d}(\omega)\\
    f_{21}(\omega) & f_{22}(\omega) & \ldots & f_{2d}(\omega)\\
    \vdots &\vdots& \ddots &\vdots\\
    f_{d1}(\omega) & f_{d2}(\omega) & \ldots & f_{dd}(\omega)\\
    \end{array}
    \right)=
    \frac{1}{2\pi}\sum_{h=-\infty}^{\infty}\bfGa(h)\exp(-ih\omega)
\end{equation}
for $0\leq\omega\leq2\pi$.
Thus, $\mf(\omega)$ is a Hermitian positive definite (Hpd) matrix and there is a one-to-one relationship between the spectral density and the covariance matrix. 
The diagonal elements of $\mf(\omega)$ are real-valued and equal to the marginal spectral densities of each of the $d$ univariate time series. The off-diagonal elements $f_{ij}$ for $i\neq j$ are complex-valued in general and referred to as the cross-spectra of $\{Z_{ti}, t=0, 1,...\}$
and  $\{Z_{tj}, t=0, 1,...\}$.
Consider a  sample $\mz_{1},...,\mz_{n}$ of $\lbrace\mZ_{t}\rbrace$ of size $n$ and denote by $\mbfz_{n}:=(\mz_{1}^{T},...,\mz_{n}^{T})^{T}\in\bbR^{nd}$ the vector of stacked observations. The discrete Fourier transform of $\mbfZ_{n}$ \begin{align}
    \mtZ_{j}:=\frac{1}{\sqrt{n}}\sum_{t=1}^{n}\mZ_{t}\exp(-it\omega_{j}),\quad\omega_{j}=\frac{2\pi j}{n}\ \text{and}\ j=0,...,\lf\frac{n}{2}\rf
\end{align} provides the \textit{Fourier coefficients} at the \textit{Fourier frequencies} $\omega_{j}$. It holds that $\mtZ_{j}=\mtZ_{n-j}^{*}$ for even $n$ and $\mtZ_{j}=\mtZ_{n-j+1}^{*}$ for odd $n$ when $j>\lfloor n/2\rfloor$, so the first $\lfloor n/2\rfloor$ Fourier coefficients contain the same information as $\mbfZ_{n}$. An important concept related to the Fourier coefficients is the \textit{periodogram matrix} (simply as \textit{periodogram} in the following) defined as $I_{nd}(\omega_{j}):=\frac{1}{2\pi}\mtZ_{j}\mtZ_{j}^{*}$. According to the well-known asymptotic properties of the Fourier coefficients \citep{brockwell1991}, the periodograms at different Fourier frequencies are  asymptotically independent, the expectations of the periodograms converge to the corresponding spectral density matrices with increasing sample size, that is \begin{align}
    \bbE I_{nd}(\omega_{j})\rightarrow\mf(\omega_{j}),\quad\text{as $n\rightarrow\infty$}
    \end{align}
    and the periodograms have  Exponential distributions with means $\mf(\omega_{j})$.
  Based on the asymptotic distribution of the Fourier coefficients, the Whittle likelihood $P_{W}^{n}$ is defined as \begin{align}\label{eq2.4}
    p_{W}^{n}(\mtz_{1},...,\mtz_{N}|\mf)=\prod_{j=1}^{N}\frac{1}{\pi^{d}|2\pi\mf(\omega_{j})|}\exp\lr-\frac{1}{2\pi}\mtz_{j}^{*}\mf(\omega_{j})^{-1}\mtz_{j}\rr
\end{align} for $j=1,...,N=\lceil n/2\rceil-1$.  For the purpose of spectral density estimation, the boundary Fourier frequencies $\omega_{0}=0$ and $\omega_{n/2}=\pi$ (the latter for $n$ even) are not considered in (\ref{eq2.4}) since the corresponding Fourier coefficients are proportional to the sample mean and the alternating sample mean, respectively. 
Under certain assumptions on the time series, the Whittle likelihood provides an approximation to the true likelihood but is exact only in the case of Gaussian white noise \citep{shaowu2007}. 
In contrast to the true Gaussian likelihood,
%which is not directly dependent on the spectral density matrix, (\ref{eq2.4}) indicates that 
the Whittle likelihood  depends directly on the spectral density matrix. Furthermore, an evaluation of the Whittle likelihood only requires the inversion of $n$ matrices of dimension $d\times d$  instead of the inversion of a  $nd\times nd$ matrix for the true Gaussian likelihood. This requires  $O(nd^{3})$ floating point operations, a significant reduction in computational costs compared to the evaluation of the true Gaussian likelihood.

\subsection{Nonparametric likelihood correction}

For {\em univariate}  time series, a nonparametrically corrected likelihood has already been proposed by \cite{kirch2019}.  
The idea is to begin with a parametric working likelihood in the time domain and  then apply a nonparametric correction  in the frequency domain. This is to ensure that the likelihood has the correct second-order dependence structure even if the parametric model is misspecified. Here we propose a nonparametrically corrected likelihood for {\em multivariate} time series.

To illustrate that the idea stems from the multivariate Whittle likelihood and thus provides an extension and refinement thereof, we first consider multivariate Gaussian white noise, i.e.\ $\mZ_{t} \stackrel{iid}{\sim} N_{d}(\mzero,\bsI_{d})$, as the parametric working model. Note that the spectral density matrix of this parametric working model is $\frac{1}{2\pi}\bsI_{d}$. However, if the model is misspecified, i.e., in the case of coloured noise, the estimate of the spectral density matrix based on this model may not be efficient. Therefore, we nonparametrically correct for potential model misspecification in the frequency domain by multiplying the Fourier coefficients by the square root of a correction matrix and back-transform to the time domain via a discrete inverse Fourier transform. In the case of a white noise working model, the \textit{correction matrix}  is defined by the block diagonal matrix \begin{align*}
    C_{nd,W}:=(2\pi)^{-1/2}\begin{pmatrix}
    \mf^{1/2}(\omega_{1}) & &\\
    & \ddots &\\
    & & \mf^{1/2}(\omega_{n})
    \end{pmatrix}.
\end{align*} 
This is illustrated in the diagram below where $\bsF_{nd}$
denotes the \textit{Fourier transform operator}  such that $\bsF_{nd}\mbfZ_{n}=\mbftZ_{n}$:
\begin{center} \begin{tikzpicture}

%n1
\node[] (n1) at (0,0){$\mbfZ_{n}\sim\text{parametric working model}$};

\draw (n1.north) 
    node[above]{\large \underline{time domain}};
 
%n2
\node [right=3cm of n1]  (n2) {$\bsF_{nd}\mbfZ_{n}$};

\draw (n2.north east) node[above]{\large\underline{frequency domain}};
 
% n3
\node [below=3cm of n2] (n3) {$C_{nd}\bsF_{nd}\mbfZ_{n}$};
 
% n4
\node [left=4cm of n3
]  (n4) {$\bsF_{nd}^{T}C_{nd}\bsF_{nd}\mbfZ_{n}$};
 
% Arrows with text label
\draw[-stealth] (n1.east) -- (n2.west)
    node[midway,above]{FT};
 
\draw[-stealth] (n2.south) -- (n3.north) 
    node[midway,right]{$C_{nd}$};

\draw[-stealth] (n3.west) -- (n4.east) node[midway,above]{$\text{FT}^{-1}$};

\end{tikzpicture}
\end{center} 
 As is readily verified, the density induced by the above linear transformation with $C_{nd}=C_{nd,W}$ under the Gaussian white noise working model turns out to be the Whittle likelihood defined in (\ref{eq2.4}), up to the boundary frequencies. 

This derivation of the Whittle likelihood as a nonparametrically corrected parametric likelihood provides a way towards a generalization by starting with a more adequate parametric working model than Gaussian white noise. We suggest using a VAR model since the dependence structure of any causal Gaussian linear process can be captured by a VAR model of sufficiently large order. However, any other parametric time series model could be used. Denote by $\mf_{\pa}$ the spectral density matrix corresponding to the parametric model.

Then the general correction matrix $C_{nd}$ is defined as \begin{align}
    \bsC_{nd}:=\begin{pmatrix}
    \mf^{1/2}(\omega_{1})\mf_{\pa}^{-1/2}(\omega_{1}) & & &\\
    & \mf^{1/2}(\omega_{2})\mf_{\pa}^{-1/2}(\omega_{2}) & &\\
    & & \ddots &\\
    & & & \mf^{1/2}(\omega_{n})\mf_{\pa}^{-1/2}(\omega_{n})
    \end{pmatrix}.
\end{align} 
Denoting the parametric likelihood of $\mbfZ_{n}$  by $p_{\pa}^{n}$, the corrected likelihood is given by \begin{align}\label{eq2.6}
    p_{C}^{n}(\mbfZ_{n}|\mf,\underline{\bsB})\propto|\bsC_{nd}\bsC_{nd}^{*}|^{-1/2}p_{\pa}^{n}(\bsF_{nd}^{*}\bsC_{nd}^{-1}\bsF_{nd}\mbfZ_{n}|\underline{\bsB}),
\end{align} with $\underline{\bsB}$ being the parameters for the parametric working model.

The following theorem states the properties of the corrected likelihood. Most importantly,   the corrected likelihood is equivalent to the parametric likelihood if the parametric model is the true data-generating model and the periodograms under the corrected likelihood are asymptotically unbiased for the true spectral density matrix regardless of whether the parametric model is correctly specified or not.

\begin{manualtheorem}{1}\label{Th1}
Let $\mbfZ_{n}\in\bbR^{nd}$ be a stationary time series with zero mean and spectral density $\mf$. Let $\bfGa_{nd,\pa}$ and $\mf_{\pa}$ be the autocovariance matrix and the spectral density matrix of a parametric model used in the correction likelihood. \begin{enumerate}[label=(\alph*)]
    \item If $\mf=\mf_{\pa}$, then $p^{n}_{C}=p_{\pa}^{n}$.
    
    \item Under the corrected likelihood, the periodogram is an asymptotically unbiased estimator for the true spectral density matrix, i.e. \begin{align*}
        \bbE_{p^{n}_{C}}I_{nd,\omega_{j}}(\mbfZ_{n})=\mf(\omega_{j})+o(1)
    \end{align*} holds for $j=0,...,N$.
    
    \item The covariance matrix of the periodogram ordinates under the corrected likelihood is related to the covariance matrix under the parametric likelihood by \begin{align*}
        &\Cov_{p_{C}^{n}}\lr I_{nd,\omega_{j}}(\mbfZ_{n}),I_{nd,\omega_{k}}(\mbfZ_{n})\rr\\
        =&\Cov\lr\bsC_{nd,j} I_{nd,\omega_{j}}(\mbfY_{n})\bsC_{nd,j}^{*},\bsC_{nd,k} I_{nd,\omega_{k}}(\mbfY_{n})\bsC_{nd,k}^{*}\rr
    \end{align*} for $j,k=0,...,N$, where $\bsC_{nd,j}$ is the $j$-th diagonal block of $\bsC_{nd}$.
\end{enumerate}
\end{manualtheorem}

Please refer to \ref{Append2} for the proof of Theorem \ref {Th1}.
\bigskip

\subsection{Mutual contiguity}

In this section, we will derive an asymptotic property of the corrected likelihood for {\em Gaussian} time series, which is useful e.g.\ for the purpose of studying posterior consistency  as in \cite{kirch2019} and \cite{meier2020}. More precisely, we will prove  mutual contiguity of the corrected likelihood, the Whittle likelihood and the full Gaussian likelihood. This is important since mutual contiguity of two probability measures defined on  measurable spaces that change with the sample size $n$ implies that those measures have the same asymptotic properties. The main usefulness of showing mutual contiguity of the corrected likelihood, the Whittle and the true Gaussian likelihood is that if one can prove consistency or find the rate of convergence of an estimator under the Whittle measure, then the estimator is also consistent and has the same rate of convergences under the true Gaussian measure.
In the univariate case, mutual contiguity between the true Gaussian likelihood and the Whittle likelihood  was proved by \cite{choudhuri2004a}, in which Le Cam's first lemma (see Lemma 6.4 in \cite{vandervaart1998}) was used. \cite{meier2020} extended the univariate contiguity to the multivariate case, so it remains to establish  mutual contiguity between the corrected likelihood and the Whittle likelihood. To this end, we first state the definition of mutual contiguity and  specify the following two assumptions, which are the generalisation of Assumption A.1 in \cite{kirch2019} to the multivariate case.

Two probability measures $P_{n}$ and $Q_{n}$ on measurable spaces  $\Omega_{n}$ are {\em mutually contiguous} if for every sequence of measurable sets $A_{n}$, $P_{n}(A_{n})\rightarrow0$ implies $Q_{n}(A_{n})\rightarrow0$ and versa vice (cf.\ Definition 6.3 in \cite{vandervaart1998}).

\begin{manassump}{1}\label{A1}
    For the parametric working model, the eigenvalues of the spectral density matrix $\mf_{\pa}(\cdot)$ are uniformly bounded and uniformly bounded away from $0$. That is, there exist positive constants $b_{0, \pa}$ and $b_{1, \pa}$ such that \begin{equation*}
        \lambda_{\min}(\mf_{\pa}(\omega))\geq b_{0,\pa},\quad \lambda_{\max}(\mf_{\pa}(\omega))\leq b_{1,\pa},\quad 0\leq\omega\leq\pi.
    \end{equation*} Moreover, the autocovariance function matrix $\bfGa_{\pa}(\cdot)$ fulfills \begin{equation*}
        \sum_{h\in\bbZ}\Vert\bfGa_{\pa}(h)\Vert|h|^{a}<\infty,
    \end{equation*} for some $a>1$.
\end{manassump}

\begin{manassump}{2}\label{A2}
    The eigenvalues of the underlying true spectral density matrix $\mf(\cdot)$ are uniformly bounded and uniformly bounded away from $0$. That is, there exist positive constants $b_{0}$ and $b_{1}$ such that \begin{equation*}
        \lambda_{\min}(\mf(\omega))\geq b_{0},\quad \lambda_{\max}(\mf(\omega))\leq b_{1},\quad 0\leq\omega\leq\pi.
    \end{equation*} Moreover, the true autocovariance function matrix $\bfGa(\cdot)$ fulfills \begin{equation*}
        \sum_{h\in\bbZ}\Vert\bfGa(h)\Vert|h|^{a}<\infty,
    \end{equation*} for some $a>1$.
\end{manassump} Assumption \ref{A2} was used by \cite{meier2020} to establish the mutual contiguity between the exact Gaussian likelihood and the Whittle likelihood. The restriction on the Frobenius norm of $\bfGa(\cdot)$ ensures the continuous differentiability of $\mf$ with derivative being H\"{o}lder of order $a-1>0$. Assumption \ref{A1} makes the same restrictions on the parametric working model. Under these two assumptions, we can establish the mutual contiguity.

\begin{manualtheorem}{2}\label{Th2}

Let $\mbfZ_{n}\in\bbR^{nd}$ be a time series with zero mean, spectral density matrix $\mf$ and autocovariance matrix $\bfGa_{nd}$ fulfilling Assumption \ref{A2}. Let $\mf_{\pa}$ and $\bfGa_{nd,\pa}$ be the spectral density matrix and autocovariance matrix of a Gaussian parametric model fulfilling Assumption \ref{A1}. Then the true joint density $P^{n}$, the joint density $P^{n}_{W}$ under the Whittle likelihood and the joint density $P^{n}_{C}$ under the corrected likelihood equipped with the Gaussian parametric working model of $\mbfZ_{n}$ are mutually contiguous.

\end{manualtheorem}

The proof of Theorem \ref{Th2} is contained in \ref{Append2}.

%% file: section3.tex
\subsection{Prior specification}
For a nonparametric Bayesian approach to spectral density estimation,  a nonparametric prior on the space of Hpd matrix-valued functions is required. To this end, we employ the Bernstein-Hpd-Gamma prior introduced by \cite{meier2020}, which combines the Bernstein polynomial and the Hpd-Gamma process. Let $\bbbS_{d}^{+}$ be the set of $d\times d$ Hermitian positive semidefinite (Hpsd) matrices with unit trace and $\bcalS^{+}_{d}\cong\bbbS_{d}^{+}\times(0,\infty)$. Let $r\in(0,\infty)$ be the radial part and $\bsU\in\bbbS_{d}^{+}$ be the spherical part such that any $\bsZ\in\bcalS_{d}^{+}$ can be decomposed as $\bsZ=r\bsU$. The Hpd-Gamma distribution $\bsX\sim\Ga_{d\times d}(\alpha,\beta)$ is defined via its Laplace transform as \begin{align*}
     \bbE\etr(\bsX\bfThe)=\exp\lb-\int_{\bbbS_{d}^{+}}\int_{0}^{\infty}(1-\etr(-r\bfThe\bsU))\frac{\exp(-\beta(\bsU)r)}{r}dr\alpha(d\bsU)\rb
\end{align*} for all $\bfThe\in\bcalS_{d}^{+}$ with $\alpha$ on $\bbbS_{d}^{+}$ and $\beta:\bbbS_{d}^{+}\rightarrow(0,\infty)$, where $\etr(\bsX):=\exp(\tr(\bsX))$. When $d=1$, the Hpd-Gamma distribution is the same as the univariate Gamma distribution, so $\alpha$ and $\beta$ can be regarded as an extension of the scale and rate parameters. Then a random process $\bfPhi\sim\text{GP}_{d\times d}(\alpha,\beta)$ is called a Hpd-Gamma process on $\calX=[0,\pi]$ if any arbitrary numbers of non-empty partitions of $\calX$ are jointly Hpd-Gamma distributed. However, instead of specifying the prior on $\mf$ directly, we specify the prior on the Hpd correction  matrix defined as \begin{align}\label{eq3.1}
    \bsQ(\omega):=\mf_{\pa}^{-1/2}(\omega)\mf(\omega)\mf_{\pa}^{-1/2}(\omega).
\end{align} For $k\in\bbN$, we equidistantly partition $[0,\pi]$ into $k$ equidistant intervals as \begin{align}\label{eq3.2}
    I_{j,k}=((j-1)\pi/k,j\pi/k],\quad j=1,..,k.
\end{align} The \textit{Bernstein-Hpd-Gamma prior} for $\bsQ(\cdot)$ is defined by \begin{align}\label{eq5.3}
    \bsQ(\pi x):=\sum_{j=1}^{k}\bfPhi(I_{j,k})b(x|j,k-j+1),\quad 0\leq x\leq1,\quad\text{and}\quad k\sim p(k),
\end{align} where $b(\cdot|j,k-j+1)$ denotes the  probability density function of a Beta$(j,k-j+1)$ distribution. The parameter $k$ is a smoothness parameter, the smaller $k$, the smoother the resulting matrix-valued function $Q$. By putting a discrete prior probability function $p(k)$ on the polynomial degree $k$, we achieve a data-driven choice of the degree of smoothness.

In the univariate case, a mixture of Bernstein polynomials with weights defined  by a Dirichlet process was used by \cite{choudhuri2004b} and \cite{kirch2019}. Such a  Dirichlet process is a normalized Gamma process (see Section 1.2 in \cite{meier2018}), so the Bernstein-Hpd-Gamma prior can be regarded as a multivariate extension of the Bernstein-Dirichlet process prior on the space of univariate spectral densities to that of spectral density matrices. Analogous to the famous stick-breaking representation of the Dirichlet process, Theorem 4.4 in \cite{meier2020} shows that under weak assumptions $\bfPhi$ can be represented by an almost surely convergent series involving i.i.d. components as \begin{align}\label{eq3.4}
    &\bfPhi\overset{\text{a.s.}}{=}\sum_{j=1}^{\infty}\delta_{x_{j}}r_{j}\bsU_{j},\quad(x_{j},\bsU_{j})\overset{i.i.d.}{\sim}\alpha^{*}\nonumber\\
   & r_{j}=\rho^{-}(w_{j}|C_{\alpha},\beta(x_{j},\bsU_{j})),\quad w_{j}=\sum_{i=1}^{j}v_{i},\quad v_{i}\overset{i.i.d.}{\sim}\Exp(1),
\end{align} where \begin{align*}
    &\alpha^{*}:=\alpha/C_{\alpha},\quad  C_{\alpha}:=\int_{\calX}\alpha(x,\bbbS_{d}^{+})dx,\\
    &\rho^{-}(w|a,b)=\inf\lbrace r>0:\rho([r,\infty]|a,b)<w\rbrace,\quad\rho(dr|a,b)=a\frac{\exp(-br)}{r}dr.
\end{align*} A truncation of the infinite series allows the design of a MCMC algorithm to sample from the posterior distribution as described in Section \ref{subsec:postcomp}.

By putting a prior on the parameters of the parametric working model, both parametric and  nonparametric components of the model can be inferred simultaneously. This has been considered in the univariate corrected likelihood procedure by \cite{kirch2019}, in which the working model is an autoregressive (AR) time series model and the prior is put on  the partial autocorrelations to ensure stationarity and causality. Similarly, in the multivariate case, we can use the VAR time series model as our parametric working model. This parametric model has also been used for the multiple hybrid bootstrap by \cite{jentsch2010}. To elaborate, recall that the equation of a VAR(p) model with $p\geq1$ and Gaussian innovation is defined as \begin{align}\label{eq3.6}
    \mZ_{t}=\sum_{j=1}^{p}\bsB_{j}\mZ_{t-j}+\underline{\varepsilon}_{t},\quad \underline{\varepsilon}_{t}\overset{i.i.d.}{\sim}N_{d}(\vzero,\bfSig),\quad t=p+1,...,n
\end{align} with innovation covariance $\bfSig\in\calS_{d}^{+}(\bbR)$ and the coefficient matrices \begin{align*}
    \bsB_{j}=\begin{pmatrix}
       B_{j,1,1} & \cdots & B_{j,1,d}\\
       \vdots & \ddots & \vdots\\
       B_{j,d,1} & \cdots & B_{j,d,d}
    \end{pmatrix}
\end{align*} $\bsB_{1},...,\bsB_{p}\in\bbR^{d\times d}$ with $\bsB_{p}\neq\mzero$. Consider a vector \begin{align*}
    \underline{\beta}=(B_{1,1,1},...,B_{1,1,d},B_{2,1,1},...,B_{2,1,d},...,B_{p,1,d},B_{1,2,1},...,B_{p,d,d})^{T}\in\bbR^{pd^{2}},
\end{align*} which stacks all the VAR coefficients of (\ref{eq3.6}). In the implementation, the VAR order $p$ is  predetermined by some order selection criteria, such as Akaike's Information Criterion (AIC) by \cite{akaike1974} or the elbow criterion (see Section \ref{Se6}). For the coefficients, we assume \begin{align}
    \underline{\beta}\sim N(\boldsymbol{\mu}_{\beta},\bsV_{\beta}),
\end{align} so that the prior is parametrised by $\boldsymbol{\beta}$ and $\bsV_{\beta}$. 
%For the priori applied to those two hyperparameters,
We use a noninformative prior  by setting $\boldsymbol{\mu}_{\beta}=\mzero\in\bbR^{pd^{2}}$ and $\bsV_{\beta}^{-1}\approx\mzero\in\bbR^{pd^{2}\times pd^{2}}$.
\bigskip

\subsection{Posterior computation}\label{subsec:postcomp}

To implement an MCMC algorithm, we truncate the representation of $\bfPhi$ in (\ref{eq3.4}) at some large $L$ such that $ \bfPhi\approx\sum_{l=1}^{L}\delta_{x_{l}}r_{l}\bsU_{l}$. 
The value of $L$ should be determined by considering the sample size and the error tolerance in practical computations. A conservative truncation choice is $L=\max\lbrace20, n^{1/3}\rbrace$. In fact, that choice is used in the stick-breaking representation of the Dirichlet process found by \cite{muliere1998} and \cite{choudhuri2004b}. As a result, the posterior distribution combined with the prior of $k$ and the corrected likelihood (\ref{eq2.6}) follows \begin{align}\label{eq3.5}
    p_{C}(\underline{\Theta}_{\bfPhi},k|\mbfZ_{n},\underline{\bsB})\propto p_{C}^{n}(\mbfZ_{n}|\underline{\Theta}_{\bfPhi},k,\underline{\bsB})p(\underline{\Theta}_{\bfPhi})p(k)p(\underline{\bsB})
\end{align} with $\underline{\Theta}_{\bfPhi}:=(x_{1},...,x_{L},r_{1},...,r_{L},\bsU_{1},...,\bsU_{L})$being the parameter space of $\bfPhi$ is composed of $3L$ parameters and $\underline{\bsB}=(\bsB_{1},...,\bsB_{p})$ containing the parameters of the VAR parametric working model with order $p\geq1$.

\begin{manremark}{1}
In the univariate case, for reasons of identifiability, it is important to use a parametric working model with standardised errors, i.e., with $\sigma=1$, because all choices for $\sigma$ lead to the same corrected likelihood. This eliminates the need of  pre-estimating or putting a prior on the error variance. However, the situation is more involved in the multivariate case as two different covariance matrices can lead to different corrected likelihoods. This is because two covariance matrices can differ by more than just a multiplicative constant and, indeed, two correlation matrices can have a very different structure. Based on some pilot simulation studies, putting a prior on the innovation term results in  a high computational demand and can cause difficulty in posterior convergence (possibly due to some additional non-identifiability beyond multiplicative constants). Therefore, we pre-estimate the covariance of the innovation term and do not update it in the MCMC algorithm. The simulation results and the case studies in Sections \ref{Se5} and \ref{Se6} show that this implementation with a fixed (pre-estimated) innovation covariance  performs well.
\end{manremark}

\subsection{A Metropolis-within-Gibbs algorithm}

We use  Metropolis-Hastings (MH) steps (\cite{hastings1970}) to update the full conditionals in the Gibbs sampler, a Metropolis-within-Gibbs algorithm (\cite{tierney1994}). To elaborate, we will use the same algorithm to infer $\bfPhi$ and $k$ as \cite{meier2020}, in which $\bsU_{j}$'s are reparametrised by some hyperspherical coordinates $\varphi_{j}$'s. The multivariate normal random walk is used for the full conditional for the coefficients of the parametric working model. Algorithm \ref{Algo1} summarises the algorithm of sampling from the posterior. For more details, please see \url{https://github.com/easycure1/vnpctest}. 

\begin{algorithm}[H]
{\scriptsize
\caption{MCMC algorithm}\label{Algo1}
\begin{algorithmic}
\State\textbf{Initialise} $k^{(1)}, (r_{1}^{(1)},...,r_{L}^{(1)}),(x_{1}^{(1)},...,x_{L}^{(1)}), (\varphi_{1}^{(1)},...,\varphi_{L}^{(1)})$ and 
\State $\qquad\qquad\ \ (\bsB_{1}^{(1)},...,\bsB_{p}^{(1)})$.
\State \textbf{Start Gibbs sampling}
\While{$i\leq N$}
\State Propose $k^{*,(i+1)}$ using random walk restricted by a sufficiently upper limit $k_{\max}$.
\State\ \ \ Update $k^{(i+1)}$ in the MH-step by calculating (\ref{eq3.5}).
\While{$l\leq L$}
\State Propose $r_{l}^{*,(i+1)}$ using the log-normal random walk with the scale parameter updated during
\State the burn-in period.
\State\ \ \  Update $r_{l}^{(i+1)}$ in the MH-step by calculating (\ref{eq3.5}).
\EndWhile
\While{$l\leq L$}
\State Propose $x_{l}^{*,(i+1)}$ using the uniform random walk with scale parameter $\delta_{x_{l}}:=\frac{\pi l}{l+2\sqrt{n}}$.
\State\ \ \ Update $x_{l}^{(i+1)}$ in the MH-step by calculating (\ref{eq3.5}).
\EndWhile
\While{$l\leq L$}
\State Propose $\underline{\varphi}_{l}^{*,(i+1)}$ using the uniform random walk with the scale parameter updated during the
\State burn-in period.
\State\ \ \ Update $\underline{\varphi}_{l}^{(i+1)}$ in the MH-step by calculating (\ref{eq3.5}).
\EndWhile
\While{$j\leq p$}
\State Propose $\underline{\bsB}_{j}^{*,(i+1)}$ using the multivariate normal random walk with the scale parameter updated
\State during the burn-in period.
\State\ \ \ Update $\underline{\bsB}_{j}^{(i+1)}$ in the MH-step by calculating (\ref{eq3.5}).
\EndWhile
\EndWhile
\end{algorithmic}}
\end{algorithm}

%% file: section4.tex
In the simulation study, we compare the corrected likelihood approach to the parametric VAR approach and the nonparametric Whittle likelihood approach proposed by \cite{meier2020} in the case of a correctly specified parametric model and a misspecified one. To this end, we first introduce some criteria used for quantifying the estimation performance, and the corresponding results for the posterior will be summarised in Section \ref{Se4.3}.  The implementation of this approach is available at \url{https://github.com/easycure1/vnpctest} while the other two approaches are included in the \texttt{R} package \texttt{beyondWhittle} on CRAN (\cite{beyondwhittle2022}).

Consider a posterior spectral density matrix sample $\mf^{(1)}(\omega),...,\mf^{(M)}(\omega)$ at any given frequency $\omega$. Let $\calS_{d}^{+}$ be the set of $d\times d$ Hpd matrices. Consider the mapping $\calH:\calS_{d}\rightarrow\bbR^{d\times d}$ \footnote{CK: It might be easier to give the formula instead of a text explanation that goes over 4 lines} such that the real parts above the diagonal of an Hpd matrix $\bsA$ will remain above the diagonal of $\calH\bsA$, the imaginary parts above the diagonal of $\bsA$ will be assigned to below the diagonal of $\calH\bsA$ and the diagonal entries of $\bsA$ stay in the same position in $\calH\bsA$. Denote the image of the sample $\mf^{(j)}(\omega)$ via $\calH$ by $\underline{h}^{(j)}:=\calH\mf^{(j)}(\omega)$ for $j=1,...,M$. Assume that $\underline{h}^{(j)}$ has entries $\lbrace h_{1}^{(j)},...,h_{d^{2}}^{(j)}\rbrace$. Let $\underline{\hat{h}}$ be the median of all $\underline{h}^{(1)},...,\underline{h}^{(M)}$ so that $\hat{h}_{k}$ is the median of $\lbrace h_{k}^{(1)},...,h_{k}^{(M)}\rbrace$ for $k=1,...,d^{2}$. Then we call $\hat{\mf}_{0}(\omega):=\calH^{-1}\underline{\hat{h}}$ the \textit{pointwise median spectral density matrix} of $\mf^{(1)}(\omega),...,\mf^{(M)}(\omega)$. In the following contents, $\hat{\mf_{0}}$ will be regularly used as a Bayes estimator for the true spectral density matrix $\mf_{0}$. To measure the goodness of $\hat{\mf}_{0}$, we consider the corresponding $\bsL_{1}$- and $\bsL_{2}$-errors.

We are also interested in the uncertainty of the posterior estimate. To this end, based on the posterior sample, we construct \textit{pointwise $90\%$ credible regions} for the spectral density matrix, where pointwise refers to the frequency.  Additionally, we construct \textit{uniform $90\%$ credible regions} as follows: Denote by $\hat{\sigma}_{k}(\omega)$ the median absolute deviation of $\lbrace h_{k}^{(1)}(\omega),..., h_{k}^{(M)}(\omega)\rbrace$ for $0\leq\omega\leq\pi$ and $k=1,...,d^{2}$. Let $\hat{\sigma}:=(\hat{\sigma}_{1},...,\hat{\sigma}_{2})$. We consider the smallest positive number $\xi_{0.9}$ such that \begin{align*}
    \frac{1}{M}\sum_{j=1}^{M}\bbInd\lb\underset{\underset{k=1,...,d^{2}}{0\leq\omega\leq\pi}}{\max}\frac{\la h_{k}^{(j)}(\omega)-\hat{h}_{k}(\omega)\ra}{\hat{\sigma}(\omega)}\leq\xi_{0.9}\rb\geq0.9.
\end{align*} Let $\underline{\hat{h}}^{[0.05]}:=\underline{\hat{h}}-\xi_{0.9}\underline{\hat{\sigma}}$ and $\underline{\hat{h}}^{[0.95]}:=\underline{\hat{h}}+\xi_{0.9}\underline{\hat{\sigma}}$. As well, let $\tilde{\mf}_{0}^{[0.05]}:=\calH^{-1}\underline{\hat{h}}^{[0.05]}$ and $\tilde{\mf}_{0}^{[0.95]}:=\calH^{-1}\underline{\hat{h}}^{[0.95]}$. Then, the region between $\tilde{\mf}_{0}^{[0.05]}$ and $\tilde{\mf}_{0}^{[0.95]}$, i.e. \begin{align*}
    \calC_{\text{uni}}(\omega|0.9):=\lb t\tilde{\mf}_{0}^{[0.05]}(\omega)+(1-t)\tilde{\mf}_{0}^{[0.95]}(\omega):\ 0\leq t\leq1\rb,\quad 0\leq\omega\leq\pi
\end{align*} is the uniform $90\%$ credible region.

\subsection{Prior choice}\label{Se4.1}

The prior probability of $k$ is chosen as $p(k)\propto\exp(-0.01k\log k)$. For practical feasibility, we employ the $A\Gamma$-process $A\Gamma(\eta,\omega,\bfSig)$ (see Remark 4.1 in \cite{meier2020} for details) as the prior for $\bfPhi$. In particular, the parameters of $A\Gamma(\eta,\omega,\bfSig)$ are chosen as $\eta=d$, $\omega=d$ and $\bfSig=10^{4}\bsI_{d}$. Furthermore, to improve the robustness at the boundary, we employ the truncated Bernstein prior \begin{align*}
    \bsQ_{\tau_{l}}^{\tau_{r}}(\pi x):=\sum_{j=1}^{k}\bfPhi(I_{j,k})b_{\tau_{l}}^{\tau_{r}}(x|j,k-j+1),\quad 0\leq x\leq1,\quad\text{and}\quad k\sim p(k)
\end{align*} with $\tau_{l}=0.1$ and $\tau_{r}=0.9$.

In the implementation, we run $80,000$ iterations for the Markov Chain, and the first $30,000$ iterations are discarded after the burn-in period. The remaining $50,000$ samples are thinned by taking every $5$-th value in order to reduce the amount of memory used.
We set $k_{\max}=300$ and $L=20$.

\subsection{Simulated data}

The following two bivariate models are used to generate the simulated multivariate time series: \begin{enumerate}[label=\alph*)]
    \item VAR2 model \begin{align}\label{eq4.1}
    \mZ_{t}=\begin{pmatrix}
       0.5 & 0\\
       0 & -0.3
    \end{pmatrix}\mZ_{t-1}+\begin{pmatrix}
       0 & 0\\
       0 & -0.5
    \end{pmatrix}\mZ_{t-2}+\underline{e}_{t},\quad\underline{e}_{t}\overset{iid}{\sim}N\lr\mzero,\begin{pmatrix}
       1 & 0.9\\
       0.9 & 1
    \end{pmatrix}\rr;
\end{align}

    \item VMA(1) model \begin{align}\label{eq4.2}
    \mZ_{t}=\underline{e}_{t}+\begin{pmatrix}
       -0.75 & 0.5\\
       0.5 & 0.75
    \end{pmatrix}\underline{e}_{t-1},\quad\underline{e}_{t}\overset{iid}{\sim}N\lr\mzero,\begin{pmatrix}
       1 & 0.5\\
       0.5 & 1
    \end{pmatrix}\rr.
\end{align}
\end{enumerate} These two models have also been considered by \cite{meier2020} for the comparison between VNP and VAR. The VMA(1) model is a simple linear multivariate time series example, and it does not belong to the family of VAR models, which are used by the VAR procedure and the parametric working model for our VNPC procedure. We use a fixed small order for the parametric working model of the VNPC procedure while  the order of the parametric VAR approach is selected by AIC.

\begin{figure}[H]
    \centering
    \begin{subfigure}[b]{.65\textheight}
        \includegraphics[width=1\textwidth, height=0.2\textheight]{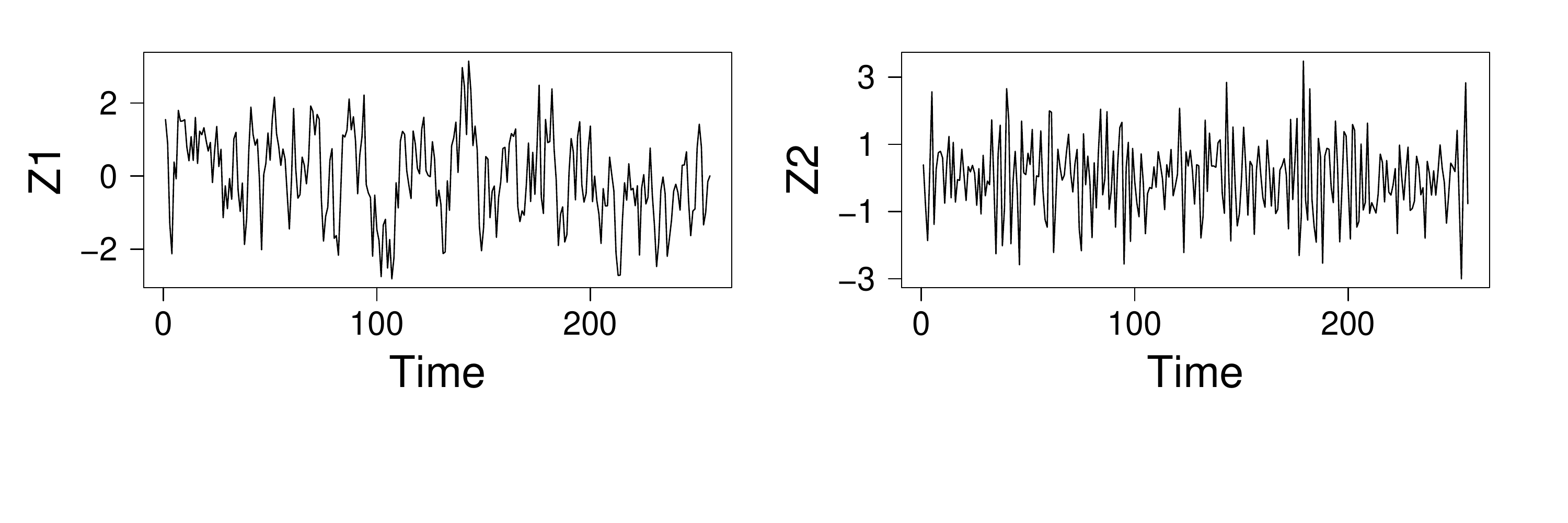}
        \vspace{-3\baselineskip}
        \caption{}
    \end{subfigure}

    \begin{subfigure}[b]{0.65\textheight}
        \includegraphics[width=1\textwidth, height=0.2\textheight]{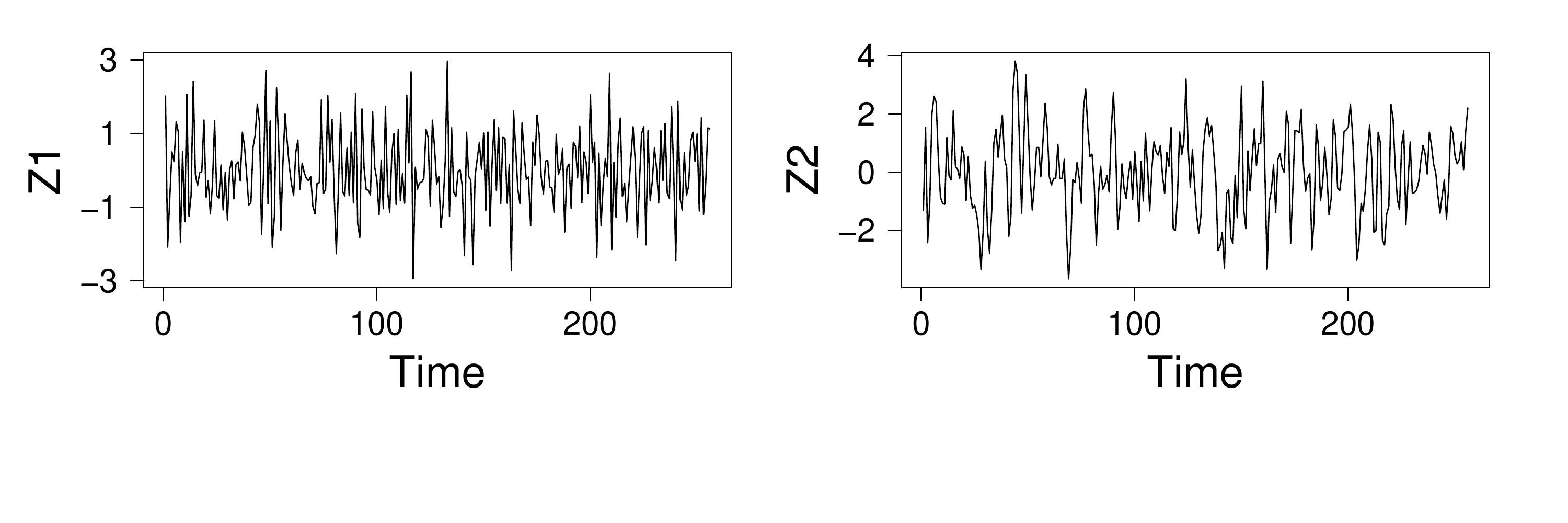}
        \vspace{-3\baselineskip}
        \caption{}
    \end{subfigure}
    \caption{Realisation of (a) the VAR(2) model (\ref{eq4.1}) and (b) the VMA(1) model (\ref{eq4.2}) of length $n=256$. The left panel displays the first component and the right panel  the second component of the time series.}
    \label{Fig4.1}
\end{figure}

We consider three different sample sizes, $n=256, n=512$ and $n=1024$. For each sample size, $N=500$ independent realisations are generated for both the VAR(2) and VMA(1) models. One realisation of length $n=256$ of each model (\ref{eq4.1}) and (\ref{eq4.2}) is given in Figure \ref{Fig4.1}. 

\subsection{Results}\label{Se4.3}

The results of the simulation study are shown in Table \ref{Tab4.1}. As to be expected, for the correctly specified VAR(2) model, the parametric VAR procedure has much smaller $L_{1}$- and $L_{2}$-errors than the other two nonparametric procedures. The VNP and the VNPC procedures show almost the same performance in this case. For the 
data generated from a VMA(1) model, a large order $p$ was  selected by AIC to fit a VAR($p$) model in this misspecified case. The estimates based on  the VNPC and the VNP model have smaller $L_{1}$- and $L_{2}$-errors than the VAR model. The VNPC model outperforms the nonparametric VNP estimate as it can benefit from the partially correct VAR(1) working model.

Table \ref{Tab4.1} also shows the empirical coverage of uniform $90\%$ credible regions and the median (among replications) of the median (over all frequencies) width of uniform $90\%$ credible regions for all procedures. Not surprisingly, for the correctly specified case, the empirical coverage of the VAR credible intervals attain the nominal level. In the misspecified VMA(1) case, the coverage of the VAR procedure is even larger due to the enlarged width. The VNP procedure produces similar coverages with smaller widths in either of the two cases. As discussed in \cite{szabo2015}, to produce  \textit{honest} (in the sense that their coverage matches (approximately) the corresponding posterior probability mass of the regions, $90\%$ in our case) credible sets in the nonparametric setting, the prior has to be selected to slightly undersmooth. However, the Bernstein polynomials used in the mixture prior in the VNP procedure tend to oversmooth the true spectral density, so it might not be appropriate for achieving an honest credible region. As a comparison, the VNPC procedure enhances the coverage sufficiently to almost $90\%$ for both  cases. 

\begin{table}[ht]
    \centering
    {\scriptsize	
    \begin{tabular}{cccccccccc}
         \hline
         \multicolumn{4}{c}{} & \multicolumn{2}{c}{VAR(2) model} & \multicolumn{4}{c}{}\\
         \multicolumn{1}{c}{} & \multicolumn{3}{c}{$n=256$} & \multicolumn{3}{c}{$n=512$} & \multicolumn{3}{c}{$n=1024$}\\
         \cline{2-10} 
         & VNPC(1) & VNP & VAR & VNPC(1) & VNP & VAR & VNPC(1) & VNP & VAR\\
         $L_{1}$-error & 0.099 & 0.106 & 0.075 & 0.076 & 0.081 & 0.051 & 0.059 & 0.063 & 0.036\\
         $L_{2}$-error & 0.130 & 0.136 & 0.099 & 0.101 & 0.106 & 0.067 & 0.080 & 0.084 & 0.047\\
         Coverage & 0.826 & 0.548 & 0.908 & 0.718 & 0.374 & 0.898 & 0.616 & 0.348 & 0.886\\
         Width $\mf_{11}$ & 0.341 & 0.314 & 0.210 & 0.177 & 0.168 & 0.121 & 0.109 & 0.104 & 0.078\\
         Width $\frakR\mf_{12}$ & 0.258 & 0.235 & 0.146 & 0.171 & 0.159 & 0.086 & 0.119 & 0.113 & 0.054\\
         Width $\frakI\mf_{12}$ & 0.212 & 0.187 & 0.122 & 0.142 & 0.128 & 0.076 & 0.097 & 0.090 & 0.051\\
         Width $\mf_{22}$ & 0.499 & 0.445 & 0.196 & 0.267 & 0.257 & 0.110 & 0.166 & 0.166 & 0.071\\
         \hline
         \multicolumn{4}{c}{} & \multicolumn{2}{c}{VMA(1) model} & \multicolumn{4}{c}{}\\
         \multicolumn{1}{c}{} & \multicolumn{3}{c}{$n=256$} & \multicolumn{3}{c}{$n=512$} & \multicolumn{3}{c}{$n=1024$}\\
         \cline{2-10} 
         & VNPC(1) & VNP & VAR & VNPC(1) & VNP & VAR & VNPC(1) & VNP & VAR\\
         $L_{1}$-error & 0.081 & 0.097 & 0.156 & 0.060 & 0.072 & 0.121 & 0.049 & 0.055 & 0.093\\
         $L_{2}$-error & 0.102 & 0.117 & 0.189 & 0.076 & 0.087 & 0.144 & 0.062 & 0.065 & 0.110\\
         Coverage & 0.888 & 0.594 & 0.980 & 0.876 & 0.518 & 0.972 & 0.690 & 0.294 & 0.966\\
         Width $\mf_{11}$ & 0.346 & 0.299 & 1.313 & 0.200 & 0.194 & 0.661 & 0.129 & 0.135 & 0.406\\
         Width $\frakR\mf_{12}$ & 0.232 & 0.211 & 0.609 & 0.139 & 0.141 & 0.414 & 0.087 & 0.099 & 0.287\\
         Width $\frakI\mf_{12}$ & 0.172 & 0.128 & 0.454 & 0.101 & 0.101 & 0.311 & 0.066 & 0.082 & 0.218\\
         Width $\mf_{22}$ & 0.498 & 0.462 & 1.890 & 0.299 & 0.301 & 1.008 & 0.193 & 0.213 & 0.625\\
         \hline
    \end{tabular}}
    \caption{Average $L_{1}$-, $L_{2}$-errors, empirical coverages and median width of uniform $90\%$ credible regions of the VNPC estimates with fixed parametric working model order $1$, the VNP,  and the VAR estimates with order selected by AIC for $N=500$ realisations of the VAR(2) model (\ref{eq4.1}) and the VMA(1) model (\ref{eq4.2}).}
    \label{Tab4.1}
\end{table}

Spectral density estimates obtained
by the three procedures for one random realisation of each of the two models are displayed in \ref{Append1}. Moreover, an additional simulation study in \ref{Append1} shows that the VNPC procedure 
using a Gaussian VAR working model performs well even in the misspecified case when the data-generating process is non-Gaussian.

%% file: section5.tex
Here we demonstrate the proposed Bayesian nonparametric approach based on the corrected multivariate likelihood on a bivariate time series  that has been extensively studied in the literature and a six-dimensional times series of windspeed measurements. Section \ref{Se5.1} first introduces a method to choose the order of the parametric VAR($p$) working model of the VNPC approach. Both studies use the same prior settings as the simulation study (see Section \ref{Se4.1}).

\subsection{Elbow criterion} \label{Se5.1}

Under some weak assumptions, the VAR($p$) model with sufficiently large order $p$ can capture the dependence structure of a Gaussian linear process to any degree of accuracy. Accordingly, using a VAR model with a sufficiently large order can be regarded as a nonparametric procedure. This result has been employed by the VAR-sieve-bootstrap method proposed by \cite{meyer2015}. In addition, that result also implies that the misspecified VAR model can still be used for the spectral density matrix estimation if the order is chosen to be sufficiently large. This explains why standard model selection techniques such as AIC tend to return a large order when misspecification occurs. 
This can lead to increased computational time, while also potentially causing the procedure to exhibit excessive confidence in the parametric model. Hence, we want to use a relatively small order for the parametric working model in our corrected likelihood procedure that is capable of describing the main features but not minute detail of the data. With this objective in mind, we follow the order selection approach used by the univariate corrected likelihood procedure in \cite{kirch2019}, which applies a rule-of-thumb rule to check for a bend or elbow in a plot of model order versus  negative maximum log-likelihood evaluated at the least squares estimates. To elaborate, in most cases we will see an obvious bend in this scree-like plot as the order increases. The penalty term of standard penalisation techniques is tuned to find the best fit but too small to find the elbow at which only the most obvious features are explained.
%might miss that bend due to the slow decay. 
In analogy to Principal Components Analysis (PCA), which uses the scree plot to reduce the dimension, we can also use the order where we see a clear bend in the negative maximum log-likelihood plot. With the corresponding autoregressive order, the main features of the data are captured and increasing the number of parameters further will contribute very little to the corrected likelihood procedure except computational costs.

\subsection{Southern oscillation index}

\begin{figure}[H]
    \centering
    \includegraphics[width=1\linewidth, height=0.18\textheight]{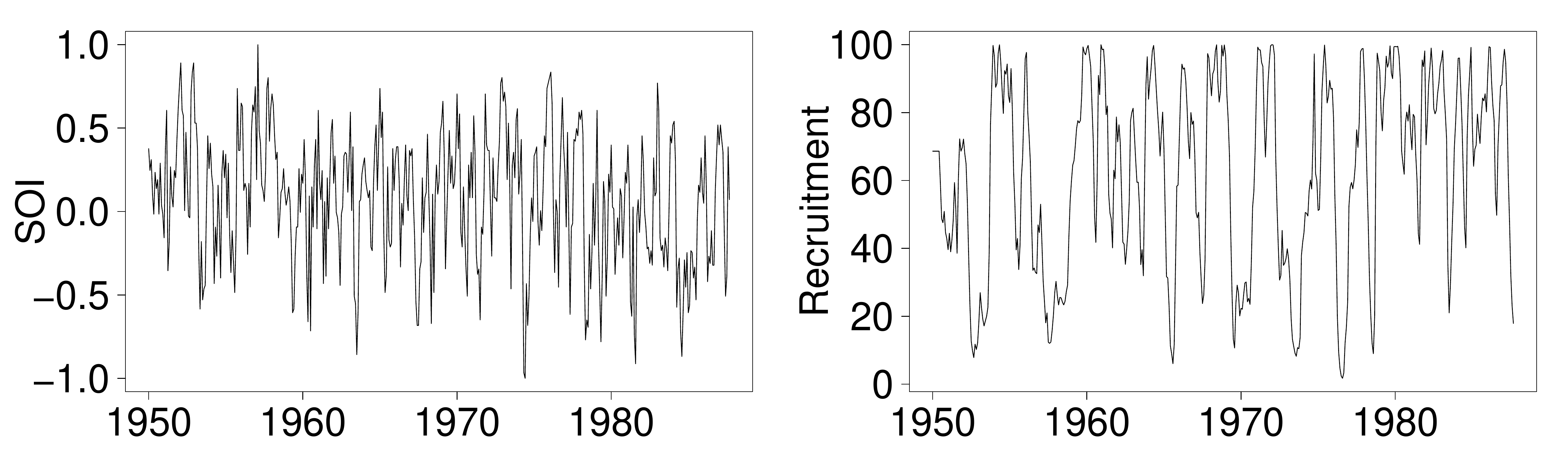}
    \caption{The SOI and recruitment series}
    \label{Fig5.1}
\end{figure}

The Southern Oscillation Index (SOI) and the associated recruitment series are two simultaneously recorded series used to explore the El Ni\~{n}o cycle (\cite{shumway2011}). The SOI is  a standardised index of sea level air pressure differences between Tahiti and Darwin, Australia. The index is related to the El Ni\~{n}o Southern Oscillation (ENSO), the movement of equatorial sea surface temperature across the Pacific Ocean and its atmospheric response. ENSO is known to be periodically occurring every 2-7 years and last 6-18 months, and it contains three phases: neutral, El Ni\~{n}o (the warm phase) and La Ni\~{n}a (the cool phase) (\cite{noaa2022}; \cite{statsnz2020}). ENSO is an extremely important meteorological movement in the Pacific Ocean as it has an impact on the weather through changes in air pressure, sea temperature and wind direction. Due to the periodic behaviour of ENSO, the observations in the SOI series are expected to have a time-dependent structure. For our analysis, we make the same simplifying assumption as \cite{rosen2007} that all effects (e.g.\ the seasonal effect) can be represented by the autocovariance structure. The recruitment series is composed of the number of newly spawned fish, 
%The data in the dataset $\texttt{rec}$ has been
re-scaled to the interval $[0,100]$ as shown in Figure \ref{Fig5.1}. It is believed that the fish spawn is associated with the water temperature so we can expect a cross-dependence between the SOI and the recruitment series. Both series consist of a period of 453 monthly values over the years 1950-1987, which are shown in Figure \ref{Fig5.1}. The data is available in the \texttt{R} package \texttt{astsa} as datasets \texttt{soi} and \texttt{rec} (\cite{stoffer2022}). 

\begin{figure}[H]
    \centering
    \includegraphics[width=.9\linewidth, height=0.4\textheight]{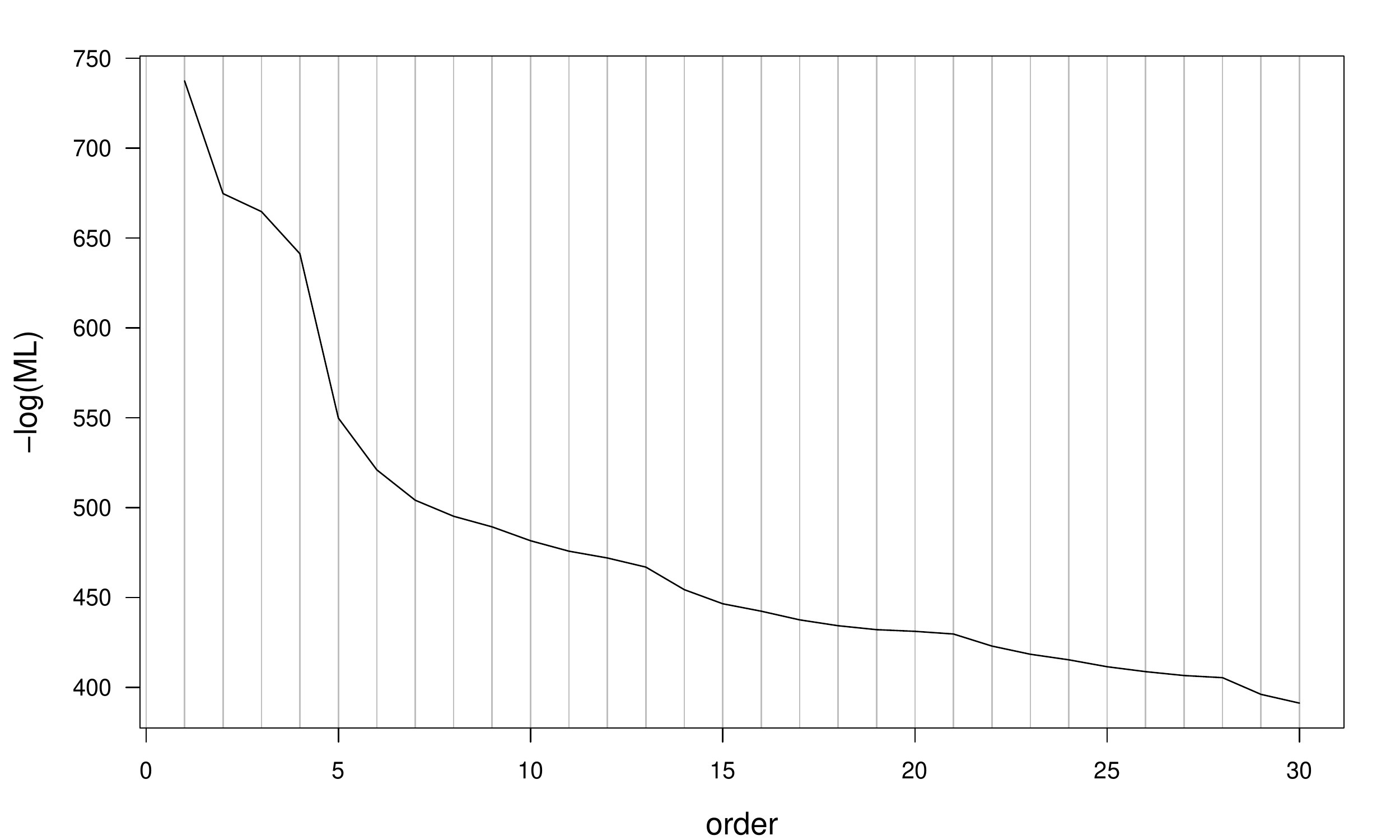}
    \caption{Negative maximum log likelihood for different VAR(p) models applied to the SOI and Recruitment series.}
    \label{Fig5.2}
\end{figure}

\begin{figure}[H]
    \centering
    \includegraphics[width=.9\linewidth, height=0.35\textheight]{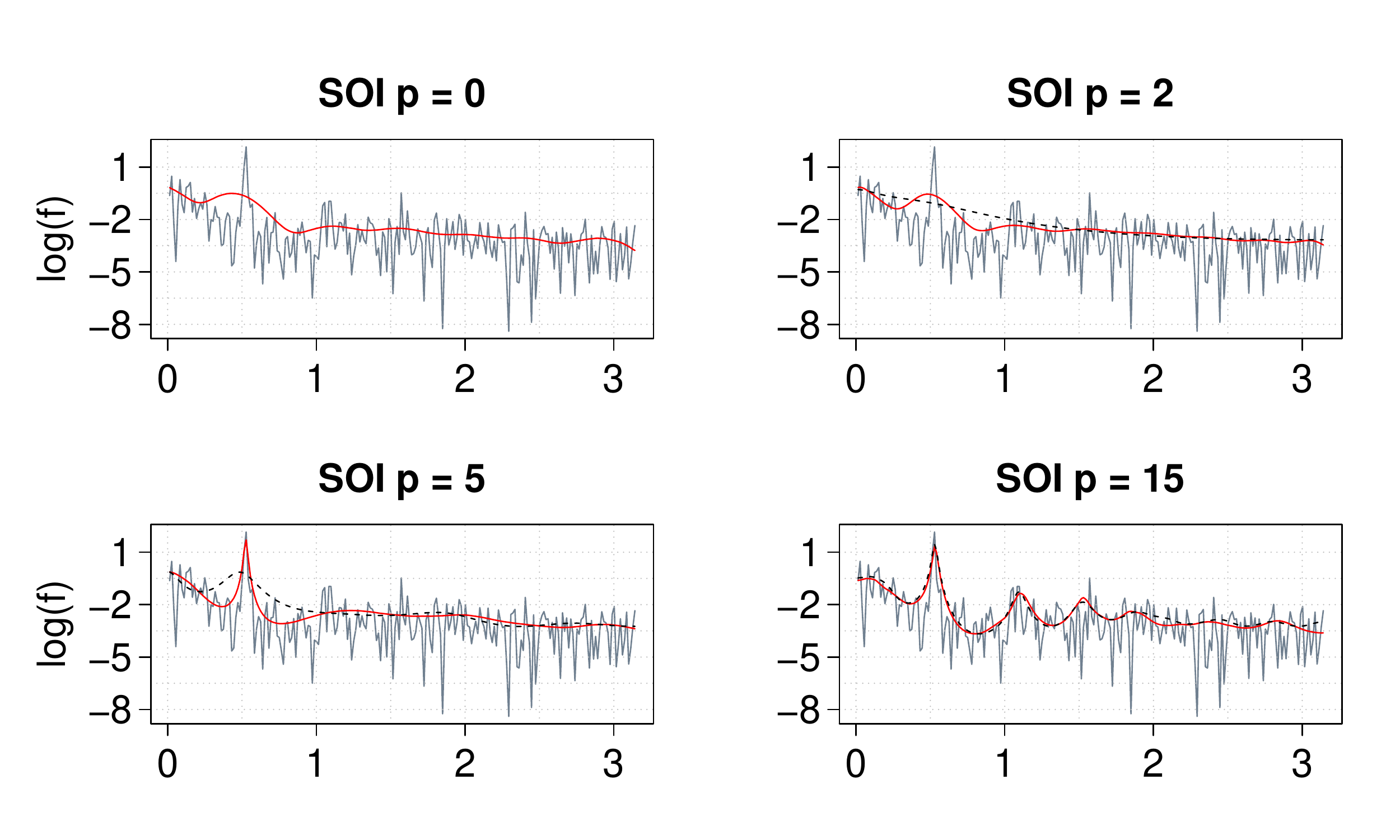}
    \caption{Logarithmic spectral estimates for the SOI series by the VNPC(p) procedure (red line) and the VAR(p) procedure (black dashed line). The periodogram is given in grey.}
    \label{Fig5.3}
\end{figure}

\begin{figure}[H]
    \centering
    \includegraphics[width=.9\linewidth, height=0.35\textheight]{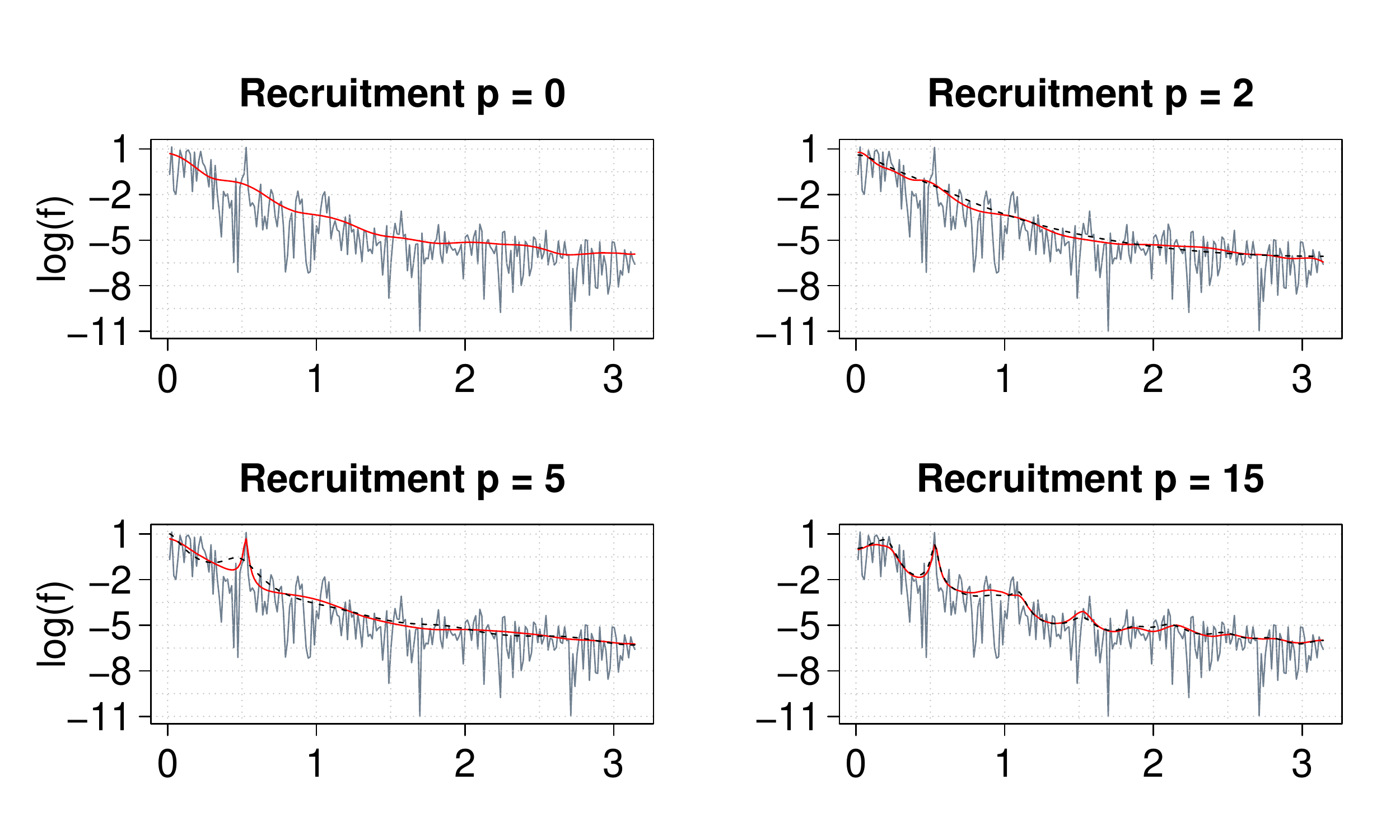}
    \caption{Logarithmic spectral estimates for the recruitment series by the VNPC(p) procedure (red line) and the VAR(p) procedure (black dashed line). The periodogram is given in grey.}
    \label{Fig5.4}
\end{figure}

Here, the standardised bivariate time series is analysed. We employ the VNPC procedure with four orders ($p=0,2,5$ and $15$) for the parametric working model and the VAR procedure with three orders ($p=2,5$ and $15$), and we compare the performance of those two procedures. In fact, when the order of the parametric working model is $0$, the VNPC procedure is the same as the VNP procedure. The order of $5$ is chosen by the elbow criterion (in Figure \ref{Fig5.2}, there is a significant bend at that order). The order of $15$ is chosen by minimising the AIC. The arrangement of the bivariate time series is the SOI to be the first series and the recruitment to be the second. To make the comparison for the three procedures with different orders, we demonstrate the results of the two series separately, the SOI in Figure \ref{Fig5.3} and the recruitment series in Figure \ref{Fig5.4}. It should be noted that since both series are monthly data, the frequency corresponding to a year period within $[0,2\pi]$ is $\omega_{\text{yearly}}:=2\pi/12\approx0.52$.

\begin{figure}[H]
    \centering
    \includegraphics[width=1\linewidth, height=0.35\textheight]{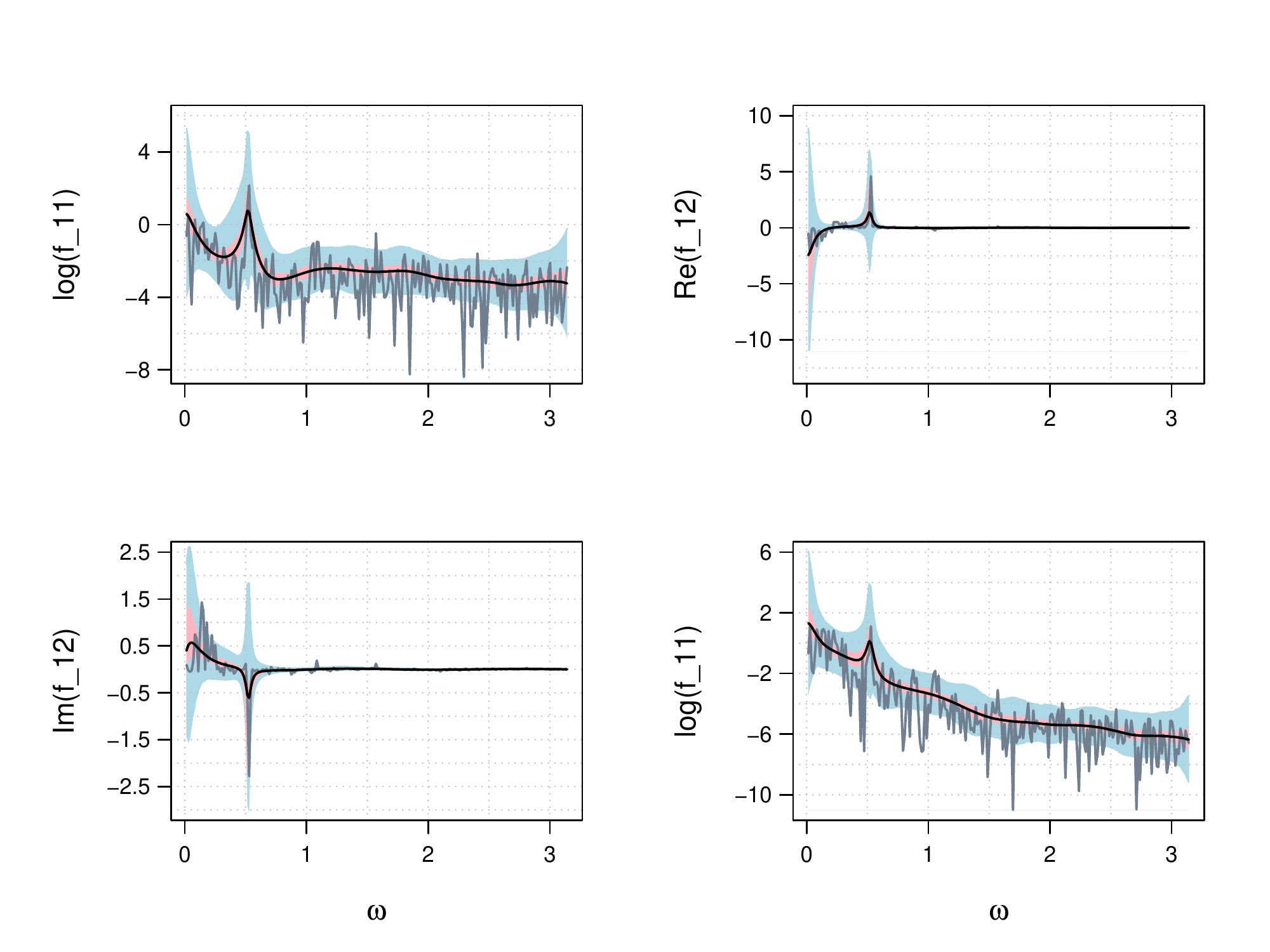}
    \caption{Posterior credible regions for the SOI and recruitment series for the VNPC procedure with a parametric working model with order $5$. The diagonal pair shows the logarithmic spectral estimate for the two series and the off-diagonal pair displays the real and imaginary parts of the cross-spectrum. Pointwise $90\%$ regions are visualised in shaded pink and uniform $90\%$ regions are in shaded blue. The posterior median is given by the black solid line and the periodogram is shown in grey.}
    \label{Fig5.5}
\end{figure}

From the periodograms of SOI in Figure \ref{Fig5.3}, it can be seen that there is an annual peak. The peak is detected by the VNP procedure, while the spectral density estimate is overly smooth. However, when $p=2$, the VAR procedure performs even worse as it is too smooth to pick up any spectral density peaks. In this case, the VNPC procedure with $p=2$ does not improve much compared to VNP, i.e., VNPC with $p=0$. When the order is increased to $5$, which is the relatively small order chosen by the elbow criterion, the VNPC procedure has successfully estimated the main annual peak, in contrast to the VAR procedure.  When using the large order given by AIC, i.e., $p=15$, the performances of the VNPC and the VAR procedures are very similar. Both estimators pick up the annual peak and its harmonics as well as a broader low-frequency that can be interpreted as the El Ni\~{n}o effect which occurs irregularly every 2--7 years. The findings for the recruitment series in Figure \ref{Fig5.4} are similar. The VNPC procedure can make use of the relatively small VAR order, $p=5$, to capture the main annual peak in that series, while the estimates obtained by the VAR procedure with that order are too smooth. Figure \ref{Fig5.5} provides the $90\%$ pointwise region for VNPC with $p=5$. The top-right plot diplays the real part of the cross-spectrum, which is called the \textit{cospectrum}; and the bottom-left plot displays the imaginary part of the cross-spectrum, which is called the \textit{quadrature spectrum}.

\begin{figure}[H]
    \centering
    \includegraphics[width=1\linewidth, height=0.4\textheight]{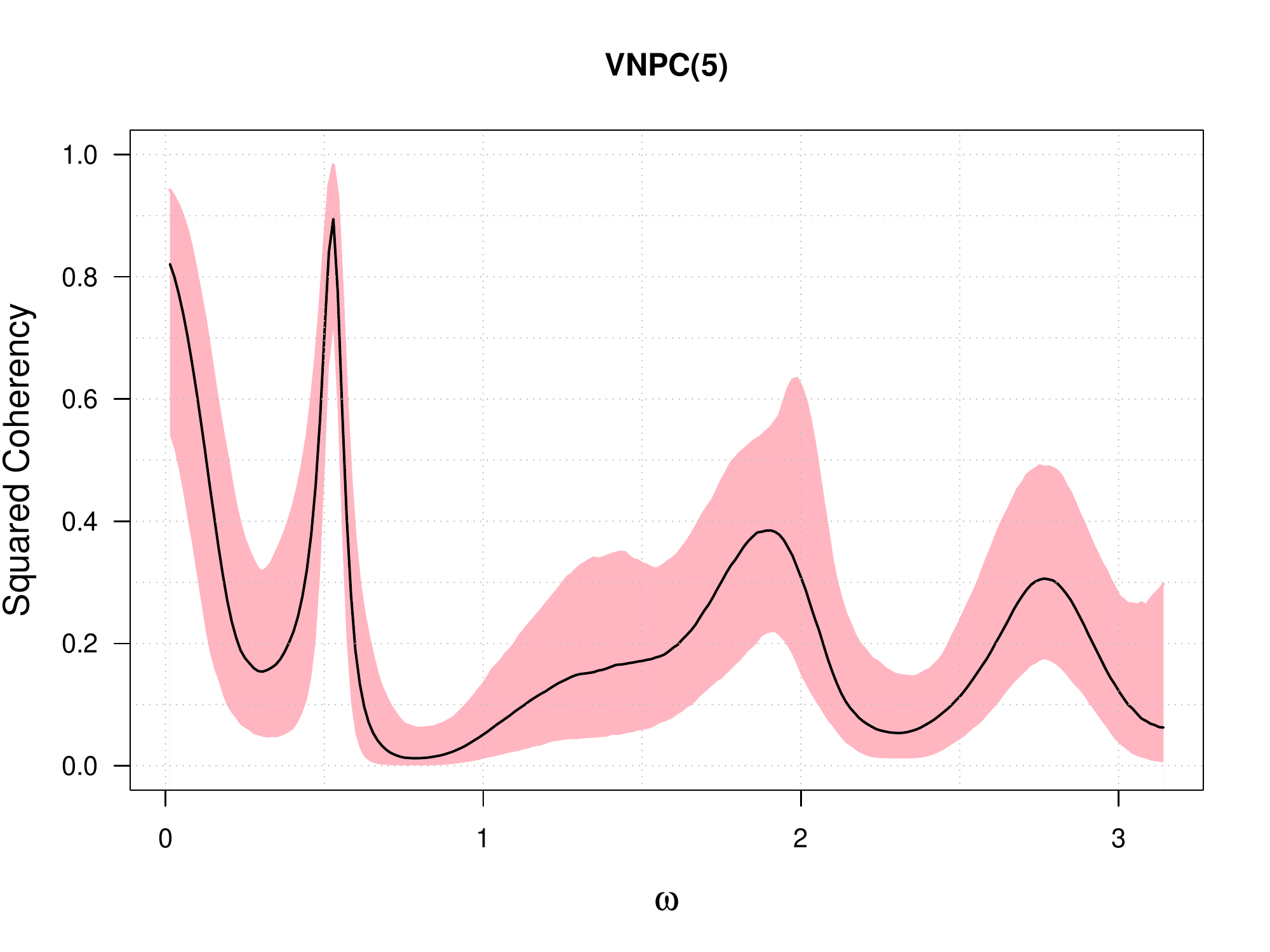}
    \caption{Estimated squared coherency for the SOI and recruitment series by the VNPC procedure with a parametric working model with order $5$. The posterior median is given by the black line and the pointwise $90\%$ credible region is in shaded pink.}
    \label{Fig5.6}
\end{figure}

The \textit{squared coherency} is more adequate for investigating the correlation between two series than the cospectrum and the quadrature spectrum. Recall that the coherency $\kappa$ between two series at frequency $\omega$ is defined as \begin{align}
    \kappa(\omega|\mf):=\frac{f_{12}(\omega)}{(f_{11}(\omega)f_{22}(\omega))^{1/2}},\quad0\leq\omega\leq\pi.
\end{align} This statistic can be viewed as the counterpart of the cross-correlation function in the time domain and it holds $|\kappa(\omega|\mf)|^{2}\leq 1$ for all $0\leq\omega\leq\pi$ (see Section 11.6 in \cite{brockwell1991}). We calculate the squared  coherency for all 5000 spectral density estimates 
and then consider the median and the $90\%$ pointwise region. The result in Figure \ref{Fig5.6} shows that there is an obvious peak at $\omega_{\text{yearly}}$, and this agrees with the findings by the cospectrum and the quadrature spectrum in Figure \ref{Fig5.6}. There are also a few minor peaks at the higher frequencies, which are the harmonics of $\omega_{\text{yearly}}$, and the two series are correlated at those frequencies as well. In addition, the strong coherency shown at the low frequencies close to $0$ might be because of the 2--7 years El Ni\~{n}o period mentioned previously. In fact, all the results in this Section are in line with the findings in \cite{rosen2007}.

\subsection{Wind speed profiles}

\begin{figure}[H]
    \begin{subfigure}{.3\textheight}
        \centering
        \includegraphics[width=0.95\textwidth, height=0.4\textheight]{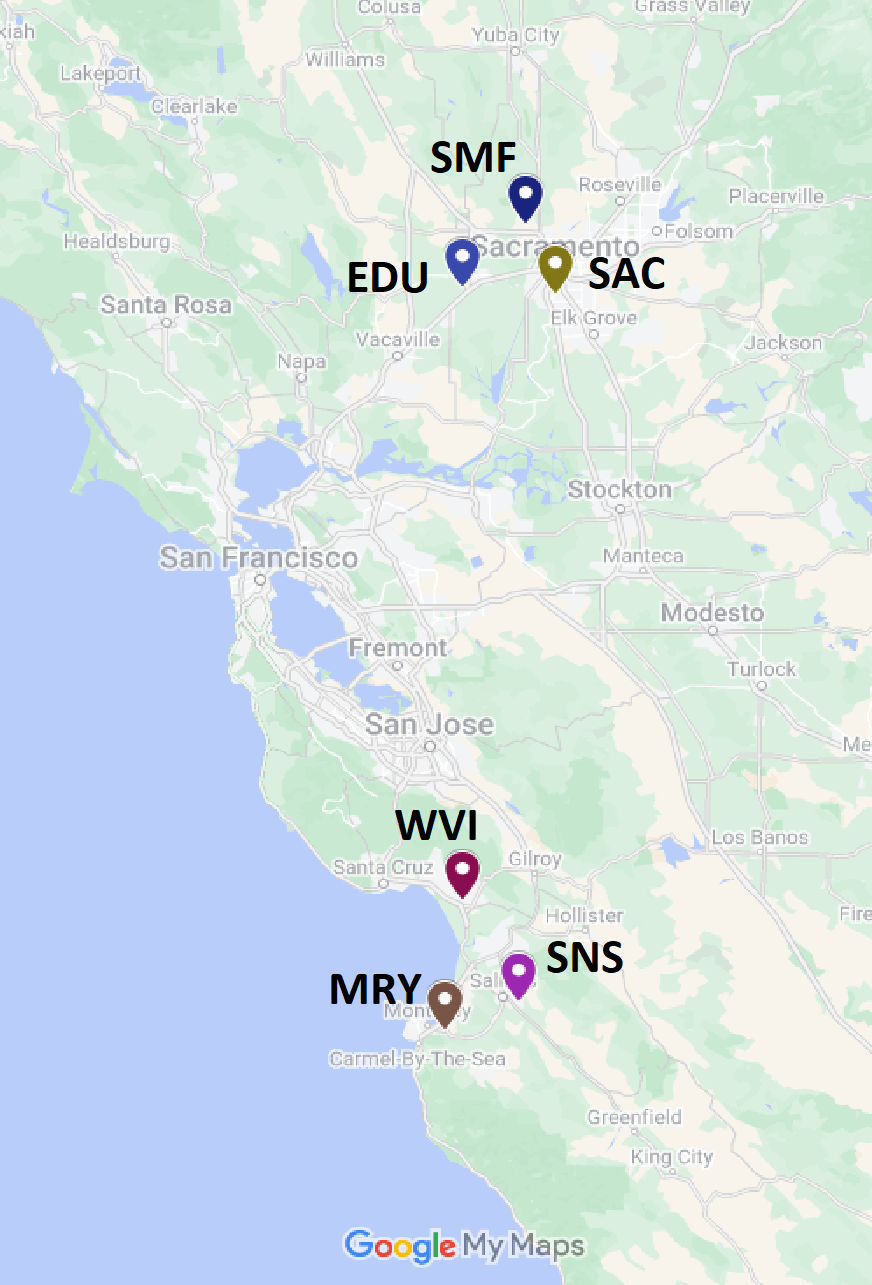}
    \end{subfigure}
    \begin{subfigure}{.55\textheight}
        \includegraphics[width=0.75\textwidth, height=0.43\textheight]{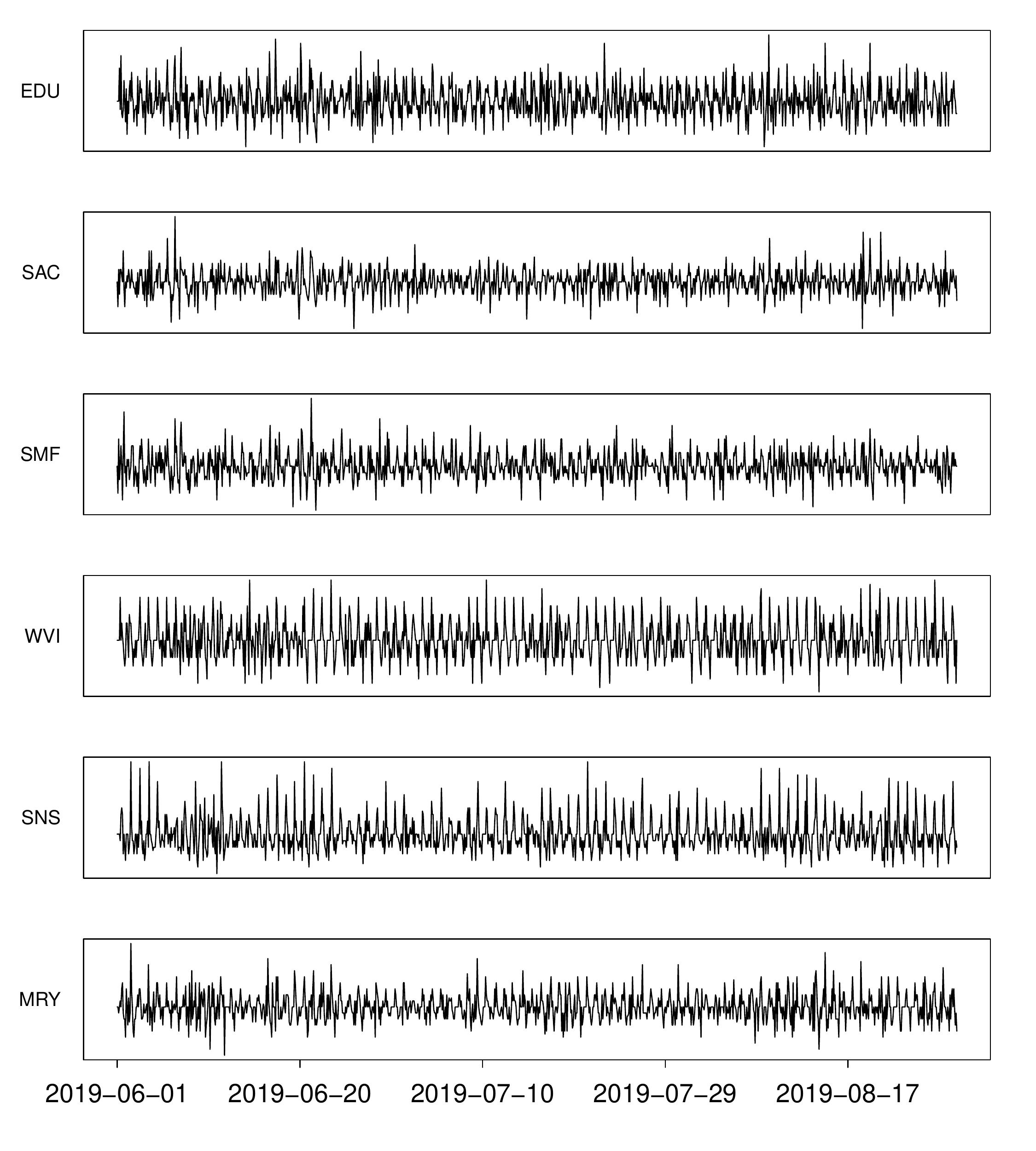}
    \end{subfigure}
    \caption{The left graph shows the geographical locations for the six airports. The right graph shows the standardised first differences of median wind speed measurements every two hours from 01-06-2019 12:00 am to 31-08-2019 11:59 pm.}
    \label{Fig5.8}
\end{figure}

\begin{figure}[H]
    \centering
    \includegraphics[width=1\linewidth, height=0.4\textheight]{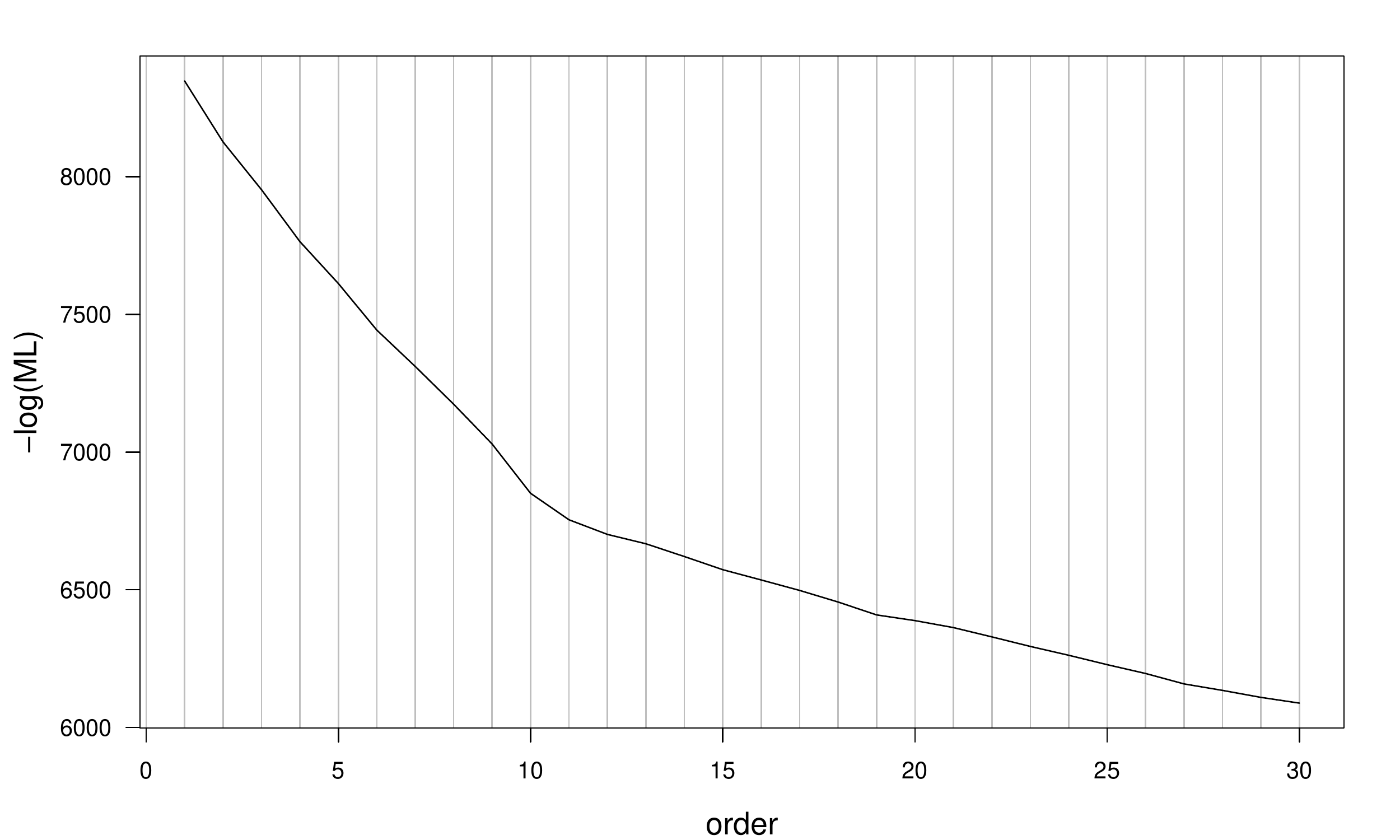}
    \caption{Negative maximum log likelihood for different VAR(p) models applied for the standardised detrended wind speed data.}
    \label{Fig5.9}
\end{figure}

 In this case study, we apply the VNPC method to the California wind speed data from the Iowa State University Environmental Mesonet (IEM) Automated Surface System (ASOS) database (\cite{iowa2022}; \cite{mannarano1998}). This example has been studied by \cite{hu2023} using a stochastic gradient variational Bayes approach for fast Bayesian inference. In their work, they consider the wind speed in knots at six airports in California from 01-06-2019 12:00 am to 31-08-2019 11:59 pm: EDU (Davis), SAC and SMF (Sacramento), MRY (Monterey), SNS (Salinas), and WVI (Watsonville). That particular time interval is used to avoid extreme values and non-stationarities caused by rainfall and storms, which occurred in other months. In particular, they apply their method to the standardised first difference of median wind speed every two hours for each location. In our study, we use the same processed data, which is accessible from the supplementary material in \cite{hu2023}. Figure \ref{Fig5.8} shows the geographical locations for those six airports and the standardised detrended data. All six series are treated as a six-dimensional time series and analysed simultaneously. The order for the parametric working model suggested by the elbow criterion is $10$ (see Figure \ref{Fig5.9}).

\begin{figure}[H]
    \centering
    \includegraphics[width=1\textwidth, height=0.5\textheight]{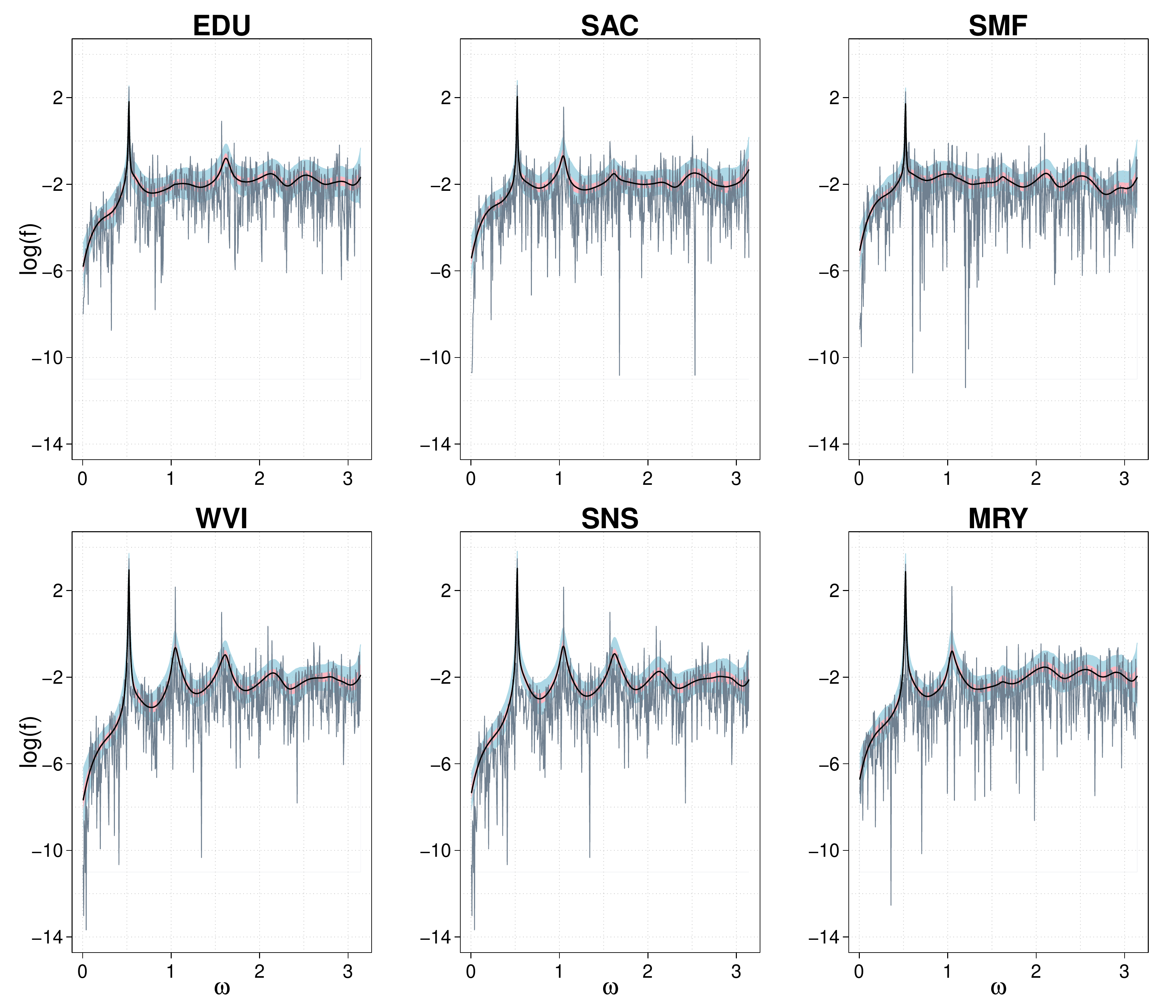}
    \caption{Logarithmic spectral estimates for the standardised detrended wind speed data by the VNPC procedure with a parametric working model with order $10$. The periodogram is given by the grey line and the posterior median by the black line. The pointwise $90\%$ posterior credible region is visualised in shaded pink and the uniform $90\%$ region is in shaded blue.}
    \label{Fig5.10}
\end{figure}

\begin{figure}[H]
    \centering
    \includegraphics[width=.9\textwidth, height=0.7\textheight]{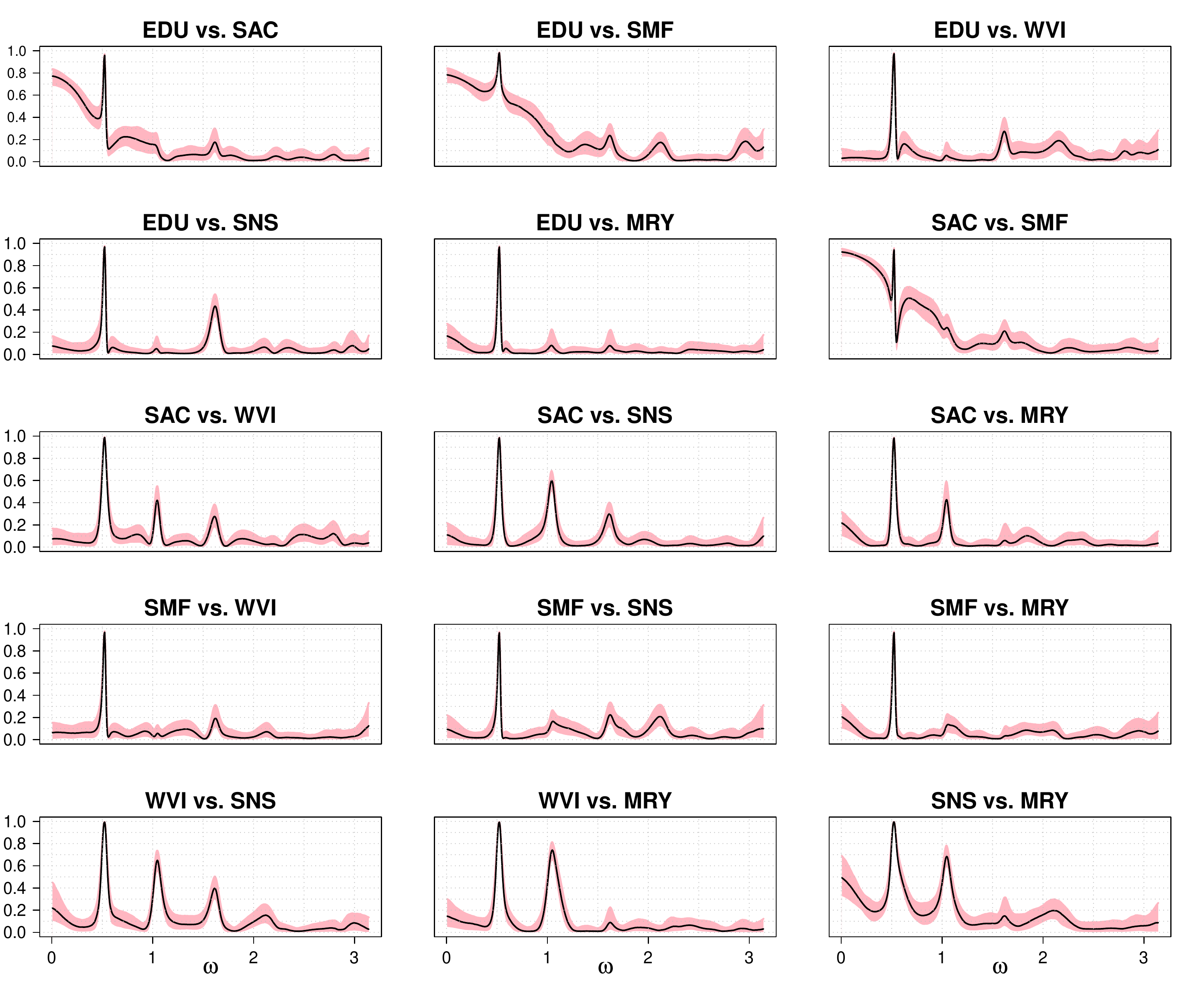}
    \caption{Estimated squared coherency for the standardised detrended wind speed data by the VNPC procedure with a parametric working model with order $10$. The posterior median is given by the black line and the pointwise $90\%$ credible region is in shaded pink.}
    \label{Fig5.11}
\end{figure}

The results in Figure \ref{Fig5.10} show that all six spectral densities have a prominent peak at around $2\pi/12\approx0.524$, which indicates a strong daily period of wind profile patterns at all locations. This is consistent with the finding in \cite{hu2023}. However, in contrast to spectral estimates obtained by variational inference in  \cite{hu2023}, peaks in the VPNC estimates are much more pronounced and sharper, for the main daily peak as well as its harmonics. The spectral estimates at $2\pi/12\approx0.524$ for the coastal locations WVI, SNS and MRY have slightly larger power than for the inland airports as also noted by \cite{hu2023}. Furthermore, we are  interested in the squared coherencies between the windspeeds at the various pairs of airports. The estimates are given in Figure \ref{Fig5.11}. Not surprisingly, all pairs have a strong daily periodic coherency, which is in accordance with the daily period found in all series in Figure \ref{Fig5.11}. Moreover, all pairs except for the ones with SMF have a strong second harmonic. 

In \cite{hu2023} correlations tend to appear locally, between locations close by, and no correlation between locations separated by 100-200 miles, i.e.,  between those pairs where one wind profile is taken from the group (EDU, SAC, SMF) and one from (WVI, SNS, MRY). In contrast, our results in Figure \ref{Fig5.11} show high squared coherencies also between locations in different groups, most pronounced at the daily frequency. Coherencies at higher frequencies drop off more quickly for between-group pairs than within-group pairs, e.g. WVI vs.\ SNS as compared to WVI vs.\ EDU. Even over geographical distances of 100--200 miles such as between the airports in the Monterey and Sacramento region, one would expect a daily correlation between the wind profiles, however hourly or minutely patterns of the wind profiles could vary greatly between distant airports but be similar for neighbouring airports. In fact, wind speed is affected by multiple factors including wind direction, topography and local climate. \cite{wang2020} analysed large-scale wind patterns in California using model simulations from the variable-resolution Community Earth System Model (VR-CESM). They described ten wind patterns by directions and strengths in the northern California and patterns are associated with geopotential height pattern, geography and climate. Wind pattern is not variable with distance necessarily and wind speed correlation may not be strictly related to distance between locations.

%% file: section6.tex
This paper presents a novel Bayesian approach for spectral density estimation of multivariate stationary Gaussian time series. A nonparametrically corrected likelihood is proposed which can be regarded as a generalisation of the Whittle likelihood. To elaborate, a parametric likelihood is specified for the data in the time domain, and then the likelihood is nonparametrically corrected in the frequency domain in order to mitigate effects of model misspecification. A multivariate extension of the
nonparametric Bernstein-Dirichlet process prior for univariate spectral densities to the space of Hermitian positive definite spectral density matrices is
specified directly on the correction matrices.
We prove contiguity of the nonparametrically corrected likelihood and the true likelihood for Gaussian time series. The results of a simulation study, exploring scenarios of a correctly specified parametric and a misspecified model,
 show that spectral density estimates based on the nonparametrically corrected likelihood  
 inherit the efficiencies of the parametric approach in the correctly specified case and the robustness of the nonparametric approach in the misspecified case. Moreover, the nonparametrically corrected Gaussian VAR model performs well even for non-Gaussian time series.  Applications demonstrate that estimates based on the nonparametrically corrected likelihood are much better in estimating sharp peaks and abrupt changes in the spectrum than nonparametric methods.
 \texttt{R}-code that implements the MCMC algorithm to sample from the posterior distribution is made publicly available at \url{https://github.com/easycure1/vnpctest}.

 The contiguity shown in this paper is useful for deriving the posterior consistency of the corrected likelihood approach for stationary multivariate Gaussian time series based on Schwartz's theorem (\cite{schwartz1965}) and its extensions. The posterior consistency of our approach for the stationary multivariate non-Gaussian time series is also an exciting topic for future work. Moreover, our proposed approach holds broad applicability for analyzing multivariate stationary time series and practical utility across various datasets and disciplines, such as ECG data in medicine and gravitational waves measurements in astrophysics.

%% file: section1_supp.tex
\begin{figure}[H]
    \centering
    \begin{subfigure}[b]{0.65\textheight}
        \includegraphics[width=1\textwidth, height=.18\textheight]{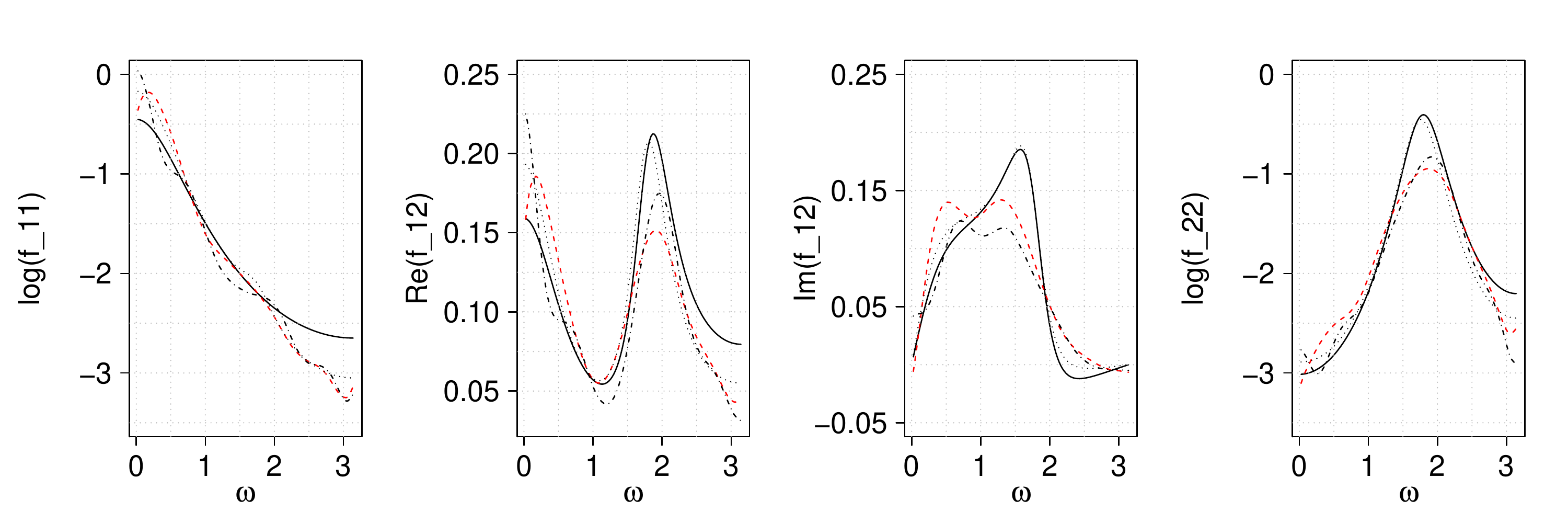}
        \vspace{-1\baselineskip}
        \caption{}
    \end{subfigure}
    \begin{subfigure}[b]{0.65\textheight}
        \includegraphics[width=1\textwidth, height=.18\textheight]{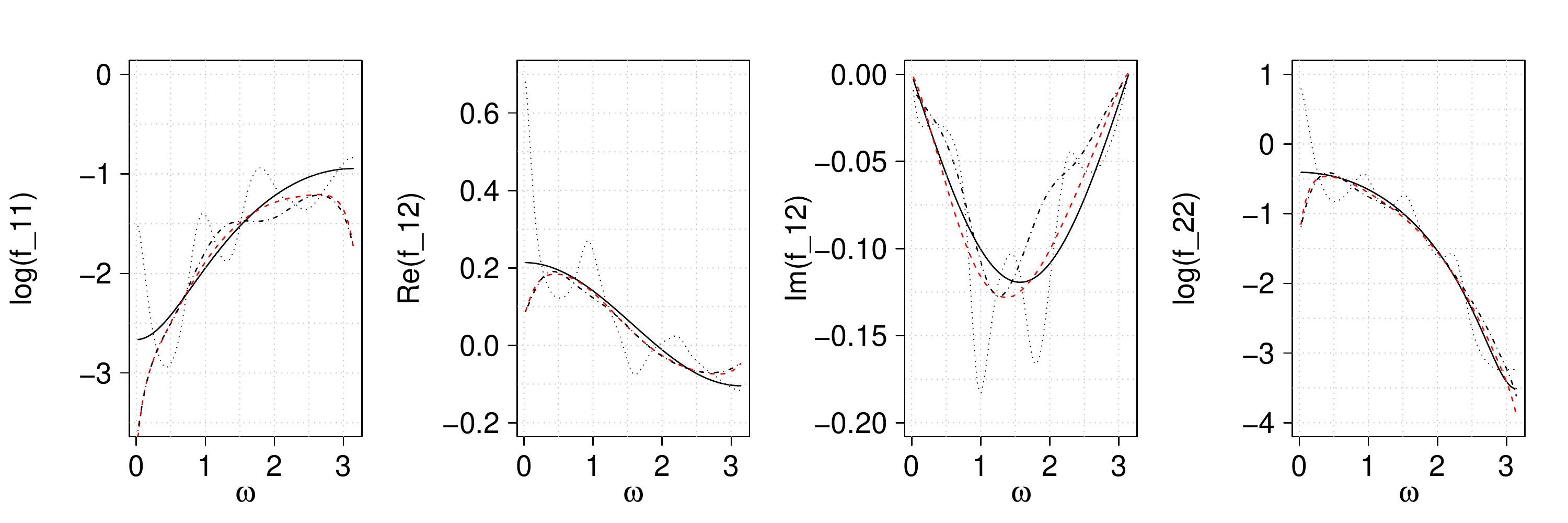}
        \vspace{-1\baselineskip}
        \caption{}
    \end{subfigure}
    \caption{Spectral estimates for the realisation of (a) the VAR(2) model and (b) the VMA(1) model in Figure 1 in the article for the VNPC procedure with fixed order $1$ for the parametric working model (red dashed), the VAR procedure with order selected by AIC (dotted) and the VNP procedure (dash-dotted), and the true spectral density is given by the solid black line.}
    \label{Fig1.1_supp}
\end{figure}

In the following, we present visual representations of the spectral density estimates obtained using the three procedures for one random realisation of the two models considered in the simulation study. These are shown  in Figure \ref{Fig1.1_supp} together the true spectral density. The individual spectra are displayed on the logarithmic scale, while real and imaginary parts of the cross spectra are  on the regular scale. Figure \ref{Fig1.1_supp} (a) shows that all three procedures have a spectral density estimate that is reasonably close to the true spectrum for the VAR(2) data. In the misspecified case, i.e., the time series generated by a VMA(1) model, the parametric VAR estimate is furthest from the true spectrum. The VNPC estimate outperforms both parametric VAR and nonparametric VNP estimate.

\begin{figure}[H]
    \centering
    \begin{subfigure}[b]{0.6\textheight}
        \includegraphics[width=1\textwidth, height=.18\textheight]{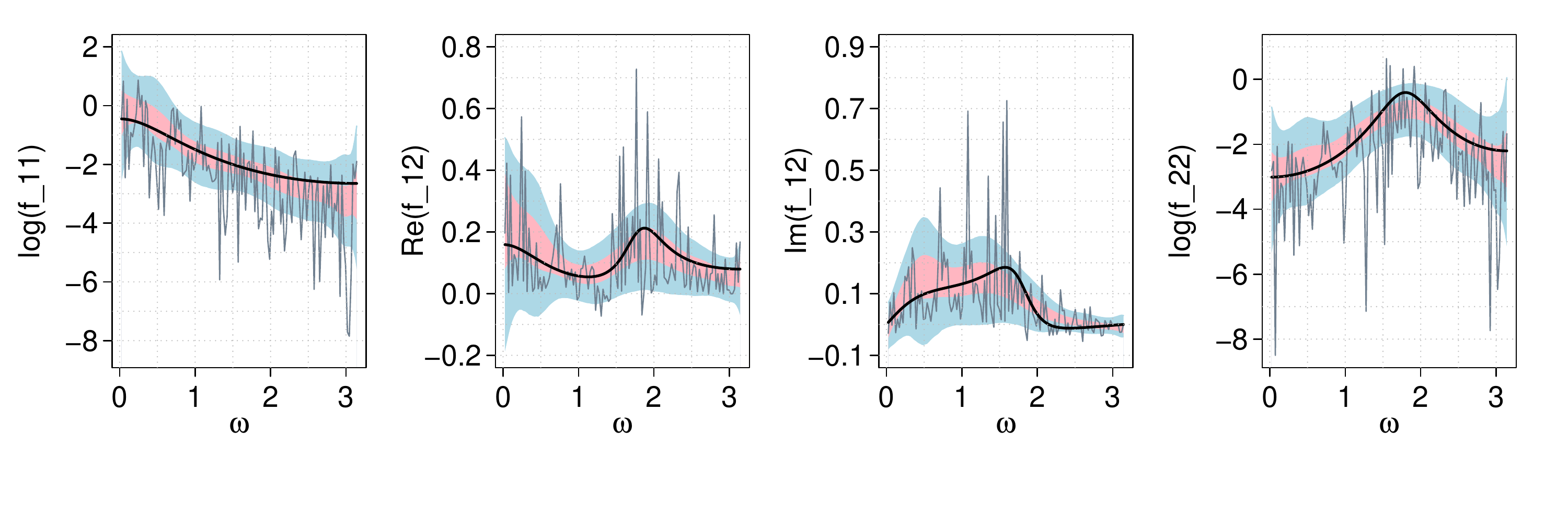}
        \vspace{-2\baselineskip}
        \caption{}
    \end{subfigure}
    \begin{subfigure}[b]{0.6\textheight}
        \includegraphics[width=1\textwidth, height=.18\textheight]{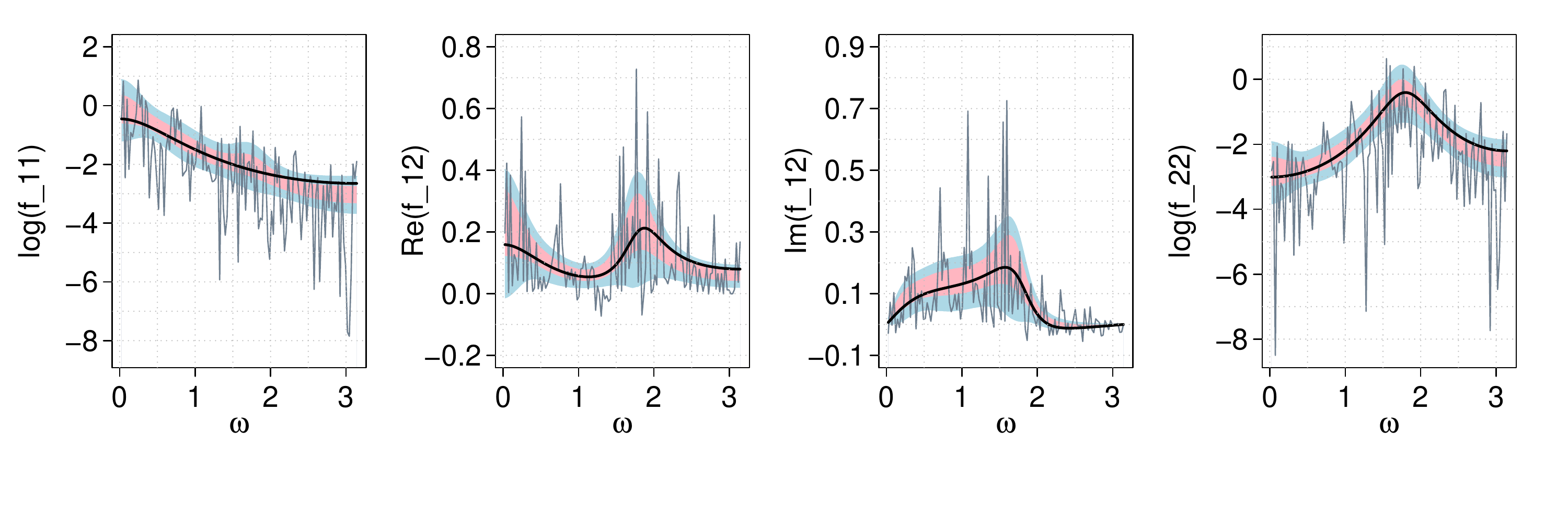}
        \vspace{-2\baselineskip}
        \caption{}
    \end{subfigure}
    \begin{subfigure}[b]{0.6\textheight}
        \includegraphics[width=1\textwidth, height=.18\textheight]{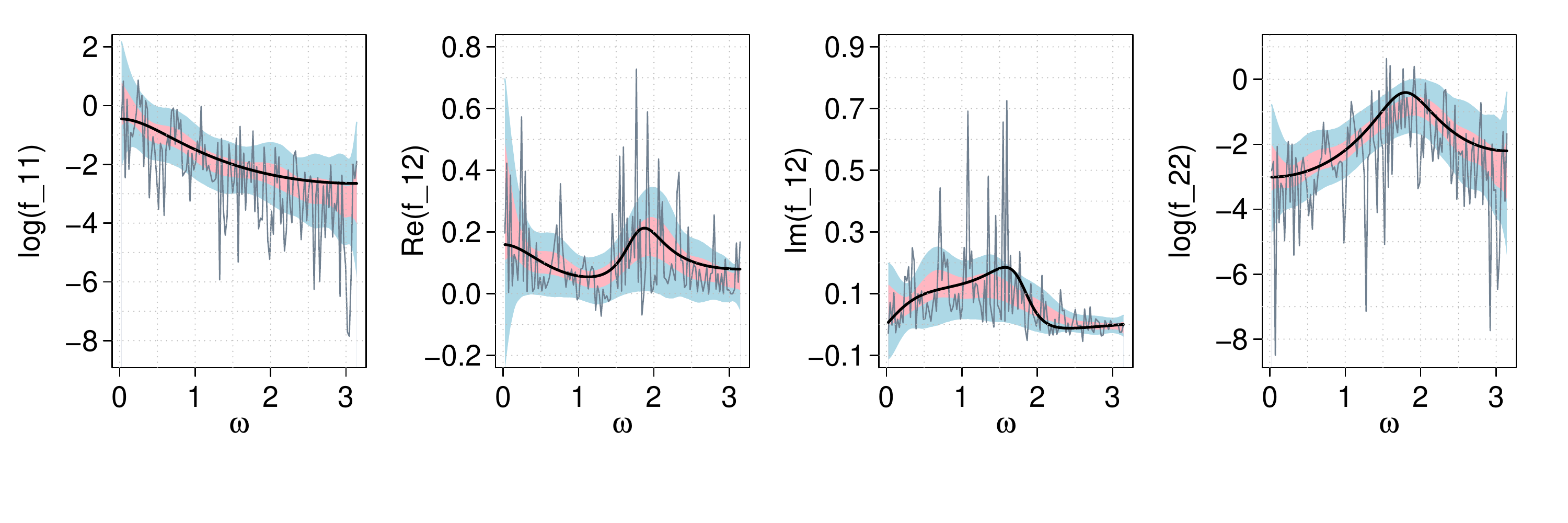}
        \vspace{-2\baselineskip}
        \caption{}
    \end{subfigure}
    \caption{Posterior credible regions for the realisation of the VAR(2) model from Figure 1a in the article for (a) the VNPC procedure with fixed order $1$ for the parametric working model, (b) the VAR procedure with order selected by AIC and (c) the VNP procedure. Pointwise $90\%$ regions are represented in shaded pink, uniform $90\%$ regions are in shaded blue,  the true spectral density is given by the black solid line and the periodogram is shown in grey.}
    \label{Fig1.2_supp}
\end{figure}

\begin{figure}[H]
    \centering
    \begin{subfigure}[b]{0.6\textheight}
        \includegraphics[width=1\textwidth, height=.18\textheight]{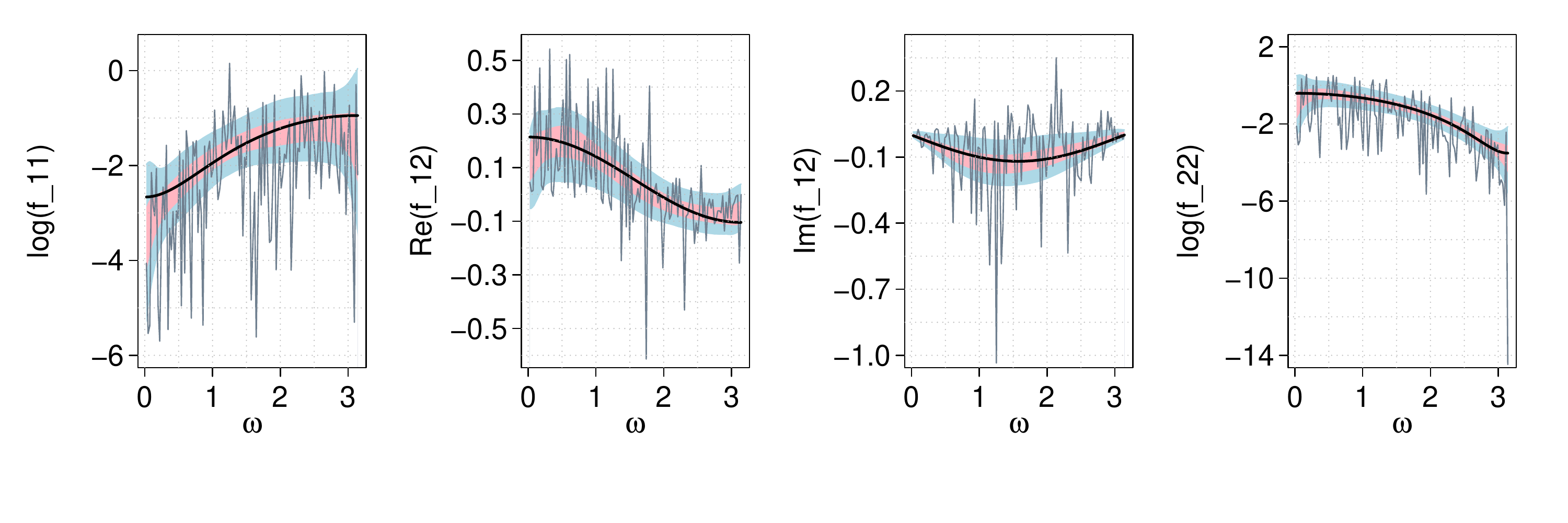}
        \vspace{-2\baselineskip}
        \caption{}
    \end{subfigure}
    \begin{subfigure}[b]{0.6\textheight}
        \includegraphics[width=1\textwidth, height=.18\textheight]{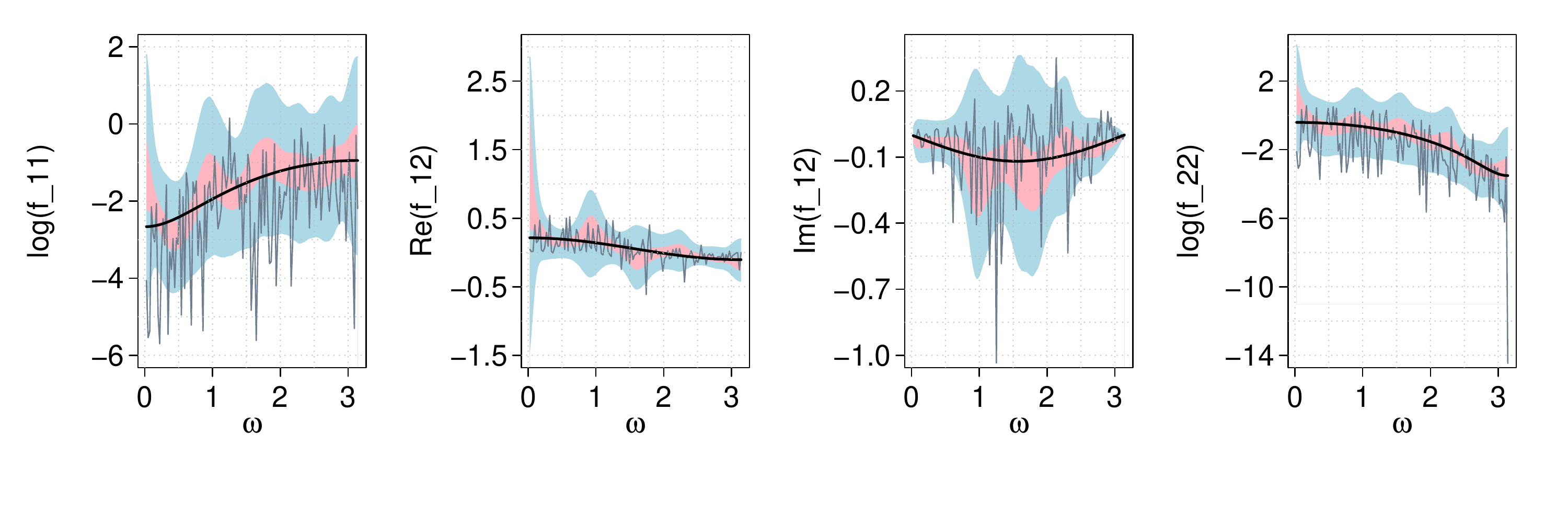}
        \vspace{-2\baselineskip}
        \caption{}
    \end{subfigure}
    \begin{subfigure}[b]{0.6\textheight}
        \includegraphics[width=1\textwidth, height=.18\textheight]{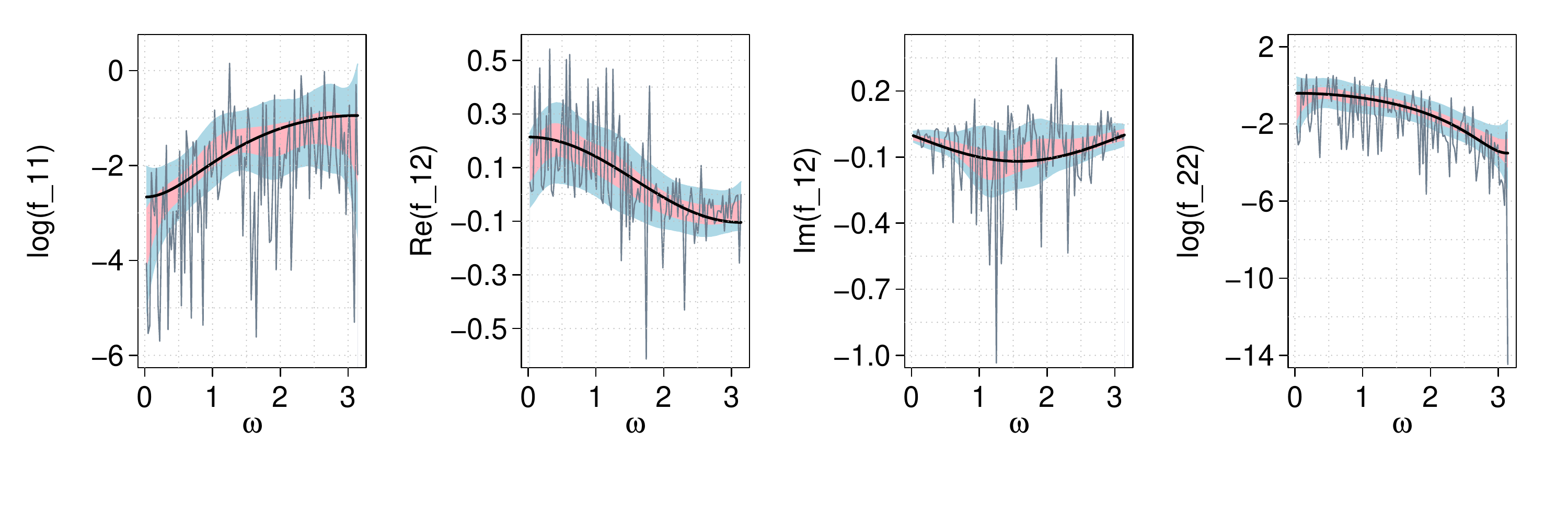}
        \vspace{-2\baselineskip}
        \caption{}
    \end{subfigure}
    \caption{Posterior credible regions for the realisation of the VMA(1) model from Figure 1b in the article for (a) the VNPC procedure with fixed order $1$ for the parametric working model, (b) the VAR procedure with order selected by AIC and (c) the VNP procedure. Pointwise $90\%$ regions are visualised in shaded pink, uniform $90\%$ regions are in shaded blue, the true spectral density is given by the black solid line and the periodogram is shown in grey.}
    \label{Fig1.3_supp}
\end{figure}

We consider the corresponding $90\%$-credible regions next, which are shown in Figure \ref{Fig1.2_supp} for the VAR(2) realisation and Figure \ref{Fig1.3_supp} for the VMA(1) realisation. It can be seen that the results are consistent with what we have found in Figure \ref{Fig1.1_supp}. To elaborate, for the VAR(2) example, Figure \ref{Fig1.2_supp} shows that all three procedures yield adequate regions, while the VAR procedure with the well-specified order appears to have slightly narrower regions than the other two procedures. Figure \ref{Fig1.3_supp} shows that the VAR estimate has much wider and more bumpy regions than the other two procedures, which demonstrates the loss of efficiency in the misspecified  VMA(1) case. The VNPC and the VNP estimates have similarly narrower and smoother regions than the VAR estimate. 

\begin{figure}[H]
    \centering
    \begin{subfigure}[b]{0.72\textheight}
        \includegraphics[width=1\textwidth, height=.2\textheight]{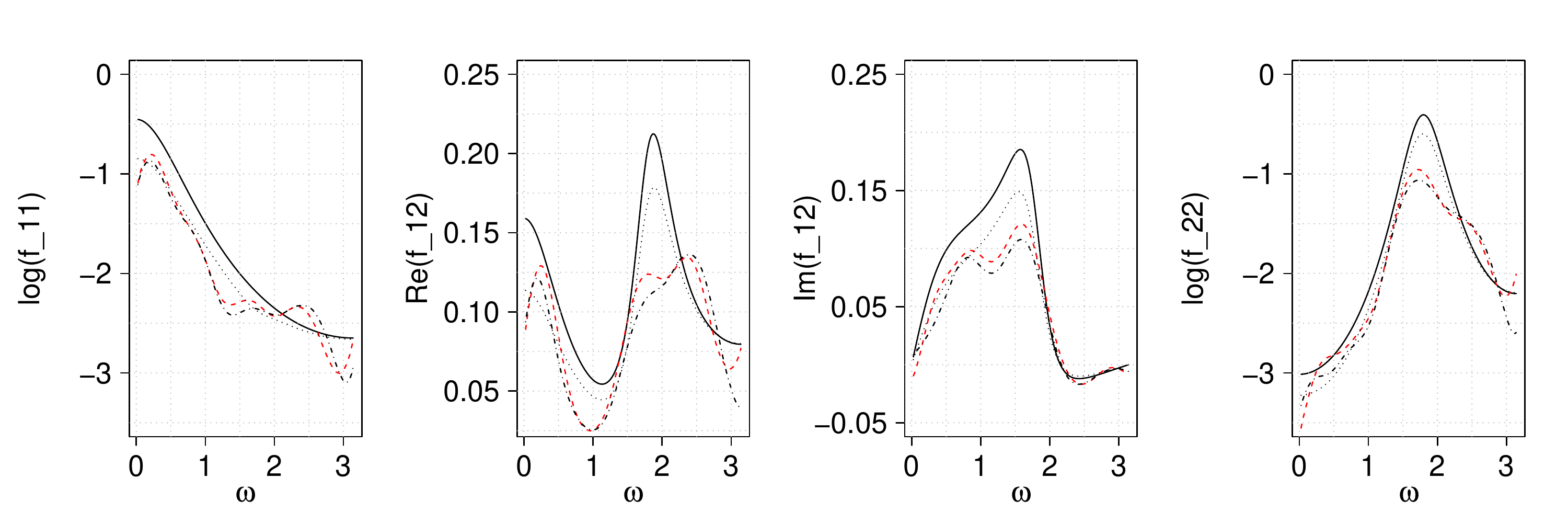}
        \vspace{-1\baselineskip}
        \caption{}
    \end{subfigure}
    \begin{subfigure}[b]{0.72\textheight}
        \includegraphics[width=1\textwidth, height=.2\textheight]{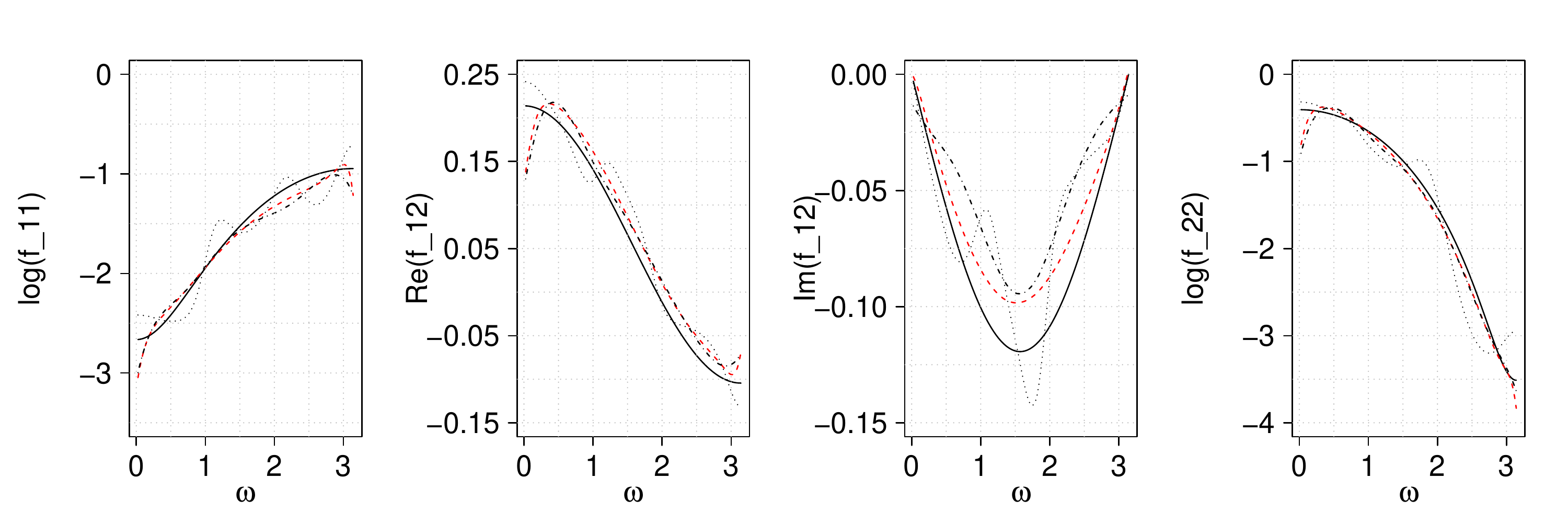}
        \vspace{-1\baselineskip}
        \caption{}
    \end{subfigure}
    \caption{Spectral estimates for one realisation of (a) the VAR(2) model and (b) the VMA(1) model with a Laplace innovation of size $n=256$ for the VNPC procedure with fixed order $1$ for the parametric working model (red dashed), the VAR procedure with order selected by AIC (dotted) and the VNP procedure (dash-dotted), and the true spectral density is given by the solid black line.}
    \label{Fig1.4_supp}
\end{figure}

\begin{figure}[H]
    \centering
    \begin{subfigure}[b]{0.7\textheight}
        \includegraphics[width=1\textwidth, height=.2\textheight]{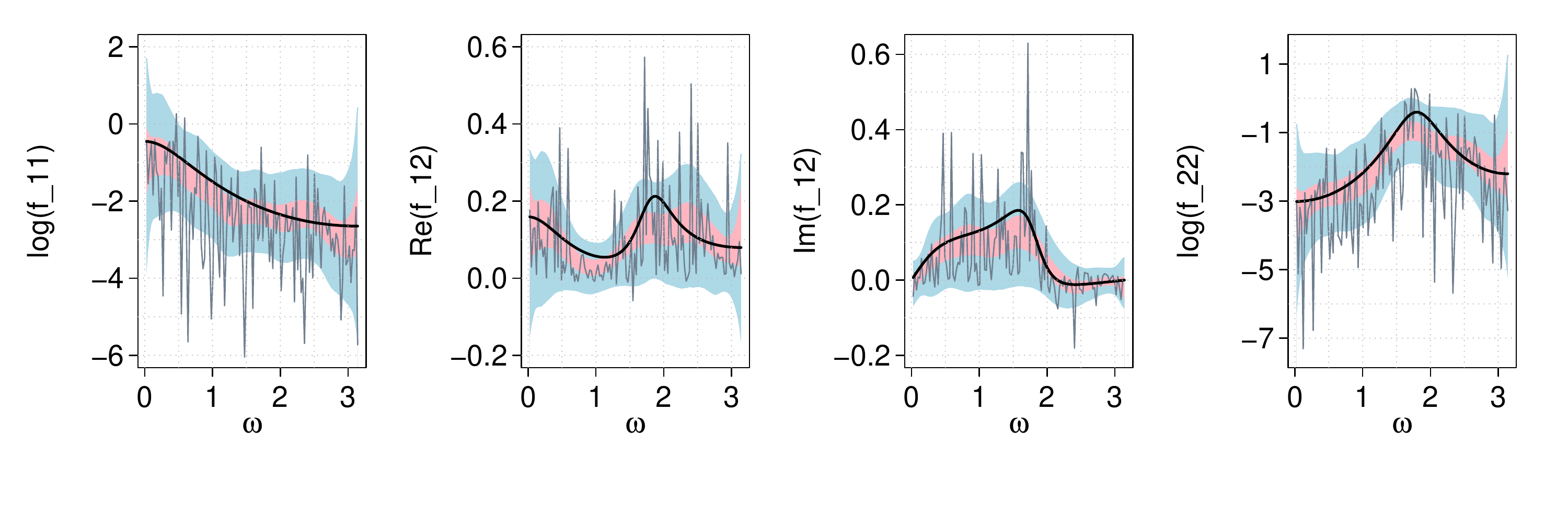}
        \vspace{-2\baselineskip}
        \caption{}
    \end{subfigure}
    \begin{subfigure}[b]{0.7\textheight}
        \includegraphics[width=1\textwidth, height=.2\textheight]{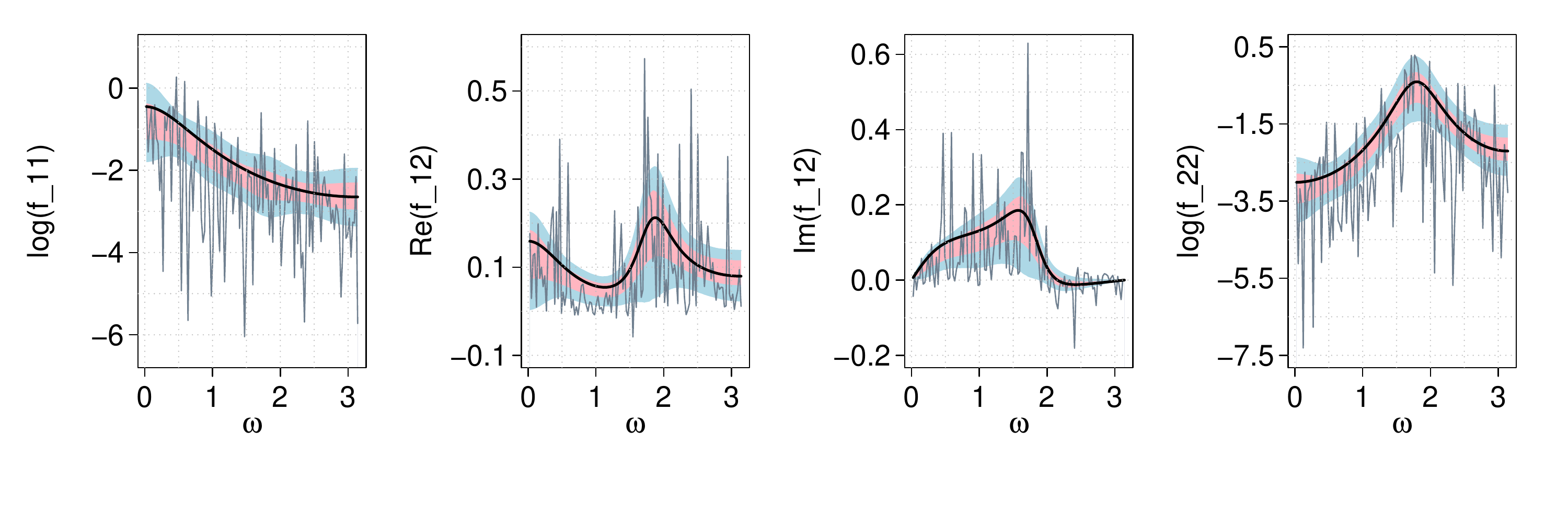}
        \vspace{-2\baselineskip}
        \caption{}
    \end{subfigure}
    \begin{subfigure}[b]{0.7\textheight}
        \includegraphics[width=1\textwidth, height=.2\textheight]{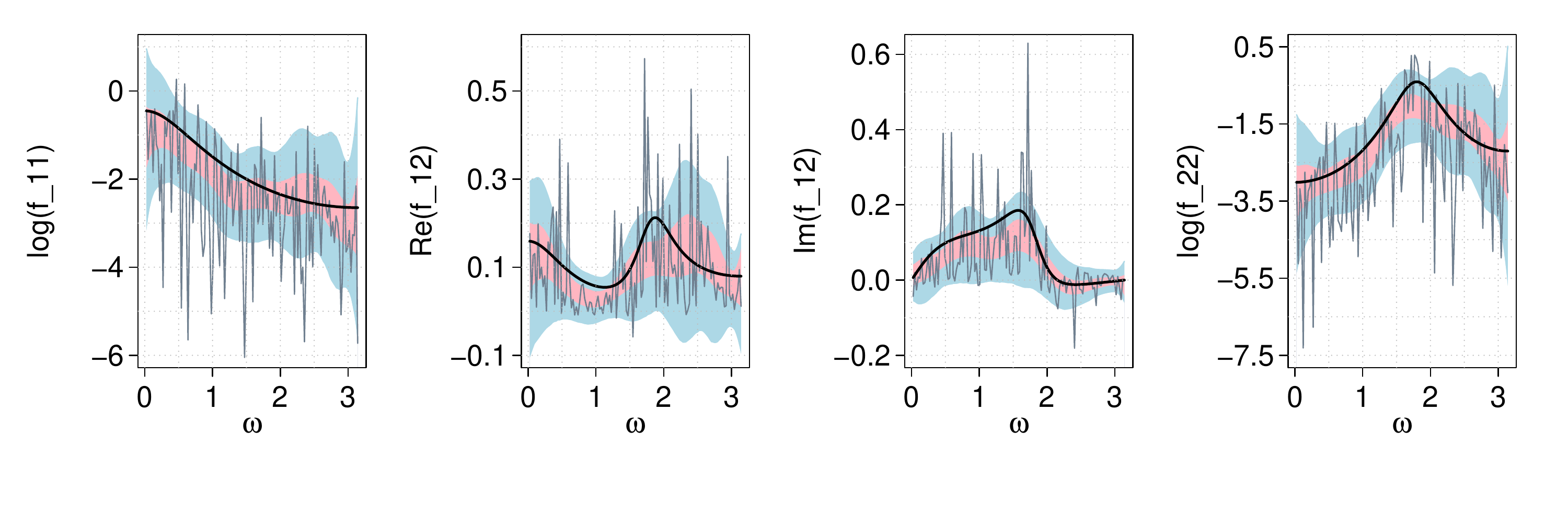}
        \vspace{-2\baselineskip}
        \caption{}
    \end{subfigure}
    \caption{Posterior credible regions for the realisation of the VAR(2) model same as the one referred by Figure \ref{Fig1.4_supp} for (a) the VNPC procedure with fixed order $1$ for the parametric working model, (b) the VAR procedure with order selected by AIC and (c) the VNP procedure. Pointwise $90\%$ regions are visualised in shaded pink, uniform $90\%$ regions are in shaded blue, the true spectral density is given by the black solid line and the periodogram is shown in grey.}
    \label{Fig1.5_supp}
\end{figure}

\begin{figure}[H]
    \centering
    \begin{subfigure}[b]{0.7\textheight}
        \includegraphics[width=1\textwidth, height=.2\textheight]{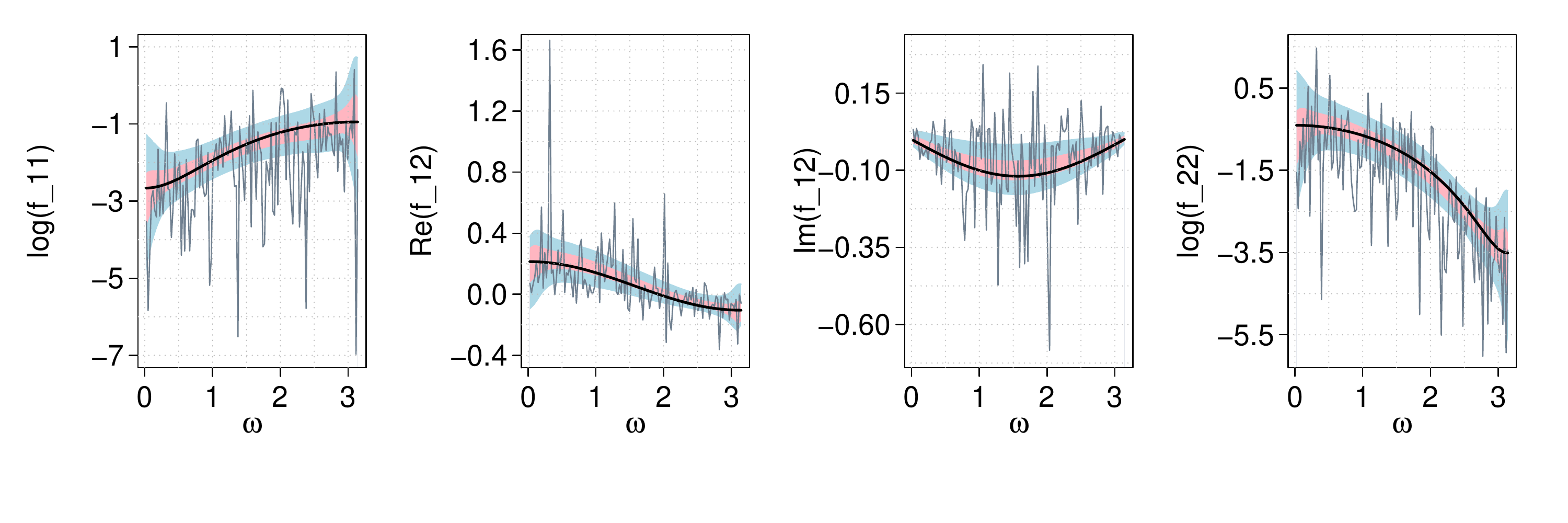}
        \vspace{-2\baselineskip}
        \caption{}
    \end{subfigure}
    \begin{subfigure}[b]{0.7\textheight}
        \includegraphics[width=1\textwidth, height=.2\textheight]{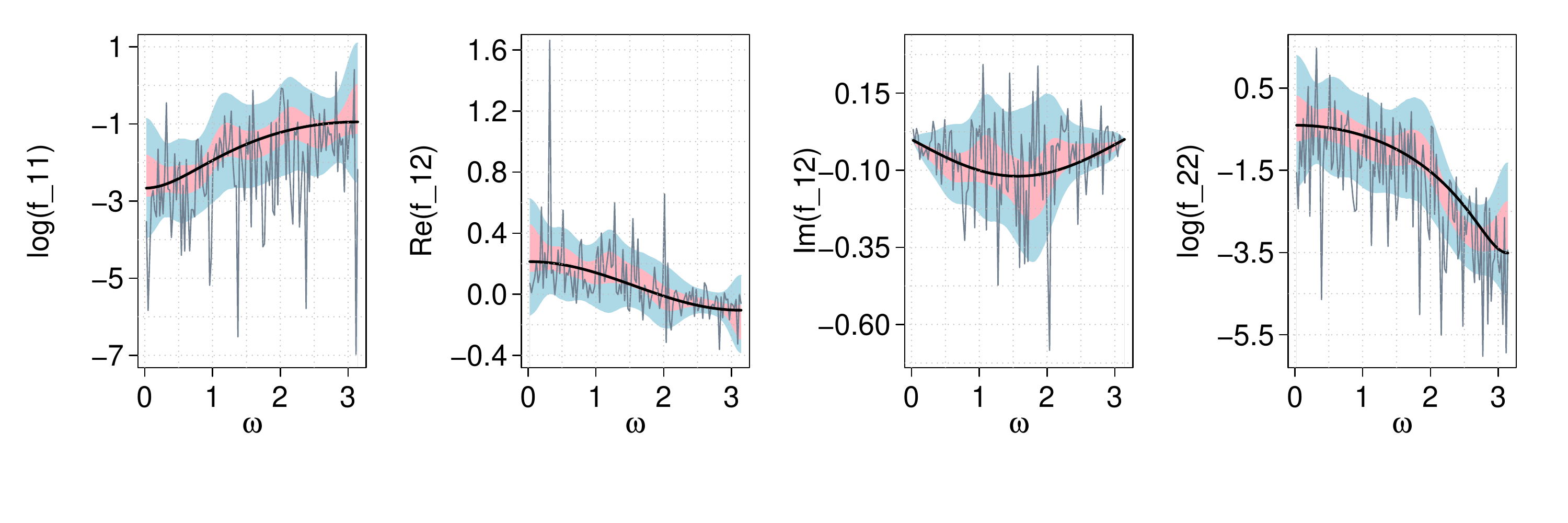}
        \vspace{-2\baselineskip}
        \caption{}
    \end{subfigure}
    \begin{subfigure}[b]{0.7\textheight}
        \includegraphics[width=1\textwidth, height=.2\textheight]{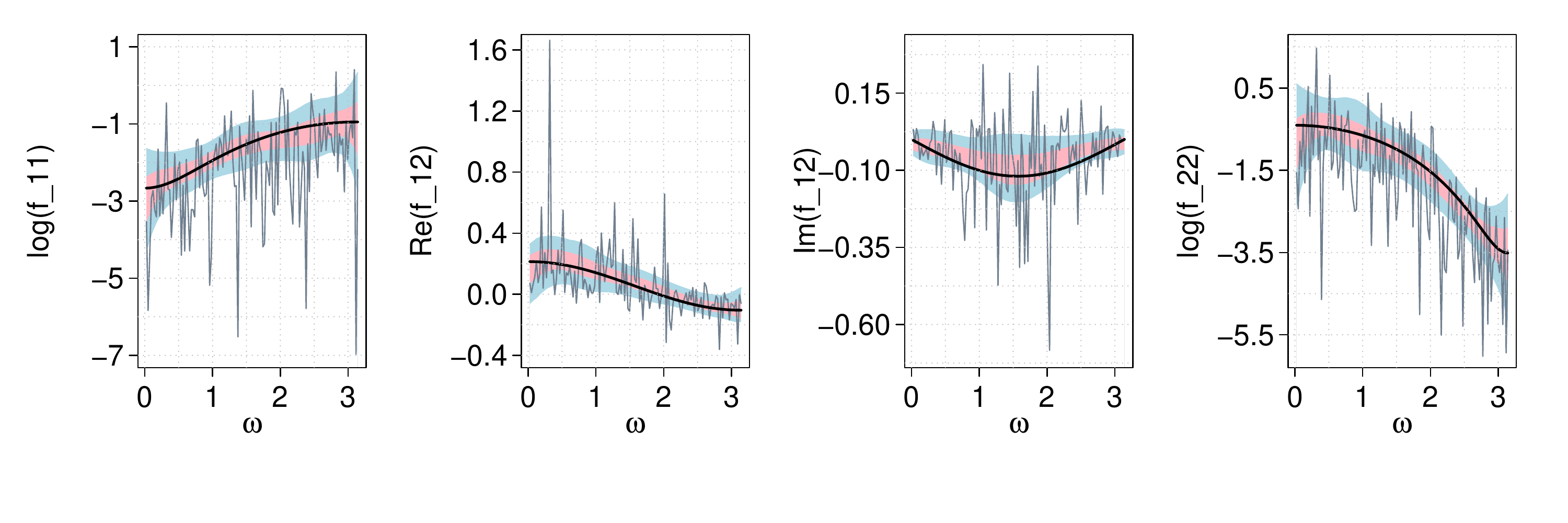}
        \vspace{-2\baselineskip}
        \caption{}
    \end{subfigure}
    \caption{Posterior credible regions for the realisation of the VMA(1) model same as the one referred by Figure \ref{Fig1.4_supp} for (a) the VNPC procedure with fixed order $1$ for the parametric working model, (b) the VAR procedure with order selected by AIC and (c) the VNP procedure. Pointwise $90\%$ regions are visualised in shaded pink, uniform $90\%$ regions are in shaded blue, the true spectral density is given by the black solid line and the periodogram is shown in grey.}
    \label{Fig1.6_supp}
\end{figure}

The following simulation study considers the VAR(2) and VMA(1) models with the same coefficients and covariance matrix as (14) and (15) in the paper, while the innovation term follows a bivariate Laplace distribution rather than the Gaussian distribution. Simulation results are provided in Table \ref{Tab1.1}. These clearly demonstrate that the VNPC procedure also performs well in the non-Gaussian case. Figures \ref{Fig1.4_supp}-\ref{Fig1.6_supp} display the estimates for one $n=256$ realisation by the three approaches.

\begin{table}[H]
    \centering
    {\scriptsize
    \begin{tabular}{cccccccccc}
         \hline
         \multicolumn{4}{c}{} & \multicolumn{2}{c}{VAR(2) model} & \multicolumn{4}{c}{}\\
         \multicolumn{1}{c}{} & \multicolumn{3}{c}{$n=256$} & \multicolumn{3}{c}{$n=512$} & \multicolumn{3}{c}{$n=1024$}\\
         \cline{2-10} 
         & VNPC(1) & VNP & VAR & VNPC(1) & VNP & VAR & VNPC(1) & VNP & VAR\\
         $L_{1}$-error & 0.104 & 0.108 & 0.080 & 0.081 & 0.085 & 0.057 & 0.062 & 0.065 & 0.039\\
         $L_{2}$-error & 0.135 & 0.138 & 0.105 & 0.107 & 0.111 & 0.074 & 0.082 & 0.087 & 0.051\\
         Coverage & 0.824 & 0.528 & 0.876 & 0.660 & 0.392 & 0.830 & 0.586 & 0.348 & 0.816\\
         Width $\mf_{11}$ & 0.339 & 0.315 & 0.202 & 0.177 & 0.171 & 0.120 & 0.110 & 0.105 & 0.076\\
         Width $\frakR\mf_{12}$ & 0.253 & 0.229 & 0.144 & 0.169 & 0.159 & 0.087 & 0.118 & 0.113 & 0.055\\
         Width $\frakI\mf_{12}$ & 0.209 & 0.183 & 0.120 & 0.139 & 0.126 & 0.074 & 0.096 & 0.089 & 0.049\\
         Width $\mf_{22}$ & 0.497 & 0.440 & 0.192 & 0.265 & 0.257 & 0.112 & 0.165 & 0.165 & 0.070\\
         \hline
         \multicolumn{4}{c}{} & \multicolumn{2}{c}{VMA(1) model} & \multicolumn{4}{c}{}\\
         \multicolumn{1}{c}{} & \multicolumn{3}{c}{$n=256$} & \multicolumn{3}{c}{$n=512$} & \multicolumn{3}{c}{$n=1024$}\\
         \cline{2-10} 
         & VNPC(1) & VNP & VAR & VNPC(1) & VNP & VAR & VNPC(1) & VNP & VAR\\
         $L_{1}$-error & 0.094 &	0.109 &	0.163 &	0.069 & 0.080 & 0.125 & 0.054 & 0.059 & 0.096\\
         $L_{2}$-error & 0.115 & 0.130 & 0.198 & 0.085 & 0.095 & 0.149 & 0.066 & 0.070 & 0.113\\
         Coverage & 0.870 & 0.574 & 0.958 & 0.798 & 0.450 & 0.954 & 0.648 & 0.314 & 0.948\\
         Width $\mf_{11}$ & 0.341 & 0.292 & 1.299 & 0.202 & 0.193 & 0.670 & 0.133 & 0.135 & 0.413\\
         Width $\frakR\mf_{12}$ & 0.228 & 0.207 & 0.603 & 0.139 & 0.140 & 0.414 & 0.088 & 0.099 & 0.291\\
         Width $\frakI\mf_{12}$ & 0.167 & 0.138 & 0.453 & 0.101 & 0.100 & 0.309 & 0.067 & 0.081 & 0.220\\
         Width $\mf_{22}$ & 0.492 & 0.457 & 1.882 & 0.298 & 0.297 & 0.989 & 0.197 & 0.209 & 0.624\\
         \hline
    \end{tabular}}
    \caption{Average $L_{1}$-, $L_{2}$-errors, empirical coverages and median width of uniform $90\%$ credible regions of the VNPC estimates with fixed parametric working model order $1$, the VNP,  and the VAR estimates with order selected by AIC for $N=500$ realisations of the VAR(2) model (14) and the VMA(1) model (15) in the paper with a bivariate Laplace distributed innovation term.}
    \label{Tab1.1}
\end{table}

%% file: section2_supp.tex
\begin{mypf}{Theorem}{1}
    Note when $\mf=\mf_{\pa}$, it holds that $\bsC_{nd}=\bsI_{nd}$ and $\bsF_{nd}^{*}\bsC_{nd}^{-1}\bsF_{nd}=\bsI_{nd}$, so (a) is proved.

    To show (b), consider a time series $\mbfX_{n}=(\mX_{1},...,\mX_{n})\sim p^{n}_{C}$. Let $\mbfY_{n}=\bsF_{nd}^{*}\bsC_{nd}^{-1}\bsF_{nd}\mbfX_{n}$. Because $(\bsF_{nd}^{*}\bsC_{nd}^{-1}\bsF_{nd})^{-1}=\bsF_{nd}^{*}\bsC_{nd}\bsF_{nd}$ and $|(\bsF_{nd}^{*}\bsC_{nd}^{-1}\bsF_{nd})^{-1}|=|\bsC_{nd}|$, we have $\mbfY_{n}\sim p_{\pa}$. Denote $\underline{\tilde{Y}}_{j}$ the Fourier coefficient of $\mbfY_{n}$ at $\omega_{j}$. By Corollary 7.2.1 of \cite{brillinger2001}, it follows that \begin{align*}
    \bbE I_{nd,\omega_{j}}(\mbfX_{n})&=\bbE I_{nd,\omega_{j}}\lr \bsF_{nd}^{*}\bsC_{nd}\bsF_{nd}\mbfY_{n}\rr\\
    &=\frac{1}{2\pi}\bsC_{nd,j}\bbE\lr\underline{\tilde{Y}}_{j}\underline{\tilde{Y}}_{j}^{*}\rr\bsC_{nd,j}^{*}\\
    &=\bsC_{nd,j}\bbE\lr I_{nd,\omega_{j}}(\mbfY_{n})\rr\bsC_{nd,j}^{*}\\
    &=\mf^{1/2}(\omega_{j})\mf_{\pa}^{-1/2}(\omega_{j})\mf_{\pa}(\omega_{j})\mf_{\pa}^{-1/2}(\omega_{j})\mf^{1/2}(\omega_{j})+o(1)\\
    &=\mf(\omega_{j})+o(1),
    \end{align*} where $\bsC_{nd,j}$ is the $j$-th diagonal block of $\bsC_{nd}$.

    Part (c) is shown by \begin{align*}
    &\Cov\lr I_{nd,\omega_{j}}(\mbfX_{n}),I_{nd,\omega_{k}}(\mbfX_{n})\rr\\
    =&\Cov\lr I_{nd,\omega_{j}}\lr\bsF_{nd}^{*}\bsC_{nd}\bsF_{nd}\mbfY_{n}\rr,I_{nd,\omega_{k}}\lr\bsF_{nd}^{*}\bsC_{nd}\bsF_{nd}\mbfY_{n}\rr\rr\\
    =&\Cov\lr\bsC_{nd,j}\lr\frac{1}{2\pi}\underline{\tilde{Y}}_{j}\underline{\tilde{Y}}_{j}^{*}\rr\bsC_{nd,j}^{*},\bsC_{nd,k}\lr\frac{1}{2\pi}\underline{\tilde{Y}}_{k}\underline{\tilde{Y}}_{k}^{*}\rr\bsC_{nd,k}^{*}\rr\\
    =&\Cov\lr\bsC_{nd,j} I_{nd,\omega_{j}}(\mbfY_{n})\bsC_{nd,j}^{*},\bsC_{nd,k} I_{nd,\omega_{k}}(\mbfY_{n})\bsC_{nd,k}^{*}\rr.
\end{align*}
\end{mypf}

We use a similar idea to the proof of Theorem 2.2 in \cite{meier2020} to show Theorem 2. In particular, we consider a corrected likelihood for real-valued random vectors, which collect the real and imaginary parts of Fourier coefficients. Using the mapping $\calB$ defined in (7.7) in \cite{meier2020}, the real-valued version of the corrected likelihood is given by \begin{align}\label{eq2.1_supp}
    p_{C}^{n}(\mbfZ_{n}|\mf)\propto|\tC_{nd}\tC_{nd}^{T}|^{-1/2}p_{\pa}^{n}(\bstF_{nd}^{T}\tC_{nd}^{-1}\bstF_{nd}\mbfZ_{n}),
\end{align} with \begin{flalign*}\scriptsize{
    \tC_{nd}:=\begin{pmatrix}
    \mf^{1/2}(0)\mf_{\pa}^{-1/2}(0) & & & &\\
    & \calB(\mf^{1/2}(\omega_{1})\mf_{\pa}^{-1/2}(\omega_{1})) & & &\\
    & & \ddots & &\\
    & & & \calB(\mf^{1/2}(\omega_{N})\mf_{\pa}^{1/2}(\omega_{N})) &\\
    & & & & \mf^{1/2}(\pi)\mf_{\pa}^{-1/2}(\pi)
    \end{pmatrix}}&&
\end{flalign*} for $n$ even and \begin{flalign*}\scriptsize{
    \tC_{nd}:=\begin{pmatrix}
    \mf^{1/2}(0)\mf_{\pa}^{-1/2}(0) & & &\\
    & \calB(\mf^{1/2}(\omega_{1})\mf_{\pa}^{-1/2}(\omega_{1})) & &\\
    & & \ddots &\\
    & & & \calB(\mf^{1/2}(\omega_{N})\mf_{\pa}^{1/2}(\omega_{N}))
    \end{pmatrix}}&&
\end{flalign*} for $n$ odd, where $\bstF_{nd}$ is a real-valued multivariate discrete Fourier transform
operator and orthogonal. 

\begin{mypf}{Theorem}{2}
    \cite{meier2020} have proved the mutual contiguity of the full Gaussian likelihood and the Whittle likelihood (see Theorem 2.2 in the article). Therefore, we only need to show the mutual contiguity of the joint densities under the Whittle likelihood and the corrected likelihood. To this end, we use the real-valued corrected likelihood (\ref{eq2.1_supp}) and follow the arguments from the proof of the mutual contiguity of the Whittle likelihood and the true Gaussian likelihood in the univariate case in \cite{choudhuri2004a}. Then, it is sufficient to show the tightness of \begin{align}\label{eq2.2_supp}
    \log\frac{P_{C}^{n}}{P_{W}^{n}}&\propto-\frac{1}{2}\lr\log|\tC_{nd}\tC_{nd}^{T}|+\log|\bfGa_{nd,\pa}|-\log|\tD_{nd}|\rr\nonumber\\
    &\qquad\qquad-\frac{1}{2}\lr\mbfZ_{n}^{T}\bstF_{nd}^{T}\tC_{nd}^{-1T}\bstF_{nd}\bfGa_{nd,\pa}^{-1}\bstF_{nd}^{T}\tC_{nd}^{-1}\bstF_{nd}\mbfZ_{n}-\mbfZ_{n}^{T}\bstF_{nd}^{T}\tD_{nd}^{-1}\bstF_{nd}\mbfZ_{n}\rr\nonumber\\
    &:=-\frac{1}{2}\bsA_{n}-\frac{1}{2}\bsB_{n},
\end{align} where \begin{align*}
    P_{W}^{n}\propto|\tD_{nd}|^{-1/2}\exp\lb-\frac{1}{2}\mbfZ_{n}^{T}\bstF_{nd}^{T}\tD_{nd}^{-1}\bsF_{nd}\mbfZ_{n}\rb
\end{align*} with $\tD_{nd}$ defined in (7.7) in \cite{meier2020} and \begin{align*}
    P_{C}^{n}\propto|\tC_{nd}\tC_{nd}^{T}|^{-1/2}|\bfGa_{nd,\pa}|^{-1/2}\exp\lb-\frac{1}{2}\mbfZ_{n}^{T}\bstF_{nd}^{T}\tC_{nd}^{-1T}\bstF_{nd}\bfGa_{nd,\pa}^{-1}\bstF_{nd}^{T}\tC_{nd}^{-1}\bstF_{nd}\mbfZ_{n}\rb.
\end{align*} Let $\tD_{nd,\pa}:=\tC_{nd}^{-1}\tD_{nd}\tC_{nd}^{-T}$. Clearly, it follows that \begin{align*}
    \log|\tD_{nd}|-\log|\tC_{nd}\tC_{nd}^{T}|=\log|\tC_{nd}^{-1}\tD_{nd}\tC_{nd}^{-T}|=\log|\tD_{nd,\pa}|.
\end{align*} In terms of Lemma 7.1 and the same arguments in \cite{meier2020}, it holds $\bsA_{n}=\log|\bfGa_{nd,\pa}|-\log|\tD_{nd,\pa}|=O(1)$ as $n\rightarrow\infty$.

Let $\tilde{\bfGa}_{nd,\pa}:=\bstF_{nd}\bfGa_{nd,\pa}\bstF_{nd}^{T}$, then $\tilde{\bfGa}_{nd,\pa}^{-1}=\bstF_{nd}\bfGa_{nd,\pa}^{-1}\bstF_{nd}^{T}$. Under $P_{C}^{n}$, it follows $\mbftZ_{n}:=\bstF_{nd}\mbfZ_{n}\sim N(\mzero, \tC_{nd}\bstF_{nd}\bfGa_{nd,\pa}\bstF_{nd}^{T}\tC_{nd}^{T})$. Following the arguments of the proofs in \cite{choudhuri2004a} and \cite{meier2020}, it holds that \begin{align*}
     \bbE_{P_{C}^{n}}\bsB_{n}&=\tr\ls\lr\tC_{nd}^{-T}\bstF_{nd}\bfGa_{nd,\pa}^{-1}\bstF_{nd}^{T}\tC_{nd}^{-1}-\tD_{nd}^{-1}\rr\tC_{nd}\bstF_{nd}\bfGa_{nd,\pa}\bstF_{nd}^{T}\tC_{nd}^{T}\rs\nonumber\\
     &=\tr\lr \bsI_{nd}-\tilde{\bfGa}_{nd,\pa}\tD_{nd,\pa}^{-1}\rr
\end{align*} and \begin{align*}
    \Var_{P_{C}^{n}}\bsB_{n}&=2\tr\ls\lr\tC_{nd}\bstF_{nd}\bfGa_{nd,\pa}\bstF_{nd}^{T}\tC_{nd}^{T}\lr\tC_{nd}^{-T}\bstF_{nd}\bfGa_{nd,\pa}^{-1}\bstF_{nd}^{T}\tC_{nd}^{-1}-\tD_{nd}^{-1}\rr\rr^{2}\rs\nonumber\\
    &=2\tr\ls\lr \bsI_{nd}-\tilde{\bfGa}_{nd,\pa}\tD_{nd,\pa}^{-1}\rr^{2}\rs.
\end{align*} Let $\bsH_{nd,\pa}:=\tilde{\bfGa}_{nd,\pa}-\tD_{nd,\pa}\in\bbR^{nd\times nd}$. By Lemma 7.3 in \cite{meier2020} and $|\tr(\bsA\bsB)|\leq\Vert\bsA\Vert\Vert\bsB\Vert$ for any $\bsA,\bsB\in\bbR^{d\times d}$ and the Frobenius norm $\Vert\cdot\Vert$, it yields that \begin{align*}
    \la\bbE_{P_{C}^{n}}\bsB_{n}\ra\lesssim\lr\lf\frac{n}{2}\rf+1\rr n^{-1}b_{0,\pa}^{-1}=O(1)
\end{align*} with $b_{0,\pa}$ from Assumption 1, where $a_{n}\lesssim b_{n}$ indicates that there exists a positive constant $C$ such that $a_{n}\leq Cb_{n}$ for all $n$. It also can be shown that \begin{align*}
    \Var_{P_{C}^{n}}\bsB_{n}\lesssim\lr\lf\frac{n}{2}\rf+1\rr^{2}n^{-2}b_{0,\pa}^{-2}=O(1).
\end{align*} Under $P_{W}^{n}$, it holds $\mbftZ_{n}=\bstF_{nd}\mbfZ_{n}\sim N(\mzero,\tD_{nd})$. Similarly, we have \begin{align*}
    \bbE_{P_{W}^{n}}\bsB_{n}&=\tr\ls\lr\tC_{nd}^{-T}\bstF_{nd}\bfGa_{nd,\pa}^{-1}\bstF_{nd}^{T}\tC_{nd}^{-1}-\tD_{nd}^{-1}\rr \tD_{nd}\rs\nonumber\\
    &=\tr\lr \tD_{nd,\pa}\tilde{\bfGa}_{nd,\pa}^{-1}-\bsI_{nd}\rr
\end{align*} and \begin{align*}
    \Var_{P_{W}^{n}}\bsB_{n}&=2\tr\ls\lr\tD_{nd}\lr\tC_{nd}^{-T}\bstF_{nd}\bfGa_{nd,\pa}^{-1}\bstF_{nd}^{T}\tC_{nd}^{-1}-\tD_{nd}^{-1}\rr\rr^{2}\rs\nonumber\\
    &=2\tr\ls\lr \tD_{nd,\pa}\tilde{\bfGa}_{nd,\pa}^{-1}-\bsI_{nd}\rr^{2}\rs.
\end{align*} By the same arguments of the proof of Theorem 2.2 in \cite{meier2020}, it holds $|\bbE_{P_{W}^{n}}\bsB_{n}|=O(1)$ and $\Var_{P_{W}^{n}}\bsB_{n}=O(1)$. Then, the proof is concluded.

\end{mypf}

%% file: main.bbl
\begin{thebibliography}{54}
\expandafter\ifx\csname natexlab\endcsname\relax\def\natexlab#1{#1}\fi
\providecommand{\url}[1]{\texttt{#1}}
\providecommand{\href}[2]{#2}
\providecommand{\path}[1]{#1}
\providecommand{\DOIprefix}{doi:}
\providecommand{\ArXivprefix}{arXiv:}
\providecommand{\URLprefix}{URL: }
\providecommand{\Pubmedprefix}{pmid:}
\providecommand{\doi}[1]{\href{http://dx.doi.org/#1}{\path{#1}}}
\providecommand{\Pubmed}[1]{\href{pmid:#1}{\path{#1}}}
\providecommand{\bibinfo}[2]{#2}
\ifx\xfnm\relax \def\xfnm[#1]{\unskip,\space#1}\fi
%Type = Article
\bibitem[{Aicher et~al.(2019)Aicher, Ma, Foti and Fox}]{aicher2019}
\bibinfo{author}{Aicher, C.}, \bibinfo{author}{Ma, Y.A.},
  \bibinfo{author}{Foti, N.J.}, \bibinfo{author}{Fox, E.B.},
  \bibinfo{year}{2019}.
\newblock \bibinfo{title}{Stochastic gradient {MCMC} for state space models}.
\newblock \bibinfo{journal}{SIAM J. Math. Anal.} \bibinfo{volume}{1(3)},
  \bibinfo{pages}{555--587}.
\newblock \bibinfo{note}{DOI: \url{10.1137/18M1214780}}.
%Type = Article
\bibitem[{Akaike(1974)}]{akaike1974}
\bibinfo{author}{Akaike, H.}, \bibinfo{year}{1974}.
\newblock \bibinfo{title}{A new look at the statistical model identification}.
\newblock \bibinfo{journal}{IEEE Trans. Automat. Contr.} \bibinfo{volume}{19},
  \bibinfo{pages}{716--723}.
\newblock \bibinfo{note}{DOI: \url{10.1109/TAC.1974.1100705}}.
%Type = Article
\bibitem[{Barigozzi and Hallin(2016)}]{barigozzi2016}
\bibinfo{author}{Barigozzi, M.}, \bibinfo{author}{Hallin, M.},
  \bibinfo{year}{2016}.
\newblock \bibinfo{title}{Generalized dynamic factor models and volatilities:
  recovering the market volatility shocks}.
\newblock \bibinfo{journal}{Econom. J.} \bibinfo{volume}{19(1)},
  \bibinfo{pages}{C33--C60}.
\newblock \bibinfo{note}{DOI: \url{10.1111/ectj.12047}}.
%Type = Article
\bibitem[{Berkowitz and Diebold(1998)}]{berkowitz1998}
\bibinfo{author}{Berkowitz, J.}, \bibinfo{author}{Diebold, F.X.},
  \bibinfo{year}{1998}.
\newblock \bibinfo{title}{Bootstrapping multivariate spectra}.
\newblock \bibinfo{journal}{Rev. Econ. Stat.} \bibinfo{volume}{80(4)},
  \bibinfo{pages}{664--666}.
\newblock \bibinfo{note}{DOI: \url{10.1162/003465398557753}}.
%Type = Book
\bibitem[{Brillinger(2001)}]{brillinger2001}
\bibinfo{author}{Brillinger, D.R.}, \bibinfo{year}{2001}.
\newblock \bibinfo{title}{Time Series: Data Analysis and Theory}.
\newblock Classics in Applied Mathematics, \bibinfo{publisher}{Society for
  Industrial and Applied Mathematics}, \bibinfo{address}{Philadelphia}.
%Type = Book
\bibitem[{Brockwell and Davis(1991)}]{brockwell1991}
\bibinfo{author}{Brockwell, P.J.}, \bibinfo{author}{Davis, R.A.},
  \bibinfo{year}{1991}.
\newblock \bibinfo{title}{Time series: Theory and methods}.
\newblock \bibinfo{edition}{2nd} ed., \bibinfo{publisher}{Springer},
  \bibinfo{address}{New York}.
%Type = Article
\bibitem[{Cadonna et~al.(2017)Cadonna, Kottas and Prado}]{cadonna2017}
\bibinfo{author}{Cadonna, A.}, \bibinfo{author}{Kottas, A.},
  \bibinfo{author}{Prado, R.}, \bibinfo{year}{2017}.
\newblock \bibinfo{title}{Bayesian mixture modeling for spectral density
  estimation}.
\newblock \bibinfo{journal}{Stat. Probab. Lett.} \bibinfo{volume}{125},
  \bibinfo{pages}{189--195}.
\newblock \bibinfo{note}{DOI: \url{10.1016/j.spl.2017.02.008}}.
%Type = Article
\bibitem[{Cadonna et~al.(2019)Cadonna, Kottas and Prado}]{cadonna2019}
\bibinfo{author}{Cadonna, A.}, \bibinfo{author}{Kottas, A.},
  \bibinfo{author}{Prado, R.}, \bibinfo{year}{2019}.
\newblock \bibinfo{title}{Bayesian spectral modeling for multiple time series}.
\newblock \bibinfo{journal}{J. Am. Stat. Assoc.} \bibinfo{volume}{114(528)},
  \bibinfo{pages}{1838--1853}.
\newblock \bibinfo{note}{DOI: \url{10.1080/01621459.2018.1520114}}.
%Type = Article
\bibitem[{Carvalho et~al.(2008)Carvalho, Chang, Lucas, Nevins, Wang and
  West}]{carvalho2008}
\bibinfo{author}{Carvalho, C.M.}, \bibinfo{author}{Chang, J.},
  \bibinfo{author}{Lucas, J.E.}, \bibinfo{author}{Nevins, J.R.},
  \bibinfo{author}{Wang, Q.}, \bibinfo{author}{West, M.}, \bibinfo{year}{2008}.
\newblock \bibinfo{title}{High-dimensional sparse factor modeling: applications
  in gene expression genomics}.
\newblock \bibinfo{journal}{J. Am. Stat. Assoc.} \bibinfo{volume}{103(484)},
  \bibinfo{pages}{1438--1456}.
\newblock \bibinfo{note}{DOI: \url{10.1198/016214508000000869}}.
%Type = Article
\bibitem[{Choudhuri et~al.(2004a)Choudhuri, Ghosal and Roy}]{choudhuri2004b}
\bibinfo{author}{Choudhuri, N.}, \bibinfo{author}{Ghosal, S.},
  \bibinfo{author}{Roy, A.}, \bibinfo{year}{2004}a.
\newblock \bibinfo{title}{Bayesian estimation of the spectral density of a time
  series}.
\newblock \bibinfo{journal}{J. Am. Stat. Assoc.} \bibinfo{volume}{99},
  \bibinfo{pages}{1050--1059}.
\newblock \bibinfo{note}{DOI: \url{10.1198/016214504000000557}}.
%Type = Article
\bibitem[{Choudhuri et~al.(2004b)Choudhuri, Ghosal and Roy}]{choudhuri2004a}
\bibinfo{author}{Choudhuri, N.}, \bibinfo{author}{Ghosal, S.},
  \bibinfo{author}{Roy, A.}, \bibinfo{year}{2004}b.
\newblock \bibinfo{title}{Contiguity of the {W}hittle measure for a {G}aussian
  time series}.
\newblock \bibinfo{journal}{Biometrika} \bibinfo{volume}{91},
  \bibinfo{pages}{211--218}.
\newblock \bibinfo{note}{DOI: \url{10.1214/18-BA1126}}.
%Type = Article
\bibitem[{Dai and Guo(2004)}]{dai2004}
\bibinfo{author}{Dai, M.}, \bibinfo{author}{Guo, W.}, \bibinfo{year}{2004}.
\newblock \bibinfo{title}{Multivariate spectral analysis using {C}holesky
  decomposition}.
\newblock \bibinfo{journal}{Biometrika} \bibinfo{volume}{91(3)},
  \bibinfo{pages}{629--643}.
\newblock \bibinfo{note}{DOI: \url{10.1093/biomet/91.3.629}}.
%Type = Article
\bibitem[{Edwards et~al.(2019)Edwards, Meyer and Christensen}]{edwards2019}
\bibinfo{author}{Edwards, M.C.}, \bibinfo{author}{Meyer, R.},
  \bibinfo{author}{Christensen, N.}, \bibinfo{year}{2019}.
\newblock \bibinfo{title}{Bayesian nonparametric spectral density estimation
  using {B}-spline priors}.
\newblock \bibinfo{journal}{Stat. Comput.} \bibinfo{volume}{29(1)},
  \bibinfo{pages}{67--78}.
\newblock \bibinfo{note}{DOI: \url{10.1007/s11222-017-9796-9}}.
%Type = Article
\bibitem[{Fiecas(2014)}]{fiecas2014}
\bibinfo{author}{Fiecas, M.}, \bibinfo{year}{2014}.
\newblock \bibinfo{title}{Data-driven shrinkage of the spectral density matrix
  of a high-dimensional time series}.
\newblock \bibinfo{journal}{Electron. J. Stat.} \bibinfo{volume}{8(2)},
  \bibinfo{pages}{2975--3003}.
\newblock \bibinfo{note}{DOI: \url{10.1214/14-EJS977}}.
%Type = Article
\bibitem[{Fox et~al.(2014)Fox, Hughes, Sudderth and Jordan}]{fox2014}
\bibinfo{author}{Fox, E.B.}, \bibinfo{author}{Hughes, M.C.},
  \bibinfo{author}{Sudderth, E.B.}, \bibinfo{author}{Jordan, M.I.},
  \bibinfo{year}{2014}.
\newblock \bibinfo{title}{Joint modeling of multiple times series via the beta
  process with application to motion capture segmentation}.
\newblock \bibinfo{journal}{Ann. Appl. Stat.} \bibinfo{volume}{8(3)},
  \bibinfo{pages}{1281--1313}.
\newblock \bibinfo{note}{DOI: \url{10.1214/14-AOAS742}}.
%Type = Article
\bibitem[{Fox et~al.(2011)Fox, Sudderth, Jordan and Willsky}]{fox2011}
\bibinfo{author}{Fox, E.B.}, \bibinfo{author}{Sudderth, E.B.},
  \bibinfo{author}{Jordan, M.I.}, \bibinfo{author}{Willsky, A.S.},
  \bibinfo{year}{2011}.
\newblock \bibinfo{title}{Bayesian nonparametric inference of switching dynamic
  linear models}.
\newblock \bibinfo{journal}{IEEE Trans. Signal Process.}
  \bibinfo{volume}{59(4)}, \bibinfo{pages}{1569--1585}.
\newblock \bibinfo{note}{DOI: \url{10.1109/TSP.2010.2102756}}.
%Type = Article
\bibitem[{Franke and Hardle(1992)}]{franke1992}
\bibinfo{author}{Franke, J.}, \bibinfo{author}{Hardle, W.},
  \bibinfo{year}{1992}.
\newblock \bibinfo{title}{On bootstrapping kernel spectral estimates}.
\newblock \bibinfo{journal}{Ann. Stat.} \bibinfo{volume}{20(1)},
  \bibinfo{pages}{121--145}.
\newblock \bibinfo{note}{DOI: \url{10.1214/aos/1176348515}}.
%Type = Article
\bibitem[{Hastings(1970)}]{hastings1970}
\bibinfo{author}{Hastings, W.K.}, \bibinfo{year}{1970}.
\newblock \bibinfo{title}{Monte {C}arki sampling methods using {M}arkov chains
  and their applications}.
\newblock \bibinfo{journal}{Biometrika} \bibinfo{volume}{57(1)},
  \bibinfo{pages}{97--109}.
\newblock \bibinfo{note}{DOI: \url{10.1093/biomet/57.1.97}}.
%Type = Article
\bibitem[{Hu and Prado(2023)}]{hu2023}
\bibinfo{author}{Hu, Z.}, \bibinfo{author}{Prado, R.}, \bibinfo{year}{2023}.
\newblock \bibinfo{title}{Fast {B}ayesian inference on spectral analysis of
  multivariate stationary time series}.
\newblock \bibinfo{journal}{Comput. Stat. Data Anal.} \bibinfo{volume}{178},
  \bibinfo{pages}{107596}.
\newblock \bibinfo{note}{DOI: \url{10.1016/j.csda.2022.107596}}.
%Type = Misc
\bibitem[{{Iowa State University}(2022)}]{iowa2022}
\bibinfo{author}{{Iowa State University}}, \bibinfo{year}{2022}.
\newblock \bibinfo{title}{Iowa environmental mesonet}.
\newblock \bibinfo{note}{\url{https://mesonet.agron.iastate.edu/}. Accessed:
  2022-12-19}.
%Type = Article
\bibitem[{Jentsch and Kreiss(2010)}]{jentsch2010}
\bibinfo{author}{Jentsch, C.}, \bibinfo{author}{Kreiss, J.P.},
  \bibinfo{year}{2010}.
\newblock \bibinfo{title}{The multiple hybrid bootstrap - resampling
  multivariate linear processes}.
\newblock \bibinfo{journal}{J. Multivar. Anal.} \bibinfo{volume}{101(10)},
  \bibinfo{pages}{2320--2345}.
\newblock \bibinfo{note}{DOI: \url{10.1016/j.jmva.2010.06.005}}.
%Type = Article
\bibitem[{Jentsch and Politis(2015)}]{jentsch2015}
\bibinfo{author}{Jentsch, C.}, \bibinfo{author}{Politis, D.N.},
  \bibinfo{year}{2015}.
\newblock \bibinfo{title}{Covariance matrix estimation and linear process
  bootstrap for multivariate time series of possible increasing dimension}.
\newblock \bibinfo{journal}{Ann. Stat.} \bibinfo{volume}{43(3)},
  \bibinfo{pages}{1117--1140}.
\newblock \bibinfo{note}{DOI: \url{10.1214/14-AOS1301}}.
%Type = Article
\bibitem[{Kalli and Griffin(2018)}]{kalli2018}
\bibinfo{author}{Kalli, M.}, \bibinfo{author}{Griffin, J.E.},
  \bibinfo{year}{2018}.
\newblock \bibinfo{title}{Bayesian nonparametric vector autoregressive models}.
\newblock \bibinfo{journal}{J. Econom.} \bibinfo{volume}{203(2)},
  \bibinfo{pages}{267--282}.
\newblock \bibinfo{note}{DOI: \url{10.1016/j.jeconom.2017.11.009}}.
%Type = Article
\bibitem[{Kastner and Huber(2020)}]{kastner2020}
\bibinfo{author}{Kastner, G.}, \bibinfo{author}{Huber, F.},
  \bibinfo{year}{2020}.
\newblock \bibinfo{title}{Sparse {B}ayesian vector autoregressions in huge
  dimensions}.
\newblock \bibinfo{journal}{J. Forecast.} \bibinfo{volume}{39(7)},
  \bibinfo{pages}{1142--1165}.
\newblock \bibinfo{note}{DOI: \url{10.1002/for.2680}}.
%Type = Article
\bibitem[{Kirch et~al.(2019)Kirch, Edwards, Meier and Meyer}]{kirch2019}
\bibinfo{author}{Kirch, C.}, \bibinfo{author}{Edwards, M.C.},
  \bibinfo{author}{Meier, A.}, \bibinfo{author}{Meyer, R.},
  \bibinfo{year}{2019}.
\newblock \bibinfo{title}{Beyond {W}hittle: nonparametric correction of a
  parametric likelihood with a focus on {B}ayesian time series analysis}.
\newblock \bibinfo{journal}{Bayesian Anal.} \bibinfo{volume}{14},
  \bibinfo{pages}{1037--1073}.
\newblock \bibinfo{note}{DOI: \url{10.1214/18-BA1126}}.
%Type = Article
\bibitem[{Kirch and Politis(2011)}]{kirch2011}
\bibinfo{author}{Kirch, C.}, \bibinfo{author}{Politis, D.N.},
  \bibinfo{year}{2011}.
\newblock \bibinfo{title}{{TFT}-bootstrap: resampling time series in the
  frequency domain to obtain replicates in the time domain}.
\newblock \bibinfo{journal}{Ann. Stat.} \bibinfo{volume}{39(3)},
  \bibinfo{pages}{1427--1470}.
\newblock \bibinfo{note}{DOI: \url{10.1214/10-AOS868}}.
%Type = Article
\bibitem[{Koop and Korobilis(2010)}]{koop2010}
\bibinfo{author}{Koop, G.}, \bibinfo{author}{Korobilis, D.},
  \bibinfo{year}{2010}.
\newblock \bibinfo{title}{Bayesian multivariate time series methods for
  empirical macroeconomics}.
\newblock \bibinfo{journal}{Found. Trends Econom.} \bibinfo{volume}{3},
  \bibinfo{pages}{267--358}.
%Type = Article
\bibitem[{Krafty and Collinge(2013)}]{krafty2013}
\bibinfo{author}{Krafty, R.T.}, \bibinfo{author}{Collinge, W.O.},
  \bibinfo{year}{2013}.
\newblock \bibinfo{title}{Penalized multivariate {W}hittle likelihood for power
  spectrum estimation}.
\newblock \bibinfo{journal}{Biometrika} \bibinfo{volume}{100(2)},
  \bibinfo{pages}{447--458}.
\newblock \bibinfo{note}{DOI: \url{10.1093/biomet/ass088}}.
%Type = Article
\bibitem[{Krampe et~al.(2019)Krampe, Kreiss and Paparoditis}]{krampe2019}
\bibinfo{author}{Krampe, J.}, \bibinfo{author}{Kreiss, J.P.},
  \bibinfo{author}{Paparoditis, E.}, \bibinfo{year}{2019}.
\newblock \bibinfo{title}{Bootstrap based inference for sparse high-dimensional
  time series models}.
\newblock \bibinfo{journal}{arXiv preprint} \bibinfo{note}{ArXiv:
  \url{1806.11083v3}}.
%Type = Article
\bibitem[{Kreiss and Paparoditis(2003)}]{kreiss2003}
\bibinfo{author}{Kreiss, J.P.}, \bibinfo{author}{Paparoditis, E.},
  \bibinfo{year}{2003}.
\newblock \bibinfo{title}{Autoregressive-aided periodogram bootstrap for time
  series}.
\newblock \bibinfo{journal}{Ann. Stat.} \bibinfo{volume}{31(6)},
  \bibinfo{pages}{1923--1955}.
\newblock \bibinfo{note}{DOI: \url{10.1214/aos/1074290332}}.
%Type = Book
\bibitem[{L\"{u}tkepohl(2005)}]{lutkepohl2005}
\bibinfo{author}{L\"{u}tkepohl, H.}, \bibinfo{year}{2005}.
\newblock \bibinfo{title}{New Introduction to Multiple Time Series Analysis}.
\newblock \bibinfo{publisher}{Springer Berlin}, \bibinfo{address}{Heidelberg}.
%Type = Misc
\bibitem[{Mannarano(1998)}]{mannarano1998}
\bibinfo{author}{Mannarano, D.}, \bibinfo{year}{1998}.
\newblock \bibinfo{title}{Automated surface observing system ({ASOS}) usrs's
  guide}.
\newblock \bibinfo{note}{\url{https://www.weather.gov/media/asos/aum-toc.pdf}}.
%Type = Article
\bibitem[{Maturana-Russel and Meyer(2021)}]{russel2021}
\bibinfo{author}{Maturana-Russel, P.}, \bibinfo{author}{Meyer, R.},
  \bibinfo{year}{2021}.
\newblock \bibinfo{title}{Bayesian spectral density estimation using
  {P}-splines with quantile-based knot placement}.
\newblock \bibinfo{journal}{Comput. Stat.} \bibinfo{volume}{36(3)},
  \bibinfo{pages}{2055--2077}.
\newblock \bibinfo{note}{DOI: \url{10.1007/s00180-021-01066-7}}.
%Type = Article
\bibitem[{McMurry and Politis(2010)}]{mcmurry2010}
\bibinfo{author}{McMurry, T.L.}, \bibinfo{author}{Politis, D.N.},
  \bibinfo{year}{2010}.
\newblock \bibinfo{title}{Banded and tapered estimates for autocovariance
  matrices and the linear process bootstrap}.
\newblock \bibinfo{journal}{J. Time Ser. Anal.} \bibinfo{volume}{31(6)},
  \bibinfo{pages}{471--482}.
\newblock \bibinfo{note}{DOI: \url{10.1111/j.1467-9892.2010.00679.x}}.
%Type = Phdthesis
\bibitem[{Meier(2018)}]{meier2018}
\bibinfo{author}{Meier, A.}, \bibinfo{year}{2018}.
\newblock \bibinfo{title}{A matrix {G}amma process and applications to
  {B}ayesian analysis of multivariate time series}.
\newblock Ph.D. thesis. Otto von Guericke University Magdeburg.
\newblock \bibinfo{note}{DOI: \url{10.25673/13407}}.
%Type = Manual
\bibitem[{Meier et~al.(2022)Meier, Kirch, Edwards and
  Meyer}]{beyondwhittle2022}
\bibinfo{author}{Meier, A.}, \bibinfo{author}{Kirch, C.},
  \bibinfo{author}{Edwards, M.C.}, \bibinfo{author}{Meyer, R.},
  \bibinfo{year}{2022}.
\newblock \bibinfo{title}{beyondWhittle: Bayesian Spectral Inference for
  Stationary Time Series}.
\newblock \bibinfo{organization}{R package version 1.1.3}.
\newblock
  \bibinfo{note}{\url{https://https://cran.r-project.org/web/packages/beyondWhittle/}}.
%Type = Article
\bibitem[{Meier et~al.(2020)Meier, Kirch and Meyer}]{meier2020}
\bibinfo{author}{Meier, A.}, \bibinfo{author}{Kirch, C.},
  \bibinfo{author}{Meyer, R.}, \bibinfo{year}{2020}.
\newblock \bibinfo{title}{Bayesian nonparametric analysis of multivariate time
  series: a matrix {G}amma process approach}.
\newblock \bibinfo{journal}{J. Multivar. Anal.} \bibinfo{volume}{175}.
\newblock \bibinfo{note}{DOI: \url{10.1016/j.jmva.2019.104560}}.
%Type = Article
\bibitem[{Meyer and Kreiss(2015)}]{meyer2015}
\bibinfo{author}{Meyer, M.}, \bibinfo{author}{Kreiss, J.},
  \bibinfo{year}{2015}.
\newblock \bibinfo{title}{On the vector autoregressive sieve bootstrap}.
\newblock \bibinfo{journal}{J. Time. Ser. Anal.} \bibinfo{volume}{36(3)},
  \bibinfo{pages}{377--397}.
\newblock \bibinfo{note}{DOI: \url{10.1111/jtsa.12090}}.
%Type = Article
\bibitem[{Muliere and Tardella(1998)}]{muliere1998}
\bibinfo{author}{Muliere, P.}, \bibinfo{author}{Tardella, T.},
  \bibinfo{year}{1998}.
\newblock \bibinfo{title}{Approximating distributions of random functionals of
  {F}erguson-{D}irichlet priors}.
\newblock \bibinfo{journal}{Can. J. Stat.} \bibinfo{volume}{26},
  \bibinfo{pages}{283--297}.
\newblock \bibinfo{note}{DOI: \url{10.2307/3315511}}.
%Type = Misc
\bibitem[{{National Oceanic and Atmospheric Administration}(2022)}]{noaa2022}
\bibinfo{author}{{National Oceanic and Atmospheric Administration}},
  \bibinfo{year}{2022}.
\newblock \bibinfo{title}{El {N}i\~{n}o/southern oscillation ({ENSO})}.
\newblock
  \bibinfo{note}{\url{https://www.ncei.noaa.gov/access/monitoring/enso}.
  Accessed: 2022-11-13}.
%Type = Article
\bibitem[{Rao and Yang(2021)}]{rao2021}
\bibinfo{author}{Rao, S.S.}, \bibinfo{author}{Yang, J.}, \bibinfo{year}{2021}.
\newblock \bibinfo{title}{Reconciling the {G}aussian and {W}hittle likelihood
  with an application to estimation in the frequency domain}.
\newblock \bibinfo{journal}{Ann. Stat.} \bibinfo{volume}{49(5)},
  \bibinfo{pages}{2774--2802}.
\newblock \bibinfo{note}{DOI: \url{10.1214/21-AOS2055}}.
%Type = Article
\bibitem[{Rodr\'{i}guez and Dunson(2011)}]{rodriguez2011}
\bibinfo{author}{Rodr\'{i}guez, A.}, \bibinfo{author}{Dunson, D.B.},
  \bibinfo{year}{2011}.
\newblock \bibinfo{title}{Nonparametric {B}ayesian models through probit
  stick-breaking processes}.
\newblock \bibinfo{journal}{Bayesian Anal.} \bibinfo{volume}{6(1)},
  \bibinfo{pages}{145--178}.
\newblock \bibinfo{note}{DOI: \url{10.1214/11-BA605}}.
%Type = Article
\bibitem[{Rosen and Stoffer(2007)}]{rosen2007}
\bibinfo{author}{Rosen, O.}, \bibinfo{author}{Stoffer, D.S.},
  \bibinfo{year}{2007}.
\newblock \bibinfo{title}{Automatic estimation of multivariate spectra via
  smoothing splines}.
\newblock \bibinfo{journal}{Biometrika} \bibinfo{volume}{94(2)},
  \bibinfo{pages}{335--345}.
\newblock \bibinfo{note}{DOI: \url{10.1093/biomet/asm022}}.
%Type = Article
\bibitem[{Rousseau et~al.(2012)Rousseau, Chopin and Liso}]{rousseau2012}
\bibinfo{author}{Rousseau, J.}, \bibinfo{author}{Chopin, N.},
  \bibinfo{author}{Liso, B.}, \bibinfo{year}{2012}.
\newblock \bibinfo{title}{Bayesian nonparametric estimation of the spectral
  density of a long or intermediate memory {G}aussian process}.
\newblock \bibinfo{journal}{Ann. Stat.} \bibinfo{volume}{40(2)},
  \bibinfo{pages}{964--995}.
\newblock \bibinfo{note}{DOI: \url{10.1214/11-AOS955}}.
%Type = Article
\bibitem[{Schwartz(1965)}]{schwartz1965}
\bibinfo{author}{Schwartz, L.}, \bibinfo{year}{1965}.
\newblock \bibinfo{title}{On {B}ayes procedures}.
\newblock \bibinfo{journal}{Z. Wahrsch. Verw. Gebiete} \bibinfo{volume}{4},
  \bibinfo{pages}{10--26}.
%Type = Article
\bibitem[{Shao and Wu(2007)}]{shaowu2007}
\bibinfo{author}{Shao, X.}, \bibinfo{author}{Wu, W.B.}, \bibinfo{year}{2007}.
\newblock \bibinfo{title}{Asymptotic spectral theory for nonlinear time
  series}.
\newblock \bibinfo{journal}{The Annals of statistics} \bibinfo{volume}{35},
  \bibinfo{pages}{1773--1801}.
%Type = Book
\bibitem[{Shumway and Stoffer(2011)}]{shumway2011}
\bibinfo{author}{Shumway, R.H.}, \bibinfo{author}{Stoffer, D.S.},
  \bibinfo{year}{2011}.
\newblock \bibinfo{title}{Time Series Analysis and Its Applications With R
  Examples}.
\newblock \bibinfo{edition}{3rd} ed., \bibinfo{publisher}{Springer},
  \bibinfo{address}{New York}.
%Type = Misc
\bibitem[{{Statistics New Zealand}(2020)}]{statsnz2020}
\bibinfo{author}{{Statistics New Zealand}}, \bibinfo{year}{2020}.
\newblock \bibinfo{title}{El {N}i\~{n}o southern oscillation}.
\newblock
  \bibinfo{note}{\url{https://www.stats.govt.nz/indicators/el-nino-southern-oscillation}.
  Accesssed: 2022-11-13}.
%Type = Manual
\bibitem[{Stoffer(2022)}]{stoffer2022}
\bibinfo{author}{Stoffer, D.}, \bibinfo{year}{2022}.
\newblock \bibinfo{title}{astsa: Applied Statistical Time Series Analysis}.
\newblock \bibinfo{organization}{R package version 1.16}.
\newblock \bibinfo{note}{\url{https://cran.r-project.org/web/packages/astsa/}}.
%Type = Article
\bibitem[{Szab\'{o} et~al.(2015)Szab\'{o}, van~der Vaart and van
  Zanten}]{szabo2015}
\bibinfo{author}{Szab\'{o}, B.}, \bibinfo{author}{van~der Vaart, A.W.},
  \bibinfo{author}{van Zanten, J.H.}, \bibinfo{year}{2015}.
\newblock \bibinfo{title}{Frequentist coverage of adaptive nonparametric
  {B}ayesian credible sets}.
\newblock \bibinfo{journal}{Ann. Statist.} \bibinfo{volume}{43(4)},
  \bibinfo{pages}{1391--1428}.
\newblock \bibinfo{note}{DOI: \url{10.1214/14-AOS1270}}.
%Type = Article
\bibitem[{Tierney(1994)}]{tierney1994}
\bibinfo{author}{Tierney, L.}, \bibinfo{year}{1994}.
\newblock \bibinfo{title}{Markov chains for exploring posterior distributions}.
\newblock \bibinfo{journal}{Ann. Stat.} \bibinfo{volume}{22(4)},
  \bibinfo{pages}{1701--1762}.
\newblock \bibinfo{note}{DOI: \url{10.1214/aos/1176325750}}.
%Type = Book
\bibitem[{van~der Vaart(1998)}]{vandervaart1998}
\bibinfo{author}{van~der Vaart, A.W.}, \bibinfo{year}{1998}.
\newblock \bibinfo{title}{Asymptotic statistics}. volume~\bibinfo{volume}{3}.
\newblock \bibinfo{publisher}{Cambridge University Press},
  \bibinfo{address}{Cambridge, UK}.
%Type = Article
\bibitem[{Wang et~al.(2020)Wang, Ullrich and Millstein}]{wang2020}
\bibinfo{author}{Wang, M.}, \bibinfo{author}{Ullrich, P.},
  \bibinfo{author}{Millstein, D.}, \bibinfo{year}{2020}.
\newblock \bibinfo{title}{Future projections of wind patterns in california
  with the variable-resolution cesm: a clustering analysis approach}.
\newblock \bibinfo{journal}{Clim. Dyn.} \bibinfo{volume}{54},
  \bibinfo{pages}{2511--2531}.
\newblock \bibinfo{note}{DOI: \url{10.1007/s00382-020-05125-5}}.
%Type = Article
\bibitem[{Whittle(1957)}]{whittle1957}
\bibinfo{author}{Whittle, P.}, \bibinfo{year}{1957}.
\newblock \bibinfo{title}{Curve and periodogram smoothing}.
\newblock \bibinfo{journal}{J. R. Stat. Soc. Series B Stat. Methodol.}
  \bibinfo{volume}{19(1)}, \bibinfo{pages}{38--63}.
\newblock \bibinfo{note}{DOI: \url{10.1111/j.2517-6161.1957.tb00242.x}}.

\end{thebibliography}
